\documentclass[12pt]{article}
\usepackage{latexsym}
\usepackage{epsfig,amssymb,euscript}
\usepackage{amsmath}
\numberwithin{equation}{section}  

\newsavebox{\ns}
\newsavebox{\dbrane}
\newsavebox{\dbshort}

\def\be{\begin{equation}}
\def\ee{\end{equation}}
\def\bea{\begin{eqnarray}}
\def\eea{\end{eqnarray}}

\DeclareFontFamily{OT1}{pzc}{}
\DeclareFontShape{OT1}{pzc}{m}{it}%
             {<-> s * [1.150] pzcmi7t}{}
\DeclareMathAlphabet{\mathscr}{OT1}{pzc}%
                                 {m}{it}

\newcommand{\nn}{\nonumber}

\newcommand{\eqn}[1]{(\ref{#1})}

\def\Dslash{\,\,{\raise.15ex\hbox{/}\mkern-12mu D}}
\def\Dbarslash{\,\,{\raise.15ex\hbox{/}\mkern-12mu {\bar D}}}
\def\delslash{\,\,{\raise.15ex\hbox{/}\mkern-9mu \partial}}
\def\delbarslash{\,\,{\raise.15ex\hbox{/}\mkern-9mu {\bar\partial}}}
\def\pslash{\,\,{\raise.15ex\hbox{/}\mkern-9mu p}}
\def\calDslash{\,\,{\raise.15ex\hbox{/}\mkern-12mu {\cal D}}}

\newcommand\R{\mathbb{R}}
\newcommand\Z{\mathbb{Z}}

\newcommand\cp{\mathbb{CP}}
\newcommand\C{\mathbb{C}}

\newcommand\diff{\mathrm{d}}
\newcommand\cod{\diff^\dagger\hspace{.7pt}}
\newcommand{\de}{\partial}
\newcommand{\e}{\mathrm{e}}
\newcommand{\vol}{\mathrm{vol}}

\newcommand{\qed}{\nobreak \ifvmode \relax \else \ifdim\lastskip<1.5em \hskip-\lastskip \hskip1.5em plus0em minus0.5em \fi \nobreak \vrule height0.75em width0.5em depth0.25em\fi}

\begin{document}
\begin{titlepage}
\begin{center}
\today

\vskip 1.9cm

{\Large \bf Symmetry-breaking vacua and\\[4.5mm]
baryon condensates in AdS/CFT}\\

\vskip 1.5cm
{Dario Martelli$^{1*}$ ~~and~~ James Sparks$^{2}$}\\
\vskip 0.8cm

1: {\em Institute for Advanced Study\\
Einstein Drive, Princeton, NJ 08540, U.S.A.}\\
\vskip 0.8cm
2: {\em Mathematical Institute, University of Oxford,\\
 24-29 St Giles', Oxford OX1 3LB, U.K.}\\

\vskip 2.4cm

\end{center}

\begin{abstract}
\noindent
We study the gravity duals of symmetry-breaking deformations of superconformal 
field theories, AdS/CFT dual to Type IIB string theory on AdS$_5\times Y$ where 
$Y$ is a Sasaki-Einstein manifold. In these vacua both conformal invariance and
baryonic symmetries are spontaneously broken. We present a detailed discussion 
of the supergravity moduli space, which involves flat form fields on asymptotically 
conical Calabi-Yau manifolds, and match this to the gauge theory vacuum moduli space.
We discuss certain linearised fluctuations of the moduli, identifying the Goldstone 
bosons associated with spontaneous breaking of non-anomalous baryonic symmetries. 
The remaining moduli fields are related to spontaneous breaking of anomalous baryonic symmetries.
We also elaborate on the proposal that computing condensates of baryon operators 
is equivalent to computing the partition function of a non-compact Euclidean D3-brane 
in the background supergravity solution, with fixed boundary conditions at infinity. 
\end{abstract}

\vfill
\hrule width 5cm
\vskip 5mm

{\noindent $^*$ {\small On leave from: \emph{Blackett Laboratory,
Imperial College, London SW7 2AZ, U.K.}}}

\end{titlepage}

\pagestyle{plain}
\setcounter{page}{1}
\newcounter{bean}
\baselineskip18pt

\tableofcontents


\section{Introduction}

The AdS/CFT correspondence \cite{Maldacena:1997re} allows one to understand gauge theory 
dynamics in terms of string theory
on some background spacetime. Properties of strongly coupled gauge theories may then be understood
geometrically, leading to non-trivial predictions on both sides of the correspondence.
In this paper we will be interested in four-dimensional supersymmetric gauge theories that
are dual to Sasaki-Einstein backgrounds in Type IIB string theory. In the infra-red (IR)
these are non-trivial interacting superconformal field theories (SCFTs).
A number of remarkable predictions of AdS/CFT have been confirmed in this case in
recent years. For example, the complete spectrum of mesonic chiral operators
may be computed purely geometrically
\cite{obstructions,Butti:2006nk,Benvenuti:2006qr,Martelli:2006vh} using the results of \cite{MSY1,MSY2}.

The AdS/CFT correspondence describes deformations of CFTs, as well as the
conformal fixed points themselves.
For instance, one may perturb a CFT in the ultra-violet (UV)
 by adding a relevant operator to the Lagrangian. Or one may consider different vacua of the theory,
leading to spontaneous symmetry-breaking. In either case the field theory then flows
 under renormalisation group (RG)
flow to the IR, where interesting dynamics may arise. In this paper we present
a study of symmetry-breaking vacua of ${\cal N}=1$ SCFTs
which are dual to Type IIB backgrounds of the
form AdS$_5\times Y$, where $Y$ is a Sasaki-Einstein five-manifold \cite{Kehagias:1998gn,KW,Acharya:1998db,MP}.
A particular emphasis will be on
the spontaneous breaking of certain global symmetries in these vacua.

The SCFTs that we discuss may be thought of as IR conformal fixed points of certain four-dimensional
supersymmetric \emph{quiver gauge theories}.
The matter content and interactions of a quiver gauge theory
are determined from combinatorial data, namely a directed graph (a quiver) together with a set of closed
paths in this graph, encoded in the superpotential. The gauge theories of interest describe the effective worldvolume theory for D3-branes
placed at an isolated conical Calabi-Yau singularity. The quiver data is then related
to the algebraic geometry of this singularity and its resolutions. There are now
many examples in the literature in which both the quiver gauge theory and
the Calabi-Yau cone metric that describes the IR SCFT are known rather explicitly
\cite{KW,paper2,Bertolini:2004xf,quiverpaper,Cvetic:2005vk,Labc, Benvenuti:2005ja,tilings,Butti:2005sw}.
However, in the paper we aim to keep the discussion as general as possible, without
resorting to specific examples.

A key point about the quiver gauge theories of interest is that they have rather large
classical vacuum moduli spaces (VMSs). The VMS is obtained by minimising the classical
potential of the theory. Since the gauge theory may be engineered by
placing pointlike D3-branes at a Calabi-Yau singularity,
it is expected that the VMS at least contains the corresponding Calabi-Yau.
This is simply because the moduli space of a pointlike object on some manifold should
at least contain that manifold. Similarly, for $N$ D3-branes one expects to find the $N$th symmetric product.
In fact the classical VMS of a
superconformal quiver gauge theory, referred to as the ``master space'' in
\cite{masterspace}, has a rather complicated structure. The centre of the
gauge group, which is a torus $U(1)^{\chi}$, dynamically decouples in the
IR and becomes a \emph{global symmetry} group of the superconformal theory. Here
$\chi\in\mathbb{N}$ denotes the number of nodes in the quiver. A $U(1)^{\chi-1}$
subgroup acts non-trivially on the VMS $\mathscr{M}$.
As we shall  see,
this VMS has a complicated fibration structure.
The fibres are themselves fibrations in which the base space
is a \emph{mesonic moduli space} and the fibres are generically tori
$U(1)^{\chi-1}$. The $U(1)^{\chi-1}$ global symmetry acts in the obvious way on these fibres,
and thus a generic point in $\mathscr{M}$ spontaneously breaks this symmetry.
For the gauge theory on $N$ D3-branes at an isolated Calabi-Yau singularity,
the mesonic moduli spaces are expected to be $N$th symmetric products of
various Calabi-Yau \emph{resolutions} of the singularity. For example,
for a D3-brane at an abelian orbifold singularity it is a rigorous result that
the moduli space $\mathscr{M}$ contains \emph{all} possible Calabi-Yau resolutions
of the orbifold \cite{craw}. The fibration structure of $\mathscr{M}$ contains all of these
resolutions, which is why it is so complicated.
Gauge-invariant chiral operators may be interpreted as the holomorphic functions
on $\mathscr{M}$. The operators carrying non-zero
charge under $U(1)^{\chi-1}$ may be written as determinants of bifundamental fields in the quiver gauge theory, and are therefore often referred to as
\emph{(di)baryon} operators.
The global symmetry group is thus a \emph{baryonic symmetry group}.

One of the main results of the paper will be the identification of the
space $\mathscr{M}$ with a moduli space of certain supergravity backgrounds, as well as the
identification of the $U(1)^{\chi-1}$ symmetry acting on this space.
 Generically, each of these
backgrounds is the gravity dual of a renormalisation group flow from a SCFT in the UV to
another\footnote{Although it is possible to flow to a SCFT with a non-trivial dual Sasaki-Einstein geometry,
as discussed in
\cite{baryonic} for example, for simplicity we will focus here on the case that the IR theory is
 ${\cal N}=4$ super-Yang-Mills (SYM).}
SCFT in the IR.
Here the ``UV theory'' is itself the IR conformal fixed point
of the ${\cal N}=1$ quiver theory  (which is defined up to Seiberg dualities).
One may then think of the RG flow as proceeding in two steps: first one flows from the far UV to a SCFT
which admits an AdS$_5\times Y$
dual description. In a vacuum which spontaneously breaks conformal and baryonic symmetries
of this theory, the RG flow
proceeds towards a new theory in the deep IR.
The supergravity backgrounds we discuss describe the second step, generalising
the analysis of the resolved conifold in \cite{KW2}.

In fact the IR theory will be richer than just a SCFT -- it will also contain additional
low-energy degrees of freedom arising
from the spontaneous breaking of the global baryonic symmetries.  For each of these symmetries we obtain
a corresponding massless Ramond-Ramond (RR) modulus in the supergravity solution,
together with its supersymmetric partner. We interpret these as fluctuations along
the flat directions in the VMS given by acting with a generator of the broken
$U(1)^{\chi-1}$ symmetry group.
An important feature of the theories
that we describe is that only a \emph{subgroup} of this global symmetry
group is an \emph{exact} quantum symmetry. The massless RR
 modes corresponding to these directions may then be identified
as \emph{Goldstone bosons}. This has been studied recently
in \cite{Klebanov:2007cx} for the conifold theory, where $\chi=2$ and the unique
baryonic $U(1)$ symmetry is non-anomalous.
More generally,  the remaining global baryonic symmetries
are anomalous, and are therefore only \emph{approximate} symmetries of the theory.
Of course, the directions in the classical moduli space that are related to
symmetries which are broken by anomalies might be lifted by possible quantum
corrections. It follows that the associated massless modes discussed above
may be lifted in the full quantum theory.
On the other hand, the Goldstone bosons are protected
since they represent motion in directions associated to true symmetries of the quantum theory;
 their existence and masslessness
is guaranteed by Goldstone's theorem. As we discuss,
the supergravity realisation of this statement is that the non-anomalous baryonic
symmetries are dual to \emph{gauge symmetries} in the bulk, which thus cannot possibly be anomalous.
The anomalous symmetries,
on  the other hand, are dual to global symmetries of the supergravity backgrounds that do no come from
any gauge symmetry.

An interesting aspect of these vacua is that baryon operators
acquire non-zero vacuum expectation values (VEVs). As discussed above, classically
the baryon operators may be identified with holomorphic functions on the VMS.
In AdS/CFT these are conjectured \cite{Witten:1998xy,Gukov:1998kn,Berenstein:2002ke} to be
 dual to D3-branes wrapped on three-submanifolds $\Sigma \subset Y$. One cannot therefore
apply the standard
AdS/CFT prescription \cite{Gubser:1998bc,Witten:1998qj}
 to compute their one-point functions in the dual gravity description.
Following a suggestion by E. Witten,
it has been proposed \cite{KM} that in order to compute baryon VEVs one should perform
a path integral over Euclidean
D3-branes with fixed boundary condition at infinity.
Several features of this idea have been succesfully verified in
\cite{baryonic}. In this paper we will investigate in more detail this prescription
for computing condensates of a reasonably large class of baryon
operators. As we will explain, these are the baryon operators whose string duals are D3-branes
wrapped on a smooth\footnote{For toric submanifolds $\Sigma$ of a toric Sasaki-Einstein manifold,
these are Lens spaces.} $\Sigma$ together
with a flat line bundle. Although this is far from being the complete set of baryon operators, it is the simplest
set of baryons which may be described in terms of classical configurations in the bulk. We will evaluate the D3-brane path integral
in the semi-classical approximation, which will reduce the computation to a sum over worldvolume gauge instantons.
A central issue is that, due to the non-compactness
of the D3-brane worldvolume, we will have to pay
particular attention to the transformation properties of the action under gauge transformations
 in the bulk. We will discuss
a prescription to obtain a gauge-invariant, and thus physically meaningful,
function on the supergravity moduli space.
We also discuss the dependence of these baryon condensates on
the Goldstone bosons and other RR moduli.

The outline of the paper is as follows. Section \ref{section2} reviews relevant aspects of quiver
gauge theories arising from D3-branes
at Calabi-Yau singularities. In particular, we include a discussion of
 anomalous $U(1)$s and a description of quiver
gauge theory moduli
spaces. In section \ref{section3} we describe the supergravity backgrounds and their moduli. Linearised fluctuations of these
 supergravity moduli are discussed in section \ref{fluctuationsection}. In section \ref{central} we
 discuss the holographic interpretation
 of the results. 
 Section \ref{condsection} presents
the calculation of baryon condensates using Euclidean D3-branes wrapped on non-compact
 four-submanifolds. In section \ref{dissection} we summarise our results and discuss some open questions and
 directions for future work. Certain technical material is relegated to several appendices.


\section{Quiver gauge theories from Calabi-Yau singularities}
\label{section2}

In this paper we are interested in vacua of field theories that in the UV
are described by superconformal field theories with AdS/CFT duals  AdS$_5\times Y$,
where $(Y,g_Y)$ is a Sasaki-Einstein five-manifold.
There is by now an extremely large class of
examples of such AdS/CFT dualities where both sides of the correspondence
are known explicitly. In these examples, the SCFT conjecturally
arises as the IR fixed point of RG flow for a
quiver gauge theory. The latter is an effective
worldvolume theory for the D3-branes at the singular point of the cone $C(Y)$, describing
the interactions of the lightest string modes. Here the metric cone
\bea
\label{cone}
g_{C(Y)} \, =\,  \diff r^2 + r^2 g_Y
\eea
is Ricci-flat and K\"ahler {\it i.e.} Calabi-Yau.

Throughout the paper we aim for as general a discussion as possible, rather
than focusing on specific examples.
We begin in this section by discussing relevant background material about Calabi-Yau singularities and D-brane quiver
gauge theories. In particular we will be interested in the general structure of the gauge theory vacuum moduli spaces.
Our comments on this will also explain some of the
recent results of \cite{counting} on counting gauge-invariant BPS operators.
Although many of the results of this section
are known, the arguments we present are somewhat more general than those typically
appearing in the literature. We also hope this section will serve as a useful introduction to
the subject.


\subsection{Resolutions of Calabi-Yau cones}\label{metricsection}

A Sasaki-Einstein five-manifold $(Y,g_Y)$ may be defined
by saying that its cone (\ref{cone}) is Ricci-flat K\"ahler.
Roughly speaking, there are currently four
known constructions of such Calabi-Yau cones:
\begin{itemize}
\item Orbifolds $\C^3/\Gamma$, where $\Gamma\subset SU(3)$ is a finite group. The Sasaki-Einstein
link is then simply a quotient $Y=S^5/\Gamma$ of the round five-sphere (one may also take
orbifolds (quotients) of some of the examples below).
\item Complex cones over del Pezzo surfaces $dP_k$, where $k=3,\ldots,8$. Recall that a del
Pezzo surface is the blow-up of $\mathbb{CP}^2$ at $k$ points in general position. Provided
$3\leq k\leq 8$ these admit
K\"ahler-Einstein metrics \cite{tian, tianyau}.
The complex
cone is obtained by taking the canonical line bundle over the del Pezzo and then collapsing
the zero section.
\item Affine toric singularities. Recently a general existence proof has been presented \cite{futakiSE}.
These include the explicit Sasaki-Einstein manifolds $Y^{p,q}$ \cite{paper1, paper2} and $L^{a,b,c}$ \cite{Lprq, Labc}
as links of the singularities.
\item Quasi-homogeneous hypersurface singularities. Again these are existence arguments.
For a review see \cite{BGreview}, or the recent book \cite{BGbook}.
\end{itemize}
Note that some examples fall into more than one of the classes above.
In all cases the cone $C(Y)$ is an affine variety. When we wish to emphasize
the algebro-geometric nature of the Calabi-Yau cone we will denote it by
$C(Y)=Z$ {\it i.e.} $Z$ is the zero set of a set of
polynomials on $\C^D$ for some $D$. $C(Y)$ has an isolated singularity at the point $p=\{r=0\}$
unless $C(Y)=\C^3$.

In this paper we will be interested in \emph{resolutions} of such conical singularities.
This requires two steps. First, we need to resolve the underlying singularity of $Z$,
as a complex variety. Second,
we need to find a Ricci-flat K\"ahler metric on that resolution
that approaches the conical metric at infinity. We now discuss these two steps in more detail.

Technically, a resolution of $Z$ is a smooth variety $X$ together with a proper birational
map $\pi:X\rightarrow Z$, such that $X\setminus E \cong Z\setminus p$
is a biholomorphism for some \emph{exceptional set} $E\subset X$.
Thus, roughly, the singular point $p$ is resolved by replacing it with $E$.
Singularities may always be resolved by a theorem of Hironaka.
However, for our purposes we require the resolution $X$ to be Calabi-Yau;
that is, to have trivial canonical bundle. For reasons that we will not
need to go in to, such a resolution is called \emph{crepant}.
In the first three examples of Calabi-Yau cones above, the
corresponding singularity $Z$ always admits a crepant resolution.
However, the case of quasi-homogeneous hypersurface singularities is quite different.
For example, the constructions reviewed in \cite{BGreview} lead to at least
68 different Sasaki-Einstein metrics on $S^5$. However, as pointed out in
\cite{obstructions}, none of the corresponding hypersurface singularities
admit a crepant (Calabi-Yau) resolution. We suspect that the dual
SCFTs do not admit a description in terms of a quiver gauge theory for these examples.
Some of these 68 metrics on $S^5$ come in quite large families, the largest
having complex dimension 5. This would mean that the dual SCFT has (at least)
a 6-dimensional space of exactly marginal deformations, including the constant
string coupling and its RR axion partner.

Assuming $Z=C(Y)$ is such that it admits a crepant resolution $X$, there are some immediate topological
consequences for $X$. We begin by noting that the singularity $Z$ is \emph{Gorenstein}.
By definition this means that it has a holomorphic $(3,0)$-form $\Omega$ on the smooth part $Z\setminus p$.
In fact existence of a Ricci-flat K\"ahler cone metric on $Z$ implies that $\Omega$ is homogeneous degree 3 under  the radial vector $r\partial/\partial r$.
Thus in particular $\Omega$ is square-integrable
\bea
\int_U \Omega\wedge\bar{\Omega}\, <\, \infty
\eea
in a small neighbourhood $U$ of the singularity $p$
at $r=0$, as one sees by writing the integral in polar coordinates. This implies that the singularity $Z$ is \emph{rational}, and hence $Z$ is a
canonical singularity, in the sense of Reid. See, for example, \cite{youngperson} for an
introduction to these concepts. It follows that for any crepant\footnote{This statement fails for non-crepant
resolutions. For example, the conifold $Z=\{x^2+y^2+z^2+w^2=0\}\subset\C^4$ has a crepant
small resolution $X=O(-1)\oplus O(-1)\rightarrow\cp^1$, which has $b_4(X)=0$, but also has
a \emph{non-crepant} resolution $X=O(-1)\rightarrow \cp^1\times\cp^1$ with $b_4(X)=1$.} resolution $X$
the cohomology groups $H^*(X;\Z)$ of $X$ depend only on $Z$;
that is, any two crepant resolutions $X$ and $X^{\prime}$ have the same cohomology groups \cite{Kollar1, Kollar2}.
For the examples listed above there are often many different choices of crepant resolution.
If we denote
\bea
b_k(X)\, =\, \dim H_k(X;\R)\eea
 the Betti numbers of $X$, then
it was shown in \cite{Caibar} (Theorem 5.2) that $b_1(X)=b_5(X)=b_6(X)=0$. Moreover, as also proven
in \cite{Caibar},
$H^2(X;\Z)$ is isomorphic to the Picard group of $X$, which in turn is generated by holomorphic line bundles.
So all of the second cohomology of $X$ is represented by closed $(1,1)$-forms; these may be
represented by curvature forms on the holomorphic line bundles.

Throughout this paper we will also make the additional \emph{assumption} that
\bea
\label{nob3}
b_3(X)\, =\, 0~.
\eea
This is satisfied by crepant resolutions of the first three types of Calabi-Yau cones
in the list above. In general we do not know of a proof that this must always hold in the
cases of interest in this paper\footnote{We
thank B. Szendr\"oi for discussions on this issue.}.
Later in the paper we will give some further physical justifications
for the assumption (\ref{nob3}).

The assumption (\ref{nob3}) has the following consequences.
Consider the long exact cohomology sequence\footnote{Recall that the relative cohomology groups $H^p(X,Y;\R)$, where $Y=\de X$,
are equivalent to compactly supported cohomology groups $H_\mathrm{cpt}^p(X;\R)$.}
\bea\label{exactly}
0\,&\cong &H^1(Y;\R) \longrightarrow H^2(X,Y;\R)
\longrightarrow H^2(X;\R)
\longrightarrow \nn \\
&&H^2(Y;\R)\longrightarrow H^3(X,Y;\R)\ \cong\ 0~.\eea
Here we have used $b_1(Y)=0$, which follows from Myers' theorem since $(Y,g_Y)$ has positive
Ricci curvature. The last isomorphism follows from  Poincar\'e duality and (\ref{nob3}).  
The exact sequence (\ref{exactly}) implies, using Poincar\'e duality
and the universal coefficients theorem, that\footnote{For toric $X$ this
relation may also be proven using Pick's theorem, by triangulating the toric
diagram \cite{Benvenuti:2005qb}. In fact (\ref{rach}) still holds even when $b_3(X)\neq 0$
\cite{Caibar}, although the argument is then much more involved.}
\bea
\label{rach}
b_3(Y) \, = \, b_2(X)-b_4(X)~.
\eea
This relation will be important throughout the paper.
We also note that the Euler characteristic of $X$ is given by
\bea
\label{Euler}
\chi=\chi(X)\equiv\sum_{i=0}^6 (-1)^i\dim\, H_i(X;\R) =  1+ b_2(X) + b_4(X) ~.
\eea

Having chosen a Calabi-Yau resolution $X$, we would like to put a
Ricci-flat K\"ahler metric on it that is asymptotically conical {\it i.e.}
approaches the cone metric (\ref{cone}) near infinity.
If $X$ were \emph{compact}, Yau's theorem \cite{yau} would imply that
$X$ has a unique Ricci-flat K\"ahler metric in each K\"ahler class
in $H^{2}(X;\R)$. Unfortunately, there is currently no general analogue of Yau's theorem
for existence and uniqueness of asymptotically
conical metrics. However,
provided the boundary conditions are right, one generally expects
results about compact manifolds to hold also for non-compact
manifolds, and we believe that being asymptotically a cone is
precisely such a good boundary condition. We state this as a conjecture
we shall assume:

\

\indent {\it Conjecture}: If $\pi:X\rightarrow Z$ is a crepant resolution
of an isolated singularity $Z=C(Y)$, where $C(Y)$ admits a Ricci-flat K\"ahler cone metric,
then $X$ admits a unique Ricci-flat K\"ahler metric in each K\"ahler class in
$H^2(X;\R)$ that is asymptotic to a cone over the Sasaki-Einstein manifold $(Y,g_Y)$.

\

Despite the lack of a general theorem, there are nonetheless several important
results that go some way in this direction\footnote{Since submitting the 
first draft of this paper to the archive, the paper \cite{craigvanC} has appeared 
giving a proof of this conjecture in the case that the cohomology class 
of the K\"ahler form $\omega$ is compactly supported, {\it i.e.} $[\omega]\in H^2_{\mathrm{cpt}}(X;\R)\subset 
H^2(X;\R)$. This substantially generalises the results of Joyce and Tian-Yau 
mentioned below.}. In \cite{joyce}, Joyce
 proves that any crepant resolution $X$ of an orbifold
$\C^3/\Gamma$, with $\Gamma\subset SU(3)$, admits a unique
asymptotically conical Ricci-flat K\"ahler metric that is
asymptotic to a cone over $S^5/\Gamma$. Thus the result we want
is true for the simplest class of Sasaki-Einstein five-manifolds, namely quotients of the round five-sphere.
On the other hand, in \cite{tianyau2} (see also \cite{bando})
it is proven that,
under certain mild technical assumptions, $X=\bar{X}\setminus D$ admits
a Ricci-flat K\"ahler metric that is asymptotic to a
cone, provided that the divisor $D\subset \bar{X}$ in the compact
K\"ahler manifold (or orbifold) $\bar{X}$ admits a K\"ahler-Einstein
metric of positive Ricci curvature. Note here that one is essentially
compactifying $X$ to $\bar{X}$ by adding a divisor $D$ at infinity.
In fact, it is a conjecture of Yau that every complete Ricci-flat K\"ahler
manifold may be compactified this way.
The metrics of \cite{tianyau2, bando} are  asymptotic to cones over regular,
or quasi-regular, Sasaki-Einstein manifolds. This result has very recently
been extended in \cite{craig} to the case that is $D$ toric, but does
not necessarily admit a K\"ahler-Einstein metric. Thus these are complete
Ricci-flat K\"ahler metrics that are asymptotic to irregular toric Calabi-Yau cones.
There are also a handful
of explicit constructions, including: cohomogeneity one
ans\"atze, including the resolved conifold metric on $O(-1)\oplus O(-1)\rightarrow \cp^1$
\cite{candelas}; the Calabi ansatz \cite{Calabi}
and its variants, for example studied in \cite{BB,PP} and more recently
extended by Futaki in \cite{Futaki}; and, finally, constructions using Hamiltonian
two-forms \cite{japanese, lupope, np1}. However, all of these
latter explicit constructions are rather special, and rely on the presence of
certain symmetries. In all cases, the Einstein equations
reduce to solving ODEs, which is why it is possible to find explicit solutions.

We regard the above paragraph as rather convincing evidence for the conjecture.
By assuming its validity we shall also obtain a consistent picture of the space of
supergravity solutions that are dual to the supersymmetric symmetry-breaking vacua
of interest.


\subsection{Quivers and fractional branes}\label{quiversection}

An interesting problem is to determine the effective worldvolume theory for D-branes placed at the
singular point $\{r=0\}$ of the cone $C(Y)$. In general this is very hard. However, if the
 Ricci-flat K\"ahler cone is either an orbifold $\C^3/\Gamma$, a complex cone
over a del Pezzo surface, or a toric variety, then this worldvolume theory is believed to be
a quiver gauge theory. The orbifold case is understood best \cite{douglas}, where
the gauge theory may be constructed via standard orbifold techniques and leads to
the Mckay quiver. For del Pezzo surfaces the quiver may be constructed
from a special type of exceptional collection of sheaves on the del Pezzo surface
-- see, for example, \cite{cachazo} and in particular \cite{herzogcollection}. For toric varieties the quiver theory is believed to
be described by a certain bipartite tiling of a two-torus called a dimer. For a
recent review, see \cite{kennaway}, \cite{yamazaki}. Note that in all these cases the
singularity $Z=C(Y)$ indeed admits a crepant resolution $X$ with $b_3(X)=0$.
In this paper we shall assume our Calabi-Yau cone singularity $C(Y)$ is such
that the worldvolume theory of $N$ D3-branes at the singularity
is described by a quiver gauge theory. This includes all of the above-mentioned cases.

A quiver is simply a directed graph.
If $V$ denotes the set of vertices and $A$ the set of arrows, then we have head and tail
maps $h,t:A\rightarrow V$. A \emph{representation} of a quiver is
an assignment of a $\C$-vector space $U_v$ to each vertex $v\in V$ and
a linear map $\phi_a:U_{t(a)}\rightarrow U_{h(a)}$ for each arrow $a\in A$.
In particular, to specify a representation we must specify a dimension
vector $\mathbf{n}\in \mathbb{N}^{|V|}$, so $\dim_\C U_v = n_v$. This data also
leads to the notion of a quiver gauge theory. This is an $\mathcal{N}=1$
gauge theory in four dimensions specified as follows:
\begin{itemize}
\item The gauge group \bea\label{Ugroup} G=\prod_{v\in V} U(n_v)\eea is a product of unitary groups.
\item To each arrow $a\in A$ we associate a chiral superfield $\Phi_a$ transforming
in the fundamental representation of the gauge group $U(n_{h(a)})$ and the anti-fundamental representation
of the gauge group $U(n_{t(a)})$. The fields are therefore often called bifundamental fields.
\item In addition one must specify a superpotential \bea W=\sum_{l=a_1\cdots a_k\in L} \lambda_l \,
\mathrm{Tr}[\Phi_{a_1}\cdots \Phi_{a_k}]~.\eea Here $L$ is a set of closed oriented paths
in the quiver, so  $l=a_1\cdots a_k$ denotes such a loop with $h(a_1)=t(a_k)$. The fact the loop is closed
allows one to take a trace to obtain a gauge-invariant object. The complex numbers $\lambda_l$ are
coupling constants.
\end{itemize}

The relation between the singularity $Z$ and the quiver is a large technical subject
which is still not very well-understood.
It is clearly very difficult
to attack the problem of determining the worldvolume theory directly since we cannot
quantise strings on $C(Y)$. However, one can circumvent this problem to some extent by replacing
the Type IIB string by the
topological string. The latter is independent of the K\"ahler class,
and so does not see any difference between $Z=C(Y)$ and the crepant resolution $(X,g_X)$.
Moreover, the topological string is sufficient for addressing certain questions
which are holomorphic in nature, such as the matter content and superpotential above.
Space-filling D-branes on $X$ are described in terms of the topological string as
coherent sheaves on $X$, or more precisely its derived category. This is
the mathematical way to understand the problem, which is then defined purely
algebro-geometrically. However, we will not go into the details of this here.

For our purposes, all that we need to know is that there conjecturally exists a special set of D-branes,
called \emph{fractional branes}, which form a basis for all other D-branes at the singularity.
Once we have resolved the singularity to a large smooth $(X,g_X)$ one may
think of D-branes as submanifolds of spacetime $\R^{1,3}\times X$ on which
open strings may end. The space-filling D-brane charges on $X$ are
then determined by their homology\footnote{More precisely they are K-theory classes, but
we will ignore this subtlety.} class in $H_*(X;\Z)$.
A complete basis therefore requires
$1+b_2(X)+b_4(X)$ fractional branes, corresponding to the charge of a
D3-brane, and wrapped D5-branes and D7-branes, respectively.
From (\ref{Euler}) this is the Euler characteristic $\chi=\chi(X)$ of $X$.
The nodes of the quiver are in 1-1 correspondence with these
fractional brane basis elements. Thus \bea |V|=\chi~.\eea The charges of any D-brane on $X$ may
then be used to expand the D-brane in terms of this basis.
The ranks of the gauge groups $\mathbf{n}\in\mathbb{N}^{|V|}$ in the quiver are
the coefficients in this expansion. More precisely, this identifies
the unitary group factor $U(n_v)$ in (\ref{Ugroup}) as
the gauge group on the $v$th fractional brane. The bifundamental
fields $\Phi_a$ describe the massless strings stretching between the
fractional branes. Thus the quiver gauge theory is essentially a description
of a D-brane at the singular point of $Z=C(Y)$ as a marginally bound state
of the fractional branes, which should be mutually supersymmetric at the singular point.

Since a quiver gauge theory is in general chiral, it will typically suffer from various
anomalies. In particular, gauge anomaly cancellation for the gauge group $U(n_v)$,
corresponding to a triangle diagram with three gluons for this gauge group,
is equivalent to
\bea
\label{gaugeanomaly}
\sum_{a\in A\mid h(a)=v} n_{t(a)} - \sum_{a\in A\mid t(a)=v} n_{h(a)}\, =\, 0
\eea
for all $v\in V$. Note this is $|V|$ equations for $|V|$ variables $n\in \mathbb{N}^{|V|}$. A gauge anomaly would of course lead to an inconsistent quantum
theory, so one may wonder where the condition (\ref{gaugeanomaly}) comes from in
string theory. As originally pointed out for del Pezzo surfaces in \cite{cachazo}, this condition
should be understood simply as charge conservation on $X$. In general,
a space-filling D-brane wrapped on a $(k-3)$-submanifold $\Sigma_{k-3}\subset X$ has a
charge in $H^{9-k}(X,Y;\R)$.
Here $k=3,5,7$ are the possible D$k$-branes.
A D$k$-brane is a magnetic source for the RR flux $G_{8-k}$, which thus satisfies
\bea\label{bianchiforG}
\diff G_{8-k} \,=\, \frac{2\pi M}{\mu_k}\,\delta(\Sigma_{k-3})~.
\eea
Here
\bea\label{Dcharge}
\mu_k \,=\, \frac{1}{(2\pi)^k\alpha'^{(k+1)/2}}
\eea
is the charge of a D$k$-brane, $\delta(\Sigma_{k-3})$ is a delta-function representative of the Poincar\'e dual to
$\Sigma_{k-3}$ in $X$, and $M$ is the number of wrapped branes.
Thus $[\delta (\Sigma_{k-3})]\in H^{9-k}(X,Y;\R)$ represents the charge
of a single space-filling D-brane on $\Sigma_{k-3}$. The modified Bianchi
identity (\ref{bianchiforG}) implies that the image of this in $H^{9-k}(X;\R)$
is zero. There is a long exact cohomology sequence
\bea\label{exactlong}
\cdots \longrightarrow H^{8-k}(Y;\R) \stackrel{\beta}{\longrightarrow} H^{9-k}(X,Y;\R)
\longrightarrow H^{9-k}(X;\R)
\longrightarrow \cdots ~.\eea
Thus the only allowed D-brane charges on $X$ are elements of $H^{9-k}(X,Y;\R)$
that are images under $\beta$ of $H^{8-k}(Y;\R)$. The latter group measures
the flux of $G_{8-k}$ at infinity.

In the case at hand, we have $k=3,5,7$. It is easy to show that $\beta(H^5(Y;\R))\cong \R$,
 $\beta(H^3(Y;\R))\cong\R^{b_3(Y)}$, since $b_3(X)=0$, and $\beta(H^1(Y;\R))\cong0$.
Here the last
relation follows since $H^1(Y;\R)=0$ by Myers' theorem.

Thus there is only a $(b_3(Y)+1)$-dimensional space of space-filling D-brane charges.
The anomaly cancellation condition (\ref{gaugeanomaly}) is identified with this
charge conservation condition. This implies there is a $(b_3(Y)+1)$-dimensional
space of solutions to the $|V|=\chi=1+b_2(X)+b_4(X)$ linear equations in (\ref{gaugeanomaly}).
In other words, the skew part of the adjacency matrix of the quiver
has kernel of dimension $(b_3(Y)+1)$. We shall use this later in the paper.


\subsection{Anomalous $U(1)$s}
\label{anomaloussection}

Throughout the paper a crucial role is played by the $|V|=\chi$ central $U(1)$ factors in the gauge group
$G=\prod_{v\in V}U(n_v)$. In the quiver gauge theory these are dynamical $U(1)$ gauge fields.
However, there are anomalies in addition to those already discussed in the previous subsection,
namely mixed $\mathrm{Tr} [SU(n_{v})^2U(1)_{v'}]$ triangle anomalies.
As is well-known, such anomalies
often occur in string theory, and are cancelled via a form of Green-Schwarz mechanism.
The anomalous combinations of $U(1)$ gauge fields become massive in the process,
and are described by the St\"uckelberg action.  This has recently been discussed in some detail
for the case of del Pezzo singularities in \cite{Morrison}.
Due to the importance of the
anomalous $U(1)$s in our later discussion of AdS/CFT,  we shall here review the salient
features. At the same time this will allow us to generalise the del Pezzo results of
\cite{Morrison} to any asymptotically conical Ricci-flat K\"ahler $(X,g_X)$.
In particular, the St\"uckelberg scalars are related to  certain $L^2$ harmonic forms on $(X,g_X)$,
which we show indeed exist by appealing to a recent mathematical result.
More practically, this section will allow us to review various properties of RR fields and their
gauge symmetries, and also introduce notation used later in the paper.

We begin by noting that on the spacetime $\mathcal{M}=\R^{1,3}\times X$ one is free to turn on
various fields without affecting the background metric. Firstly, there is the constant dilaton field $\phi$,
which determines the string coupling constant $g_s=\exp \phi$. This is
paired under the $SL(2;\R)$ symmetry of Type IIB supergravity with
a constant axion field $C_0$. These combine into the axion-dilaton
\bea
\label{axdil}
\tau\, =\, C_0 + i\exp(-\phi)~.
\eea

Secondly, we may turn on various flat form fields. In particular, we may turn on a
flat $B$ field, and flat RR fields $C_2$ and $C_4$ on $X$. The classification
of such fields, up to gauge equivalence, is discussed in appendix \ref{appendixB}.
However, in the presence of a non-trivial $B$ field the gauge transformations of
the RR fields are twisted.
Recall that the gauge symmetries of string theory require that
gauge transformations of RR potentials are  also
accompanied by transformations of higher rank potentials. Consider, for instance, the $SL(2;\Z)$, or equivalently large gauge,
transformation
\bea\label{shiftC0}
C_0 \rightarrow C_0 + 1~.
\eea
This must be accompanied by the transformations
\bea
C_2 \rightarrow C_2 + B~, \qquad \qquad
C_4 \rightarrow C_4 + \frac{1}{2} B \wedge B~.
\eea
One way to see this is to note that the gauge-invariant RR field strength may be written as
\bea
\tilde{G} = \diff C - H_3 \wedge  C = \e^{B}\, \diff (\e^{-B} C)~.
\label{multiRR}
\eea
$C$ is a formal sum of RR form potentials
\bea
C = \sum_{p\geq 0} C_{2p}~.
\eea
The combination $C\, \e^{-B}$, which appears in the Chern-Simons
action of D-branes to be discussed below, transforms as
\bea\label{twistme}
C\, \e^{-B}\rightarrow C\, \e^{-B} +\diff\Lambda
\eea
where $\Lambda$ is a formal sum of odd-degree forms.
General gauge transformations may  also be written this way,
if one allows $\diff\Lambda$ to be any closed form with appropriately
quantised periods\footnote{$\Lambda$ is then
roughly instead a formal sum of connection forms on
gerbes.}.

In the case at hand, $\mathcal{M}=\R^{1,3}\times X$ is contractible to $X$. Thus we  may turn on the following
(non-torsion) flat fields
\bea\label{flat2}
C_2\, =\, \frac{2\pi}{\mu_1}\sum_{M=1}^{b_2(X)}c_2^M \Upsilon^M~,\qquad \qquad
B\, = \,\frac{2\pi}{\mu_1}\sum_{M=1}^{b_2(X)} b^M\Upsilon^M\eea
\bea\label{flat4}
C_4 \,=\, \frac{2\pi}{\mu_3}\sum_{A=1}^{b_4(X)}c_4^A\, \Xi^A~.
\eea
The factors of $\mu_k$ are related to the normalisation of large gauge transformations, which are in turn
determined by the D-brane Wess-Zumino couplings.
The $\Upsilon^M$ are closed two-forms with integer periods,  generating
the lattice $H^2_{\mathrm{free}}(X;\Z)\cong H^2(X;\Z)/H^2_{\mathrm{tors}}(X;\Z)$. Similarly,
the $\Xi^A$ are closed four-forms with integer periods, generating
the lattice $H^4_{\mathrm{free}}(X;\Z)$. Before taking into account large gauge transformations we may view
the flat RR fields as a vector
\bea
(C_0,[C_2],[C_4])\in H^0(X;\R)\oplus H^2(X;\R)\oplus H^4(X;\R)\cong \R\oplus \R^{b_2(X)}\oplus \R^{b_4(X)}\cong \R^{\chi}~.
\eea
The lattice of large gauge transformations is
\bea
\Lambda_B^X &=& \Bigg\{\left(n,\frac{2\pi}{\mu_1}\sigma+nB,\frac{2\pi}{\mu_3}\kappa+\frac{2\pi}{\mu_1}\sigma\wedge B+ \frac{1}{2}nB\wedge B\right)\mid n\in\Z, \nn\\
&&\sigma\in H^2_{\mathrm{free}}(X;\Z), \kappa\in H^4_{\mathrm{free}}(X;\Z)\Bigg\}\subset\R^{\chi}~.
\eea
Thus the  flat RR fields, parameterised by the $\chi$ constants
$C_0$ and $c_2^M$, $c_4^A$ in (\ref{flat2}), (\ref{flat4}), respectively, live in the twisted torus
\bea
(C_0,[C_2],[C_4])\in \R^{\chi}/\Lambda_B^X~.
\eea

Once we have resolved the singularity to a large smooth $(X,g_X)$, the $\chi$
fractional branes may be described as certain space-filling D3-D5-D7 bound states.
In the remainder of this subsection we study the dynamics of the $U(1)$
gauge fields on $\R^{1,3}$ using this large-volume description, essentially
following \cite{Morrison}.
We focus on a single fractional brane, and assume for simplicity
that it has a non-zero D7-brane charge. We shall denote the compact four-submanifold
in $X$ that the brane wraps by $\Sigma_4$, which gives rise to a homology
class $[\Sigma_4]\in H_4(X;\Z)$. At large volume the worldvolume theory
of the fractional brane is described by the Born-Infeld and Chern-Simons actions.
The Born-Infeld action is
\bea
S_{BI}\,=\, -T_7\int_{\mathbb{R}^{1,3}\times \Sigma_4} \diff \sigma^8\, \mathrm{Tr}\sqrt{-\det(h+2\pi\alpha' F -B)}~.
\eea
Here $T_k$ is the D$k$-brane tension, related to the charge (\ref{Dcharge}) by
\bea\label{Dtension}
g_sT_k\, =\, \mu_k~,
\eea
and $\sigma_{\alpha}$, $\alpha=0,\ldots,7$ denote worldvolume coordinates.
$h$ denotes the induced metric on the worldvolume from its embedding into
spacetime $\mathcal{M}=\R^{1,3}\times X$, $B$ is the pull-back of the NS two-form,
and $F$ denotes the curvature of a $U(n)$ gauge field for a worldvolume gauge bundle $E$ of rank $n$.
The induced metric and the $B$ field are understood to be multiplied by a unit
$n\times n$ matrix in these formulae.  Recall that B is not gauge-invariant, but rather transforms as $B\rightarrow B+\diff\lambda$
where $\lambda$  is a one-form.  In fact large gauge transformations
may also be included if $\mu_1\lambda$ is taken to be a connection one-form on
some line bundle over the spacetime $\mathcal{M}$. Thus $[\mu_1\diff\lambda/2\pi] \in
H^2(\mathcal{M};\Z)$. At the same time the worldvolume gauge field $F$
transforms as
\bea
F\rightarrow F+\mu_1\iota^*\diff\lambda ~,\eea
where $\iota$ denotes the embedding. Again, for non-abelian $F$ a unit $n\times n$  matrix is understood in this formula.

The Chern-Simons terms are given by\footnote{See, for example, \cite{johnson}. The normal
bundle couplings are given in \cite{MM}.}
\bea\label{CScoupling}
S_{CS}=\mu_7\int_{\R^{1,3}\times \Sigma_4} C \, \mathrm{Tr}\, \e^{2\pi\alpha'F-B}\,
\sqrt{\frac{\hat{\mathcal{A}}(4\pi^2\alpha'R_T)}{\hat{\mathcal{A}}(4\pi^2\alpha'R_N)}}~.\eea
Here $\mu_7$ is the D7-brane charge (\ref{Dcharge})
and the curvature terms will play no role in our discussion, so we shall ignore them. The topology of the
 gauge bundle $E$ over $\Sigma_4$ induces D5-brane
 and D3-brane charges on the D7-brane via (\ref{CScoupling}).

The worldvolume gauge field $A$, with field strength $F$, dimensionally reduces to a $U(n)$ gauge field on
$\mathbb{R}^{1,3}$. Since we are only interested in the $U(1)$s
we study here only the abelian part of this gauge field, which we denote $\mathcal{A}$. Its
field strength is $\mathcal{F}$. At low energies this has a standard kinetic term
\bea
-\frac{1}{4g^2}\int_{\R^{1,3}} \mathcal{F}\wedge *_4\mathcal{F}
\eea
where the gauge coupling $g$ may be related to the Born-Infeld volume of $\Sigma_4$.
The flat background fields (\ref{flat2}), (\ref{flat4}), together
with the topology of the gauge bundle $E$ over $\Sigma_4$, also induce an effective $\theta$-angle term
\bea
\frac{1}{32\pi^2} \int_{\R^{1,3}} \theta\, \mathcal{F}\wedge\mathcal{F}
\eea
where
\bea\label{thetaangle}
\frac{\theta}{8\pi} &= &\int_{\Sigma_4} \Bigg\{C_0\left[\mathrm{ch}_2(E)-\mathrm{ch}_1(E)\wedge b^M\Upsilon^M
+\frac{1}{2}\mathrm{ch}_0(E)b^Mb^N\Upsilon^M\wedge\Upsilon^N\right]\nn\\
&& + \, c_2^M\Upsilon^M\wedge\left[\mathrm{ch}_1(E)-\mathrm{ch}_0(E)b^N\Upsilon^N\right]+c_4^A\mathrm{ch}_0(E)\Xi^A\Bigg\}~.
\eea
Summation is understood over repeated indices. $\mathrm{ch}(E)$ denote the Chern characters of the bundle $E$, so in particular $\mathrm{ch}_0(E)=n$
is the rank of the gauge bundle, or equivalently number of D7-branes. Note that the above, slightly
technical, discussion of large gauge transformations in string theory is crucial
for seeing that the expression (\ref{thetaangle}) is a well-defined angle.

Now consider fluctuations of the background form fields. If one has a $k$-form field $C_k$
on $\R^{1,3}\times X$ then one will obtain a massless dynamical scalar
field $\varphi$ on $\R^{1,3}$ via an ans\"atz
\bea
C_k \, =\,  \varphi \, \psi
\eea
provided the $k$-form $\psi$ is closed, co-closed and $L^2$ normalisable on $(X,g_X)$.
The last condition ensures that the kinetic energy of $\varphi$ is finite.
Thus in particular $\psi$ is an $L^2$ harmonic $k$-form on $(X,g_X)$.
We denote the space of such forms by $\mathcal{H}^k_{L^2}(X,g_X)$.
For $(X,g_X)$ asymptotically conical, the number of such
harmonic forms is known
\cite{Hausel}. Theorem 1A of the latter reference says that for a complete
asymptotically conical manifold $(X,g_X)$ of real dimension $m$ with boundary $\partial X$ the following
natural isomorphisms\footnote{Here $f$ is the forgetful map
$H^k(X,\de X;\R) \stackrel{f}{\longrightarrow} H^k(X;\R)$, that forgets that a class
has compact support.} hold:
\bea
\label{tamas}
\mathcal{H}^k_{L^2}(X,g_X) \cong \left\{
\begin{array}{ll} H^k(X,\partial X;\R), & k<m/2 \\
f(H^{m/2}(X,\partial X;\R))\subset H^{m/2}(X;\R), & k=m/2 \\
H^k(X;\R), & k>m/2\end{array}\right.~.
\eea
Thus the space of $L^2$ harmonic forms is topological.
It follows that the only $L^2$ harmonic forms on $(X,g_X)$
are  $\mathcal{H}^2_{L^2}(X,g_X)\cong H^2(X,\partial X;\R)\cong H_4(X;\R)$ and
$\mathcal{H}^4_{L^2}(X,g_X)\cong H^4(X;\R)$. There are hence
$b_4(X)$ $L^2$ harmonic two-forms and four-forms
on $(X,g_X)$, respectively. Since $X$ is complete and asymptotically a cone, these
forms are also closed and co-closed.

Thus only $b_4(X)$ of the $b_2(X)$ constants in (\ref{flat2}) may be interpreted as VEVs
of massless dynamical axions in $\R^{1,3}$,  whereas all of the
constants $c_4^A$ in (\ref{flat4}) are VEVs of massless dynamical axions.
We focus in the following only on the RR fields, and write the dynamical fields
\bea
C_2 \,=\, \frac{2\pi}{\mu_1}c_2^A \Upsilon^A~, \qquad \qquad C_4\, =\, \frac{2\pi}{\mu_3}
c_4^A\, \Xi^A
\eea
where now $c_2^A$ and $c_4^A$ are massless scalar fields on $\R^{1,3}$, and
$\Upsilon^A \in \mathcal{H}^2_{L^2}(X,g_X)$, $\Xi^A\in \mathcal{H}^4_{L^2}(X,g_X)$.
In fact since the Hodge dual of an $L^2$ harmonic form is also an $L^2$ harmonic form, we may clearly take
\bea
\Xi^A = R^{AB}*_6\Upsilon^B
\eea
for some constant matrix $R=(R^{AB})\in GL(b_4(X);\R)$, where $*_6$ denotes the Hodge operator
on $(X,g_X)$. Recall that
RR fields are self-dual, so we must also turn on
\bea
C_6 \,=\, \frac{2\pi}{\mu_5} \tilde{c}_2^A \wedge\Xi^A
\eea
and more correctly write
\bea
C_4 \,=\, \frac{2\pi}{\mu_3} (c_4^A \, \Xi^A + \tilde{c}_4^A \wedge\Upsilon^A)~.
\eea
Here $\tilde{c}_2^A$ and $\tilde{c}_4^A$ are dynamical two-form potentials on $\R^{1,3}$. Self-duality then requires
\bea\label{dualboys}
\diff c_2^A \, =\,  \frac{\mu_1}{\mu_5}R^{BA}*_4\diff \tilde{c}_2^B,\qquad \qquad \diff \tilde{c}_4^A \,=\,  R^{BA}*_4\diff c_4^B~.
\eea
Note also that $C_2$ itself describes a
 two-form on $\R^{1,3}$. Altogether these terms produce a coupling
\bea
\int_{\R^{1,3}} c\wedge \mathcal{F}
\eea
where
\bea\label{c}
c &= &\frac{\mu_1}{2\pi} C_2\int_{\Sigma_4} \left[\mathrm{ch}_2(E)-\mathrm{ch}_1(E)\wedge b^M\Upsilon^M
+\frac{1}{2}\mathrm{ch}_0(E)b^Mb^N\Upsilon^M\wedge\Upsilon^N\right]\nn\\
&&+\, \tilde{c}_4^A\int_{\Sigma_4} \Upsilon^A\wedge\left[\mathrm{ch}_1(E)-\mathrm{ch}_0(E)b^M\wedge\Upsilon^M\right]
+\tilde{c}_2^A\int_{\Sigma_4}\mathrm{ch}_0(E)\Xi^A~.
\eea

The interesting part of the effective Lagrangian for the $U(1)$ gauge field $\mathcal{A}$ on $\R^{1,3}$ is then
\bea\label{lagrange}
\mathcal{L}=-\frac{1}{4g^2}\mathcal{F}\wedge *_4\mathcal{F}+c\wedge\mathcal{F}+\frac{\theta}{32\pi^2}\mathcal{F}\wedge\mathcal{F}
\eea
where the two-form $c$ and scalar $\theta$ are given by (\ref{c}) and (\ref{thetaangle}), respectively.
The precise formulae are not particularly important. The important point to notice is that
$c$ is linear in the $1+b_4(X)+b_4(X)=\chi-b_3(Y)$ variables $C_2$, $\tilde{c}_2^A$ and $\tilde{c}_4^A$,
respectively.  The field $C_2$ is non-dynamical since  $(X,g_X)$ has infinite volume.
Thus, for fixed $B$ field flux $b^M$ on $X$, $c$ depends linearly on $2b_4(X)$ dynamical
two-forms.
There are $\chi$ different fractional branes, wrapping different cycles in $X$ or with different gauge bundles $E$, each with a
gauge field $\mathcal{A}_s$, $s=1,\ldots,\chi$. By taking linear combinations of fractional $U(1)$s with
no $C_2$ term in $c$  we obtain generically $2b_4(X)$ linearly independent gauge fields with an effective Lagrangian of the form (\ref{lagrange}) with $c\neq0$,
and also a finite  kinetic term for $c$. Here the kinetic term for $c$ comes from the
bulk IIB supergravity kinetic terms for the RR fields.
A standard change of variable then shows
that the dual scalar to $c$ in $\R^{1,3}$ is a St\"uckelberg field, and thus (\ref{lagrange})
describes the action for a \emph{massive gauge field}. Indeed, the equation of
motion for $c$, in the presence of a $c\wedge \mathcal{F}$ coupling, is given by
\bea\label{cEOM}
\diff *_4\diff c = \mathcal{F}~.
\eea
A duality transformation to a scalar field $\rho$ involves interchanging equations of
motion and Bianchi identities. Thus one defines $\rho$ satisfying
\bea\label{defnrho}
\diff\rho = *_4\diff c-\mathcal{A}
\eea
so that (\ref{cEOM}) is automatic. The equation of motion for $\rho$ is then
\bea
\diff *_4(\diff\rho+A)=0~.
\eea
The definition (\ref{defnrho}) implies that a gauge transformation $\mathcal{A}\rightarrow \mathcal{A}+\diff\lambda$ must be
accompanied by a transformation $\rho\rightarrow\rho-\lambda$. Thus $\rho$ is a St\"uckelberg
scalar, and the gauge field $\mathcal{A}$ is in fact massive. The dual scalar is,
from (\ref{dualboys}), a linear combination of the dynamical scalars $c_2^A$, $c_4^A$,
which in turn enter the expression for the $\theta$-angle (\ref{thetaangle}).
Thus the $2b_4(X)$ St\"uckelberg fields that give masses to $2b_4(X)$ of the fractional brane
$U(1)$s are linear combinations of $\theta$-angles. It is precisely this fact that
allows the additional triangle anomalies to be cancelled.
On the other hand, the
$\chi-2b_4(X)=1+b_3(Y)$ linear combinations of fractional brane $U(1)$s with no
$\tilde{c}_2^A$ and $\tilde{c}_4^A$ fields in $c$ remain massless.
Note that the overall $U(1)$, often referred to as the centre of mass $U(1)$,
should essentially decouple from everything.

We end this section with a comparison to the quiver gauge theory formula for anomaly
cancellation.
Let $q\in\Z^{|V|}$ denote a vector of charges specifying a subgroup $U(1)_{q}\subset U(1)^{|V|}\subset G$.
Recall that $|V|=\chi$. Then cancellation of the triangle
anomaly $\mathrm{Tr}[U(1)_{q} SU(n_{v})^2]$ is equivalent to
\bea\label{triangle}
\sum_{a\in A\mid h(a)=v} n_{t(a)}q_{t(a)}-\sum_{a\in A\mid t(a)=v} n_{h(a)} q_{h(a)}=0~.
\eea
The following neat argument is due to \cite{herzog}. Recall that anomaly cancellation (\ref{gaugeanomaly}) for the gauge group $G=\prod_{v\in V} U(m_v)$
requires
\bea\label{gaugeanomalym}
\sum_{a\in A\mid h(a)=v} m_{t(a)} - \sum_{a\in A\mid t(a)=v} m_{h(a)}=0
\eea
for all $v\in V$. As explained in the previous subsection, there exist $b_3(Y)+1$ linearly independent solutions to
(\ref{gaugeanomalym}). We may thus solve (\ref{triangle}) by setting
\bea
q_v=\frac{m_v}{n_v}
\eea
assuming $n_v\neq 0$ for all $v\in V$ for the D3-brane worldvolume theory of interest.
A discussion of why this should be the case may be found in \cite{herzog}.
Note, however, that taking $\mathbf{m}=\mathbf{n}$  leads to $q=(1,1,\ldots,1)$.
This corresponds to the overall diagonal $U(1)$ under which nothing is charged.
Thus one has $b_3(Y)$ non-anomalous $U(1)$s, precisely as we have argued
above using a large-volume description of the fractional branes.


\subsection{Marginal couplings and superconformal quivers}
\label{marginalsection}

In the IR all $U(1)$s dynamically decouple, anomalous or otherwise. The overall diagonal $U(1)$
completely decouples from everything as nothing is charged under it.
The massless non-anomalous $U(1)$s decouple since their gauge coupling goes to zero in the IR,
while the massive $U(1)$s decouple because they are massive.
In the IR theory we will therefore encounter only global $U(1)$ symmetries, and these likewise split as
non-anomalous and
anomalous, as global symmetries. The IR gauge group will thus be
\bea\label{SUgroup}
SG\equiv\prod_{v\in V}SU(n_v)~.
\eea

A necessary condition for an IR fixed point of the quiver gauge theory is that the $\beta$ functions
of all coupling constants vanish. For a quiver gauge theory the vanishing of the NSVZ $\beta$ functions, which are exact in perturbation theory,
is given by\footnote{The expressions $\hat{\beta}$ in these formulae are not the actual NSVZ $\beta$ functions, but are rather proportional to them. }
\bea\label{betafunctions}
0&=&\hat{\beta}_{1/g^2_v} \, \equiv \, 2n_v + \sum_{a\in A\mid h(a)=v}(R_a-1)n_{t(a)} + \sum_{a\in A\mid t(a)=v}(R_a-1)n_{h(a)}\\
0&=& \hat{\beta}_{\lambda_l}\,  \equiv\,  -2+\sum_{a\in \mathrm{loop} \ l} R_a~.
\eea
Here $g_v$, $v\in V$, are the gauge couplings while $\lambda_l$, $l\in L$, are the superpotential couplings.
Recall that $L$ is a set of oriented loops in the quiver. We have also defined
\bea
R_a=\frac{2}{3}(1+\gamma_a)
\eea
where $\gamma_a=\gamma_a(\{g_v\}_{v\in V},\{\lambda_l\}_{l\in L})$ is the anomalous dimension of the
bifundamental field $\Phi_a$. Thus $\gamma_a$, and hence $R_a$, are functions of $|V|+|L|$ couplings.

Setting the $\beta$ functions (\ref{betafunctions}) to zero gives $|V|+|L|$ linear equations
in the variables $\{R_a\}_{a\in A}$. However, notice from (\ref{triangle}) that
if $\{R_a^*\}$ is a zero of the $\beta$ functions, then so is $\{R_a^*+\mu Q_a\}$ for any
real number $\mu\in\R$, where
\bea\label{baryoniccharges}
Q_a=q_{t(a)}-	q_{h(a)}~.
\eea
If we instead regard the $\beta$ functions as functions of the \emph{couplings}, then this simple argument
shows that, generically, the space of \emph{marginal} couplings will be at least $b_3(Y)$-dimensional.
Indeed, the non-anomalous $U(1)$ symmetries are directly related to the number of marginal couplings
via the above argument. We will see how this happens in the dual AdS description in section \ref{AdSsection}.

For the theory on $N$ D3-branes, we also expect in general a linear relation
\bea\label{couplingbeta}
N\sum_{l\in L}\hat{\beta}_{\lambda_l}=\sum_{v\in V} \hat{\beta}_{1/g^2_v}
\eea
leading to another marginal direction. In particular, we conjecture that the
relation (\ref{couplingbeta}) should correspond
to the constant string coupling in the dual AdS background. One can show that
(\ref{couplingbeta}) is indeed an identity for toric quiver gauge theories as follows
(for further discussion, see the recent paper \cite{yama}). For a toric quiver gauge theory on $N$ D3-branes
at a toric Calabi-Yau singularity
one has $n_v=N$ for all $v\in V$. Also, each field $\Phi_a$ appears precisely twice in the superpotential $W$.
These statements  imply
\bea
N\sum_{l\in L}\hat{\beta}_{\lambda_l} &=& 2N\sum_{a\in A}R_a - 2N|L|\\
\sum_{v\in V}\hat{\beta}_{1/g^2_v}&=& 2N|V|+2\sum_{a\in A} (R_a-1)N~.
\eea
Thus
\bea
\sum_{v\in V} \hat{\beta}_{1/g^2_v}-N\sum_{l\in L}\hat{\beta}_{\lambda_l}= 2N(|V|-|A|+|L|)=0~,
\eea
where the last relation follows from Euler's theorem applied to the brane tiling \cite{Brianandfriends}.

Finally, note that coupling constants in $\mathcal{N}=1$ gauge theories are always
complex. In particular, the gauge couplings $g_v$ are paired with $\theta$-angles.
 Thus one expects \emph{at least} a $(b_3(Y)+1$)-dimensional space of complex marginal
couplings.


\subsection{Classical vacuum moduli space}
\label{VMSsection}

In this section we review the classical vacuum moduli
space $\mathscr{M}$ of a quiver gauge theory with gauge group (\ref{SUgroup}). This may be
referred to\footnote{The terminology is apparently
due to A. Bertram.} as the ``master space'' \cite{masterspace}.  The main purpose of this subsection is to
describe a (singular) fibration structure of $\mathscr{M}$  in which the fibres are constructed
from mesonic moduli spaces.  We shall
see in section \ref{section3} that much of the structure of this classical VMS is reproduced
in the dual supergravity solutions.
There exist two complementary descriptions of the VMSs of interest, namely the \emph{K\"ahler
quotient}
description and the \emph{GIT quotient} description,  and we will discuss both below. The former
is perhaps more familiar to physicists and indeed is the most convenient for the purposes
of the  paper.

As discussed in the previous subsection, in the IR the gauge group is given by (\ref{SUgroup}).
The classical VMS of such a quiver gauge theory
is the space of constant matrix-valued fields $\Phi_a$ minimising the potential. This is equivalent to setting the  F-terms
to zero
\bea\label{Fterms}
\frac{\partial W}{\partial \Phi_a}\, =\, 0~, \quad a\in A
\eea
and also the D-terms to zero
\bea\label{Dterms}
\mu_v\, =\, 0~,\quad v\in V~.
\eea
Here $\mu$ denotes the vector of D-terms
\bea\label{liono}
\mu_v  &= & - \sum_{a \in A \mid
\ t(a)=v} \left[ \Phi_a^\dagger \Phi_a - \frac{1}{n_v}\mathrm{Tr}(\Phi_a^\dagger \Phi_a) 1_{n_v\times n_v}\right] \nn\\
&&+ \sum_{a\in A \mid \ h(a)=v}\left[ \Phi_a \Phi_a^\dagger- \frac{1}{n_v}\mathrm{Tr}(\Phi_a\Phi_a^{\dagger}) 1_{n_v\times n_v}\right] ~.
\eea
Finally, one must identify configurations related by the action of the
gauge group $SG$ given by (\ref{SUgroup}).

A bifundamental field $\Phi_a$ in vacuum is just an $n_{h(a)}\times n_{t(a)}$ matrix. The space of all such fields may then be thought of
as
\bea
\C^D & = & \bigoplus_{a\in A} \C^{n_{h(a)} \times n_{t(a)}}~.\eea
Since the superpotential $W$ is a polynomial in the $\Phi_a$, the F-term equations (\ref{Fterms}) cut out an affine
algebraic set
\bea
\mathcal{Z}\,=\,\left\{\diff W=0\right\}\subset\C^D~.
\eea
Since $W$ is invariant under $G$, and thus also $SG\subset G$,
it follows that $SG$ acts on $\mathcal{Z}$. The process of setting
the D-terms to zero and quotienting by the action of the gauge group
is then by definition the \emph{K\"ahler quotient}
\bea\label{Kahlerquotient}
\mathscr{M}=\mathcal{Z}//SG~.
\eea
Indeed, the D-terms (\ref{Dterms}) are, up to a factor of $i$, the moment map
for the action of $SG$ on $\C^D$ equipped with its standand
flat K\"ahler structure\footnote{This assumes that one takes
canonical kinetic terms for the bifundamental fields.}. Note the
subtraction of the trace terms ensures that $\mu_v$ is traceless,
as required for an element of the (dual) Lie algebra of $SU(n_v)$.
The vacuum moduli space $\mathscr{M}$ inherits a K\"ahler metric from the flat metric on $\C^D$.

We may alternatively construct $\mathscr{M}$ algebro-geometrically. It is useful to introduce the complexified gauge group
\bea
SG_{\C}=\prod_{v\in V} SL(n_v;\C)~.
\eea
In Geometric Invariant Theory (GIT) it is natural to define the quotient of $\mathcal{Z}$
by $SG_{\C}$ in terms of the ring of invariants of $SG_\C$
\bea\label{GITquotient}
\mathscr{M}=\mathcal{Z}//SG_\C = \mathrm{Spec}\ \C[\mathcal{Z}]^{SG_\C}~.
\eea
This is also the ring of \emph{semi-invariants} of $G_\C$. The construction (\ref{GITquotient}) realises $\mathscr{M}$ as an affine set. In more detail,
the ring of invariants of $SG_\C$ is finitely generated as the group $SG_\C$
is reductive (it is the complexification of a compact Lie group). One may thus pick a set of $d$ generators, for some
$d$. This realises $\mathscr{M}$ as an affine set in $\C^d$, the relations among the generators being the defining equations.
It then follows from a  general theorem
that the GIT quotient (\ref{GITquotient}) is isomorphic to the K\"ahler quotient (\ref{Kahlerquotient}),
as complex manifolds defined as the complement of the singular points.

Such moduli spaces,
for certain examples of simple quiver gauge theories on a
D3-brane at a Calabi-Yau singularity, have recently been investigated in
detail in \cite{masterspace}. Very little is known
in general about the detailed structure of these moduli spaces.
A notable point is that, in general, $\mathscr{M}$ is reducible.
Also, so far in the paper we have ignored the fact that the quiver gauge theory is
usually far from unique: the different quiver theories are valid in
different regions of K\"ahler moduli space. However, they are all
in the same universality class,
flowing in the IR to the same superconformal fixed point. The quivers are
then related by a form of Seiberg duality. As first pointed out for
the complex cone over $dP_3$ in
\cite{beas}, the moduli space $\mathscr{M}$ is not always invariant under
Seiberg duality. However, it was conjectured in \cite{masterspace}
that there is a top-dimensional irreducible component of $\mathscr{M}$ that
is invariant. The discussion below should probably be applied to this irreducible component
of $\mathscr{M}$. Having said this, the structure of $\mathscr{M}$ that we wish
to describe is so general that these precise details will not be important for our purposes.

The global symmetry group $U(1)^{\chi}\subset G$
acts holomorphically on $\mathscr{M}$, preserving its K\"ahler structure.
In fact, recall that no field is charged under the
diagonal $U(1)_{\mathrm{diag}}\subset U(1)^{\chi}$. The effectively acting
group is in fact the torus
\bea\label{baryonictorus}
\mathbb{T}\, =\, U(1)^{\chi-1}\, \cong \,U(1)^{\chi}/U(1)_{\mathrm{diag}}~.
\eea
We may thus in particular take the K\"ahler quotient of
$\mathscr{M}$ by $\mathbb{T}$. Since the dual Lie algebra of
$U(1)^{\chi}$ is isomorphic to $\R^{\chi}$, we may also pick a non-zero
moment map level, or FI parameter in physics language, $\zeta\in\R^{\chi}$ satisfying
\bea
\sum_{v\in V}\zeta_v\, =\, 0~.
\eea
This is equivalent to quotienting the space of F-term solutions $\mathcal{Z}$ by the
 gauge group $G=\prod_{v\in V}U(n_V)$
with moment map
\bea\label{ted}
\mu_v(\zeta)  &\equiv & - \sum_{a \in A \mid
\ t(a)=v} \Phi_a^\dagger \Phi_a + \sum_{a\in A \mid \ h(a)=v}\Phi_a
\Phi_a^\dagger \, + \frac{1}{n_v}\,\zeta_v \, 1_{n_v\times n_v}~.
\eea
We shall denote this quotient by
\bea\label{mesonic}
\mathscr{M}(\zeta)=\mathcal{Z}//_{\zeta} G~.
\eea
This is usually called the mesonic moduli space with FI parameters given by the vector $\zeta$.

It is perhaps worth stressing that the above quotient by $G$ (as opposed to that by $SG$) is not
 physically relevant; it may be regarded as a mathematical trick that is useful for describing the global
 structure of the physical moduli space, which is $\mathscr{M}$.
Indeed, picking a point $p\in\mathscr{M}$ determines a vector $\zeta\in\R^{\chi-1}\subset
\R^{\chi}$ via setting $\mu_v(\zeta)=0$ in (\ref{ted}). Notice that the sum of quadratic
terms in the bifundamental fields is necessarily proportional to the identity matrix
$1_{n_v\times n_v}$ since any point in $\mathscr{M}$ satisfies $\mu_v=0$ in (\ref{liono}). This gives
a well-defined map
\bea\label{pimap}
\Pi: \mathscr{M}\rightarrow \R^{\chi-1}~.
\eea
The mesonic moduli space $\mathscr{M}(\zeta)$ is then $\Pi^{-1}(\zeta)/U(1)^{\chi-1}$.
For a generic (smooth) point, the group $U(1)^{\chi-1}$ acts freely and thus
$\Pi^{-1}(\zeta)$ is a $U(1)^{\chi-1}$ fibration over $\mathscr{M}(\zeta)$.
Thus $\mathscr{M}$ fibres over $\R^{\chi-1}$ where the fibres are themselves
fibrations with base space $\mathscr{M}(\zeta)$ and generic fibre $U(1)^{\chi-1}$.

In fact, not all values of $\zeta\in\R^{\chi-1}$ are realised. The set of $\zeta$ for which $\Pi^{-1}(\zeta)$
is non-empty
correspond to points in a convex cone in $\R^{\chi-1}\subset \R^{\chi}$ \cite{dolgachev}. This cone
in $\R^{\chi-1}$ is further subdivided into a set of \emph{chambers} $C$, which are
the open interiors of convex rational polyhedral cones, with
boundaries between chambers being known as \emph{walls}.
Mesonic moduli spaces with FI parameters inside the same chamber $C$ are all isomorphic
to the same complex manifold $\mathscr{M}_C$, although they have
distinct K\"ahler forms. The K\"ahler class locally varies linearly with the FI parameters. As one crosses a wall from one chamber into another,
the mesonic moduli space undergoes a form of
small birational transformation called a \emph{flip} \cite{dolgachev}. The moduli spaces
corresponding to FI parameters on the walls are singular. Thus, strictly speaking,
the map $\Pi$ in (\ref{pimap}) is not a fibration, since the fibres are only locally
isomorphic. Across the walls in $\R^{\chi-1}$ the fibres change topology.

For applications to AdS/CFT, where the
quiver gauge theory describes the theory on $N$ coincident D3-branes transverse to $X$, one \emph{expects}\footnote{When there is an equivalence of derived categories, there is always an irreducible component of the vacuum 
moduli space
isomorphic to the original cone -- see \cite{proudfoot}.} that
\bea
\mathscr{M}_N(\zeta) \, \cong \, \mathrm{Sym}^N X \, = \, X^N/S_N\eea
is the $N$th symmetrised product of $X$. Here
the set of dimension vectors $\mathbf{n}$, which we have suppressed in the notation (\ref{mesonic}), are of course fixed in terms of $N$. The space $\mathscr{M}_1(\zeta)\equiv X(\zeta)=X$ is naturally
the vacuum moduli space of a single pointlike D3-brane on $X$. Thus the dual geometry
is expected to arise as the classical vacuum moduli space\footnote{If the singularity $\{r=0$\} is not isolated
the situation may be more complicated. In this case the mesonic moduli space
contains the dual geometry, but typically also has other branches.} for the gauge theory on
a D3-brane. The singular cone geometry $C(Y)$ corresponds to
the zero moment map level $\zeta=0$. Thus we identify
\bea\label{conemoduli}
C(Y) \, = \, \mathscr{M}_1(0)~.
\eea
From the above discussion regarding convex cones and chambers in FI space, setting $\zeta\neq0$ in the quiver gauge theory
with gauge group $G$ corresponds to (partially) resolving the moduli space (\ref{conemoduli}). Indeed, in \cite{Ueda} it was proven
that for toric quiver gauge theories described by dimers
the identification (\ref{conemoduli}) indeed holds, and
moreover $\mathscr{M}_1(\zeta)$ is a toric \emph{crepant}
resolution of $C(Y)$ for generic $\zeta$. These results
are also known to hold for orbifolds $C(Y)=\mathbb{C}^3/\Gamma$
with $\Gamma\subset SU(3)$ a finite group \cite{bridgeland}. Indeed, for orbifold quiver gauge theories
it is known that all crepant resolutions $\pi:X\rightarrow \C^3/\Gamma$ arise in this
way \cite{craw}. The results of \cite{counting} for certain examples of toric theories
strongly suggest this is true in general for toric quivers.

In the preceding discussion we have reviewed the description of $\mathscr{M}$ as a certain
fibration over the space of ``FI paramaters''. This will be sufficient for comparing with the dual gravity
VMS that we discuss in the next section. In the remainder of this subsection we will describe
in more detail the GIT quotient point of view. Although this is slightly technical, and will not affect
most of the rest of the paper, it provides a description of baryon operators as holomorphic functions
on $\mathscr{M}$ which sheds light on the recent counting results presented in \cite{counting}.

Thus, consider the GIT quotient by the \emph{complexification}
$\mathbb{T}_{\C}\cong (\C^*)^{\chi-1}$ of the torus (\ref{baryonictorus}). This also acts on $\mathscr{M}$, which we now consider
as an affine set. In order
to obtain the analogue of a non-zero moment map level $\zeta$, we need to pick a \emph{character} of $\mathbb{T}$.
The character lattice $\Lambda\subset \mathtt{t}^*$ is by definition the set of all
one-dimensional representations of $\mathbb{T}$. On picking a basis, this is
\bea
\mathbb{Z}^{\chi-1} \, \cong \, \Lambda \subset \mathtt{t}^* \, \cong \, \mathbb{R}^{\chi-1}~.\eea
Picking a character $q\in \Lambda$ specifies an action of $\mathbb{T}_{\C}$ on
$\mathscr{M}\times\C$, namely
\bea
\mathscr{M}\times\C \ni (p,z) \mapsto (\lambda\cdot p, \chi_{q}(\lambda) z), \qquad \lambda\in\mathbb{T}_{\C}~.\eea
In a basis, we may write $\lambda=(\lambda_1,\ldots,\lambda_{\chi-1})\in (\C^*)^{\chi-1}$,
$q=(q_1,\ldots,q_{\chi-1})\in\Z^{\chi-1}$
and then
\bea
\chi_q(\lambda) = \prod_{i=1}^{\chi-1} \lambda_i^{q_i}~.\eea
We may now perform the GIT quotient for the action of $\mathbb{T}_{\C}$ on
$\mathscr{M}\times\C$, using the character $q$ (or rather $\chi_q$).
This picks out the
set of holomorphic functions on $\mathscr{M}$ of charge $kq$, with $k\in\Z_+=\{0,1,2,\ldots\}$.
To see this, note that
the invariant regular functions on $\mathscr{M}\times \C$ are spanned by functions\footnote{The
\emph{function} $f(p,z)=z$ transforms with the opposite weight to the \emph{coordinate} $z$.}
of the form $f_{kq}z^k$, where
$f_{kq}$ is a holomorphic function on $\mathscr{M}$ of charge $kq\in\Lambda$. Thus
\bea\label{whoyagonnacall}
\C[\mathscr{M\times\C}]^{\mathbb{T}_{\C}(q)} = \bigoplus_{k\in\Z_+} \{f_{kq}\}~.
\eea
This is a graded ring, graded by $k$, and we may thus take
the projective Proj, rather than the affine Spec, of (\ref{whoyagonnacall}). This may be defined concretely as follows.
One first takes a finite set of generators $w_i$, $i=1,\ldots,d$, of the ring (\ref{whoyagonnacall}), which may be taken to be homogeneous under the grading. From this we could construct
the corresponding affine variety in $\C^d$. However, instead
we do something different. Let $w_a$, $a=1,\ldots,m$,
be the generators of homogeneous degree zero. Then we define
Proj of (\ref{whoyagonnacall}) to be the zero set of the
relations between the generators in $\C^m\times (\C^{d-m}\setminus 0)\subset\C^d$,
quotiented by the $\C^*$ action given by the grading on the generators.
This produces the quotient
space $\mathscr{M} \ //_q \ \mathbb{T}_\C$ together with an ample line bundle
over it. The holomorphic sections of this line bundle are by definition the charge $q$
holomorphic functions on $\mathscr{M}$, which is the degree one piece
 $k=1$ of the ring (\ref{whoyagonnacall}).
It is then a fairly standard result (see, for
example, \cite{dolgachev} and references therein) that the K\"ahler quotient $\mathscr{M}(q)$ with FI parameter
$q\in\mathtt{t}^*$ is the same as the GIT quotient using the lattice point $q$:
\bea\label{thesame}
\mathscr{M}(q) \, \cong \, \mathscr{M} \ //_q \ \mathbb{T}_\C \, \cong \, \mathscr{M}_C~.\eea
Here $q\in C\subset\mathtt{t}^*$ lies in the chamber $C$, and recall that
the underlying space $\mathscr{M}_C$ depends only on the chamber: the choice of
point $q\in\Lambda\cap C$ determines an ample line bundle over $\mathscr{M}_C$ in the
GIT quotient, and a (quantised) K\"ahler form in the K\"ahler quotient.
One also naturally gets a morphism
\bea\label{res}
\pi:\mathscr{M}(q)\rightarrow \mathscr{M}(0)
\eea
via the inclusion of the invariant functions on $\mathscr{M}$ in
(\ref{whoyagonnacall}). For the gauge theory on a single D3-brane,
$\mathscr{M}_1(0)= C(Y)$. For a toric quiver gauge theory, corresponding
to a D3-brane at the singular point of an affine toric singularity,
it was proven in \cite{Ueda} that (\ref{res}) is indeed a toric crepant resolution of $C(Y)$.

The description of the mesonic moduli spaces in terms of a \emph{geometric quotient}
 of $\mathscr{M}$ by the complex torus $\mathbb{T}_{\C}\cong (\C^*)^{\chi-1}$ is standard \cite{richard}.
The points of $\mathscr{M}$ under the group action are separated into unstable, semi-stable and stable points,
 where the stable points are a subset of the semi-stable points. The unstable points, which we denote
$S_{q}$, are thrown away in the quotient. On the other hand for \emph{generic} $q$ one expects
all other points to be stable -- see \cite{Ueda} for a discussion of this for toric quiver moduli spaces.
Then the statement is that $\mathscr{M}\setminus S_q$ is a $\mathbb{T}_{\C}\cong (\C^*)^{\chi-1}$
fibration over $\mathscr{M}(q)\cong\mathscr{M}_C$. When there are semi-stable
but not stable points $\mathscr{M}\setminus S_q$ is no longer a fibration over the mesonic moduli
space. For example, this is certainly true when $q=0$.

In the abstract language above,
the gauge-invariant BPS operators are classically just the holomorphic functions
on $\mathscr{M}$. These form the coordinate ring $\C[\mathscr{M}]$. The BPS
meson operators are the subset of these that have zero
charge $q=0$ under the baryonic torus $\mathbb{T}$. Alternatively, these
are the ring of invariants $\C[\mathcal{Z}]^{G_\C}$ of $G_\C$, rather than $SG_\C$.
On the other hand, the baryon operators are by definition the gauge-invariant
BPS operators with non-zero change $q$ under $\mathbb{T}$.
For $N=1$, a baryonic
operator of charge $q\in\Lambda$ is, by the above GIT construction,
the same thing as an ample divisor of the mesonic
moduli space $X=\mathscr{M}(q)$. These statements explain the counting
of baryon operators presented in \cite{counting}, since an ample
divisor may be identified with a quantised K\"ahler class on $X$.
Thus counting baryon operators according to their baryonic charge
indeed involves summing over mesonic moduli spaces $X$
that are resolutions of the Calabi-Yau cone, and on each $X$
summing over quantised K\"ahler classes (ample line bundles).

\subsubsection*{Example: the conifold}

Since the discussion above is all rather abstract, we include here a simple example.
The gauge theory on a D3-brane at the singular point of the conifold\footnote{Ordinary
double point singularity, in mathematical terminology.}
consists of four chiral fields $A_i$, $B_i$, $i=1,2$. The gauge group
is $G=U(1)^2$, with the fields $A_i$, $B_i$ carrying charges $(1,-1)$ and
$(-1,1)$, respectively.
The
superpotential is zero, and thus the classical VMS is simply
$\mathscr{M}=\C^4$, parameterised by the VEVs of the above bifundamental fields.
We introduce standard coordinates $z_1,z_2,z_3,z_4$ on $\C^4$.
The overall $U(1)$ decouples, as always, and the charges under the remaining
$U(1)$ are $(1,1,-1,-1)$. The moment map is then, up to normalisation,
\bea
\mu = |z_1|^2+|z_2|^2-|z_3|^2-|z_4|^2~.
\eea
Picking a point $p\in\mathscr{M}=\C^4$ thus determines a real number
$\mu(p)\in\R$. This is the same as the map $\Pi$ in (\ref{pimap}). By a slight abuse of terminology, we refer to
$\mu(p)$ as the value of the FI parameter. The mesonic moduli spaces
are then given by
\bea\label{conifoldquotient}
\mathscr{M}(\zeta)=\C^4//_\zeta U(1)=\{p\in\C^4\mid \mu(p)=\zeta\}/U(1)~.
\eea
The underlying complex variety $X$ of $\mathscr{M}(\zeta)$ depends on the sign of $\zeta$:
for $\zeta>0$
one obtains the resolved conifold $X_+=\mathcal{O}(-1)\oplus\mathcal{O}(-1)\rightarrow
\mathbb{CP}^1$; for $\zeta=0$ one obtains the conifold $X_0=\{u,v,x,y\in\C^4\mid
u^2+v^2+x^2+y^2=0\}$; whereas for $\zeta<0$ one obtains the other small
resolution of the conifold $X_-$, obtained by flopping the $\mathbb{CP}^1$ in $X_+$.
Thus the space of FI parameters in $\R$ is fan consisting of two
one-dimensional chambers $\R_\pm$, together with a point: $\R=\R_-\cup\{0\}\cup\R_+$.
The K\"ahler class of $\mathscr{M}(\zeta)$, equipped with its induced K\"ahler
metric, depends linearly on $\zeta$. Roughly, one may interpret $|\zeta|$
as the size of the $\mathbb{CP}^1$. Picking a point $p\in\mathscr{M}=\C^4$ hence
determines, via (\ref{conifoldquotient}), a point in a $U(1)$ fibre over a point in
one of $X_0$ or $X_{\pm}$, together with
a K\"ahler class on $X_{\pm}$. The $U(1)$ fibre is non-degenerate everywhere,
except over the singular point of the conifold $X_0$.

The unstable points, in the GIT quotient description, are
$S_+=\{z_1=z_2=0\}\cong\C^2$ and $S_-=\{z_3=z_4=0\}\cong\C^2$. Thus
we may define $\mathscr{M}_\pm=\mathscr{M}\setminus S_{\pm}$.
 Then $\mathscr{M}_\pm$ is a $\C^*$
fibration over $X_\pm$.


\section{Gravity backgrounds}\label{section3}


\subsection{AdS$_5$ backgrounds}
\label{AdSsection}

We begin our discussion by recalling the well-known AdS$_5\times Y$
solutions of Type IIB supergravity, where $(Y,g_Y)$ is a Sasaki-Einstein five-manifold.
In particular we present a discussion of the
various background moduli described by flat form fields.
These will play a crucial role in our subsequent discussion of deformations of the conformal backgrounds.

Consider placing $N$ D3-branes at the tip of
the  Ricci-flat K\"ahler cone (\ref{cone}). The corresponding
solution of Type IIB supergravity is given by
\bea\label{D3metric}
g_{10} &=& H^{-1/2}g_{\R^4} + H^{1/2} g_{C(Y)} \\ \label{Gflux}
G_5 & = & (1+*_{10})\diff H^{-1}\wedge \vol_4
\eea
where the function $H$ is given by
\bea\label{Hfull}
H = 1+ \frac{L^4}{r^4}~.
\eea
Here $g_{\R^4}$ is four-dimensional Euclidean space, with volume form $\vol_4$, and
$L$ is a constant given by
\bea\label{Lrel}
L^4 = \frac{(2\pi)^4g_s(\alpha')^2N}{4\vol(Y)}~.
\eea

The near-horizon limit of this system of D3-branes may be obtained by simply dropping
the additive constant from the function (\ref{Hfull}), which results in the product background
AdS$_5\times Y$. Here AdS$_5$, or rather its Euclidean version which is hyperbolic
space, is realised in horospherical coordinates. Specifically, the metric (\ref{D3metric}) becomes
\bea\label{AdSmetric}
g_{10}\, =\, \frac{L^2}{r^2}\diff r^2 + \frac{r^2}{L^2} g_{\R^4} + L^2 g_Y~.
\eea
The AdS/CFT correspondence conjectures that Type IIB string theory on this
background is dual, in the large $N$ limit, to a four-dimensional $\mathcal{N}=1$ superconformal
field theory. The latter may be regarded as living on the conformal boundary
of the five-dimensional hyperbolic space in (\ref{AdSmetric}). The conformal
compactification of hyperbolic space may be described topologically as adding an $S^4$
conformal boundary to a five-dimensional open ball. In the horospherical coordinates
above, this $S^4$ boundary of AdS$_5$ is the union of $r=\infty$, which is a copy of $\R^4$,
with the point $r=0$.

Actually, more precisely, for fixed spacetime metric (\ref{AdSmetric}) there will
in general be a family of corresponding AdS backgrounds, obtained by
turning on various flat background fields. These correspond
to exactly marginal directions in a family of $\mathcal{N}=1$ superconformal
field theories. In the remainder of the subsection we give a careful
summary of these marginal deformations.

Firstly there is the constant axion-dilaton $\tau$ in (\ref{axdil}).
Secondly, we may turn on a flat RR $C_2$ field and its $SL(2;\R)$ partner, a flat NS $B$ field.
In the current set-up, with $\mathcal{M}=\mathrm{AdS}_5\times Y$,
the spacetime is contractible to $Y$.
Note that before taking into account large gauge transformations, we may view the non-torsion
flat RR fields as a vector $(C_0,[C_2])\in
\R\oplus\R^{b_3(Y)}\cong\R^{b_3(Y)+1}$, where $[C_2]\in H^2(Y;\R)\cong \R^{b_3(Y)}$.
The lattice of large gauge transformations is then given by
\bea
\Lambda_B^Y = \left\{\left(n,\frac{2\pi}{\mu_1}\sigma + nB\right)\mid n\in\Z, \sigma\in  H^2_{\mathrm{free}}(Y;\Z)\right\}\subset \R^{b_3(Y)+1}~.
\eea
The flat RR fields thus live in the torus
\bea\label{RRYtorus}
([C_0],[C_2])\in \R^{b_3(Y)+1}/\Lambda_B^Y\,\cong \, U(1)^{b_3(Y)+1}~.
\eea
The $C_2$ field and $B$ field pair naturally into the complex combination $\tau B - C_2$.
Note that when $H^3_{\mathrm{tors}}(Y;\Z)$ is non-trivial it is possible to turn
on torsion $G_3$ and $H$ fields. These should correspond to discrete parameters labelling
the dual SCFTs.

In principle we might also have been able to turn on a flat RR $C_4$ field, in addition to
the background flux (\ref{Gflux}). However, since $b_1(Y)=0$ by Poincar\'e duality we have $H^4(Y;\R)=0$. There is
hence no room for such a flat field.
Such fields will play an important role once we deform the AdS background geometry
in the next subsection.

The above flat fields, including the axion-dilaton, may be identified with the
$(b_3(Y)+1)$-dimensional space of marginal couplings discussed in section \ref{marginalsection},
in the case that the dual SCFT has a quiver gauge theory description. Note, however, that
both $B$ and $C_2$  are periodic variables. Although there is some field theory understanding of
this for simple examples, such as the conifold with $(Y,g_Y)=T^{1,1}$, a general account seems
to be lacking at present.
Marginal deformations also arise if there is a non-trivial moduli space of Sasaki-Einstein
metrics on $Y$, as often occurs in the constructions of Sasaki-Einstein manifolds as links of
hypersurface singularities in \cite{BGreview}. There may also be metric deformations that take us outside
the class of Sasaki-Einstein backgrounds, notably the $\beta$-deformations of
\cite{ML} for toric Sasaki-Einstein manifolds.  We will not consider either of these possibilities in the present paper.


\subsection{Symmetry-breaking backgrounds}\label{defsection}

The quantum field theories dual to the above backgrounds are in vacua in which all scalar operators have zero VEVs. Indeed, a non-zero VEV will break conformal invariance,
leading to a renormalisation group flow via an associated Higgs effect.
The aim of this paper is to consider more general field theory
vacua in which various operators have non-zero VEVs. This was first
discussed for $\mathcal{N}=4$ Yang-Mills theory in \cite{Finn}, and for orbifolds
and the conifold theory in \cite{KW2}. Here we wish to extend the discussion to
general Sasaki-Einstein backgrounds with dual field theories described by quiver gauge theories.
We discussed the \emph{classical} space of such vacua in
section \ref{VMSsection}. In the remainder of the section we would like to construct the corresponding
dual supergravity solutions.

At energies well above the highest scale set by the VEVs, one
expects the physics to be well-described by the original $\mathcal{N}=1$ superconformal field
theory. The latter is thus the UV theory in this set-up. As usual in AdS/CFT, one may describe field theories
that are conformal at high energies by a dual gravitational
background that is \emph{asymptotic} to an AdS solution.
One should therefore look for supergravity solutions that are asymptotic to
AdS$_5 \times Y$. However, as emphasized in \cite{KW2}, if the dual field theory
is defined on $S^4$ one does not expect to find vacua of the type discussed in
section \ref{VMSsection}: the conformal coupling of scalar fields to the positive
scalar curvature of $S^4$ prevents them from acquiring a VEV. Instead, one
should regard the ``boundary'' of AdS$_5$ to be $\R^4$, given by $r=\infty$
in the horospherical coordinates (\ref{AdSmetric}), so that the dual field theory
is defined on flat $\R^4$. We thus seek supergravity
solutions that have an asymptotic region which approaches the large $r$ region of (\ref{AdSmetric}).
The solutions of interest will also have other asymptotic regions, as we shall
describe momentarily.

There are two natural ways of deforming
the AdS backgrounds in section \ref{AdSsection} in this manner:
\begin{itemize}
\item \emph{Mesonic deformations}: where one moves
some or all of the stack of $N$ D3-branes away from the singularity $r=0$ of $C(Y)$.
\item \emph{Baryonic deformations}: where one de-singularizes the Calabi-Yau cone
$C(Y)$, replacing it by a (possibly still singular) Ricci-flat K\"ahler manifold $(X,g_X)$ that
is asymptotic to a cone over the Sasaki-Einstein manifold $(Y,g_Y)$.
\end{itemize}
Actually these names are slightly misleading, since generically meson
and baryon operators obtain VEVs in both types of vacua. However, the space
of mesonic deformations is naturally isomorphic to the gauge theory mesonic moduli
space at zero FI parameter. Also, for certain baryonic deformations no meson
operator obtains a VEV. To be more precise, if $\pi:X\rightarrow Z$ is a
crepant resolution of the singularity $Z=C(Y)$, then in backgrounds
where all of the D3-branes are located on the exceptional set (the set of points in $X$ mapping
to the singular point $r=0$ of $C(Y)$) one expects that no meson operator
obtains a VEV. Classically this is because the meson operators
are the holomorphic functions $\C[Z]$, which, if not constant, vanish
at $r=0$. For example, the backgrounds discussed in
\cite{baryonic} are all of this form.

For any such $(X,g_X)$ above we may construct a family of supersymmetric
Type IIB backgrounds, asymptotic to AdS$_5\times Y$ in the above sense, as follows. The
ten-dimensional metric is
\bea\label{deformedmetric}
g_{10} \, =\,  H^{-1/2} g_{\R^4} + H^{1/2} g_{X}~,
\eea
with $G_5$-flux still given by (\ref{Gflux}). We pick $m$ points $x_i$, $i=1,\ldots,m$,
and place $N_i$ D3-branes at the $i$th point. Thus
\bea\label{conservation}
\sum_{i=1}^m N_i\, =\, N
\eea
and the function $H$, which is sourced by the D3-branes, satisfies
\bea\label{Heqn}
\Delta_x H \, =\,  -\frac{(2\pi)^4g_s(\alpha')^2N}{\sqrt{\det g_X}}\sum_{i=1}^m \frac{N_i}{N}\delta^6(x-x_i)~.
\eea
Here $\Delta$ is the Laplacian on $(X,g_X)$. The warp factor $H$ thus satisfies
the Laplace equation on $X\setminus\{x_1,\ldots,x_m\}$. The boundary conditions
may be described as follows. Since $(X,g_X)$
is asymptotic to a cone over $Y$ we may require the solution for $H$ to approach
$H_{\mathrm{AdS}}=L^4/r^4$ for large $r$. This, together with the D3-brane charge relation (\ref{conservation}),
precisely ensures that the Type IIB background is asymptotic to the large $r$
region of (\ref{AdSmetric}), with $L$ given by (\ref{Lrel}). Near to the $i$th stack of
D3-branes $x_i\in X$, the function $H$ behaves as
\bea\label{Hpoints}
H(x) \,=\, \frac{L_i^4}{\rho(x,x_i)^4}(1+o(1))~.
\eea
Here $\rho(x,x_i)$ is the geodesic distance from $x_i$ to $x$, and
\bea\label{gary}
L_{i}^4 \,=\, \frac{(2\pi)^4g_s(\alpha')^2 N_i}{4\vol(S^5)}~,
\eea
provided that $x_i$ is a smooth point of $X$. As discussed in \cite{baryonic}, if
$(X,g_X)$ has a conical singularity at $x_i$, with corresponding Sasaki-Einstein
link $(Y_i,g_{Y_i})$, then $\vol(S^5)$ is replaced by $\vol(Y_i)$ in (\ref{gary}). The singular nature of $H$
at $x_i$ implies that the metric (\ref{deformedmetric}) develops a
``throat'' near to this point. In fact it
approaches the metric in a neighbourhood of $r=0$ in the AdS background
\bea\label{AdSmetrici}
g_{10} \,=\, \frac{L^2_i}{r^2}\diff r^2 + \frac{r^2}{L^2_i} g_{\R^4} + L^2_i g_{S^5}~.
\eea
Again, if the point $x_i$ is a conical singularity, the round $S^5$ is replaced
by $(Y_i,g_{Y_i})$.

Notice that when all $N$ of the D3-branes
are placed at the same point, so $m=1$ in the above notation, the function $H$
is simply the Green's function on $(X,g_X)$.
Provided $(X,g_X)$ is smooth and complete, we argued in \cite{baryonic}
that there always exists a unique positive solution to (\ref{Heqn}) with the required
boundary behaviour -- this
follows from standard theorems about Green's functions on manifolds with
non-negative Ricci curvature and appropriate volume growth. For $m>1$ stacks
of D3-branes we may then simply take an appropriate linear combination of
Green's functions as solution to (\ref{Heqn}).
More generally, when $(X,g_X)$ contains singularities (such as a mesonic background with $X=C(Y)$), we
do not know of any general theorems that guarantee existence
of a unique solution to (\ref{Heqn}). However, it is very reasonable to conjecture
this to be true, at least when $(X,g_X)$ contains only
conical singularities. Indeed, for the homogeneous case of $(Y,g_Y)=T^{1,1}$ one may construct
\cite{KM} explicit solutions to (\ref{Heqn}) on the conifold.
We will nevertheless
focus on the case that $(X,g_X)$ is smooth in the present paper.

The supergravity backgrounds with $m=1$ may be interpreted as
a renormalisation group flow from the initial $\mathcal{N}=1$ superconformal field theory to
$\mathcal{N}=4$ Yang-Mills in the IR, with gauge group $SU(N)$ \cite{KW2}. As we explain later in the paper,
 there may be additional light particles in the IR, namely Goldstone bosons
associated to the spontaneous breaking of non-anomalous baryonic symmetries.
When the
branes are separated, $m>1$, the interpretation of the background is a little more
subtle: there are $m$ regions in which the supergravity
solution approaches a neighbourhood of $r=0$ in the AdS solution (\ref{AdSmetrici}).
The natural interpretation is thus
that the theory flows, in the extreme IR, to a non-trivial
fixed point that is a \emph{product} of $m$ superconformal field theories.
When the points $x_i$ are all smooth, the factors in this product are $\mathcal{N}=4$
Yang-Mills with gauge groups $SU(N_i)$, as suggested in \cite{KW2}.
More generally it is natural to conjecture that the IR theory is a
product of the $\mathcal{N}=1$ superconformal field theories dual to
$(Y_i,g_{Y_i})$. Such theories have been discussed in
\cite{Aharony}, where the IR theory itself was conjectured to be
dual to the union of $m$ AdS$_5$ spaces, with conformal boundaries
identified. Note also that the supergravity approximation
is valid only when all $N_i$ are large, or equivalently the AdS
radii $L_i$ are large compared to the string scale.

Naively ignoring this last point, the space of supergravity metrics
for fixed $(X,g_X)$ is naturally given\footnote{In fact an exception to
this is when $X=\C^3$. In this case the translational
symmetry of $\C^3$ may be used to fix the centre of mass of
the D3-brane positions at the origin \cite{Finn, KW2}, resulting
in $\mathrm{Sym}^{N-1}\C^3$.} by the
symmetric product $\mathrm{Sym}^N X$, describing the positions
of the $N$ D3-branes. Of course, such symmetric products arise
in the classical gauge theory as mesonic moduli spaces, as
we reviewed in section \ref{VMSsection}. Fixing
a non-zero FI parmeter $\zeta\in\R^{\chi-1}$ for the gauge theory
on a single D3-brane, the corresponding mesonic moduli space $\mathscr{M}_1(\zeta)=X(\zeta)$
is a resolution $\pi:X(\zeta)\rightarrow Z$ of the Calabi-Yau singularity $Z=C(Y)$. Indeed,
this is known to be a crepant resolution
for orbifold \cite{bridgeland} and toric \cite{Ueda} quiver gauge theories.
We expect this to be true in general. Note also that the K\"ahler class
in $H^2(X;\R)$ of the induced metric on $X(\zeta)$ varies linearly with $\zeta$.
Thus, provided $X\cong X(\zeta)$ for some FI parameter $\zeta$,
the space of supergravity metrics obtained by varying the positions of the D3-branes is
the same as the corresponding mesonic moduli space. Of course,
the caveat to this statement is that the supergravity solutions
are strictly valid only when the D3-branes are in large ``clumps''.

The above discussion raises the question of how to characterise those
$X$ which are of the form $X\cong X(\zeta)$ for some FI parameter $\zeta$.
Certainly not all Calabi-Yau's $(X,g_X)$, asymptotic
to a cone over $(Y,g_Y)$, are of this form. Firstly $X$ must be a
crepant resolution of $Z$. For example, the deformed conifold
is a de-singularization of the conifold, but clearly this cannot
arise as a mesonic moduli space. The deformed conifold is therefore,
at least for generic couplings, not relevant for the
vacua of interest \cite{KW2}. Note in this example one has $b_3(X)=1$.
Another, more physical, justification for the
assumption (\ref{nob3}) is that if there are odd-dimensional cycles on $X$ then one may wrap
D-branes over these cycles to obtain
topologically stable domain walls in $\R^4$ -- see, for exampe, \cite{KS,Herzog:2002ih}.
In particular, if $b_3(X)\neq 0$ one may wrap D5-branes to obtain such domain walls.
These connect different vacua of the theory.
Such backgrounds therefore have qualitatively different physics
from those without odd-dimensional cycles.
Even for toric quiver gauge theories it is not known whether
\emph{all} toric crepant resolutions $X$ of $C(Y)$ are of the form $X(\zeta)$ for some $\zeta$.
For abelian orbifolds $C(Y)=\C^3/\Gamma$ this is true \cite{craw}, and
the baryon counting results of \cite{counting} certainly suggest that it is true in general for
toric theories. Thus, at least for orbifold and toric quiver theories, it seems
that all crepant resolutions of the conical singularity should
arise as dual descriptions of the supersymmetric vacua of interest.

\subsection{Form field moduli}
\label{flatsection}

As discussed in section \ref{AdSsection}, for the AdS background $\mathrm{AdS}_5\times Y$
 one is free to turn on various flat background fields, corresponding to a choice of
marginal couplings in the dual field theory. The supergravity
backgrounds discussed in section \ref{defsection}
are asymptotic to
AdS$_5\times Y$, in the sense that there is an asymptotic region
that approaches a neighbourhood of $r=\infty$ in (\ref{AdSmetric}).
Thus $\R^4\times Y$ is a boundary component of
the full spacetime $\mathcal{M}$. One must extend the
fields on this boundary over $\mathcal{M}$, and thus in particular over $X$, to obtain
a solution to supergravity.
Note that the spacetime $\mathcal{M}$, with metric (\ref{deformedmetric}), is globally of the form
$\R^4\times (X\setminus\{x_1,\ldots,x_m\})$.
Thus $\mathcal{M}$
also has $m$ asymptotic regions that look like a neighbourhood of $r=0$ in
(\ref{AdSmetrici}). Near each such region the set $r=\epsilon$, with $\epsilon>0$ small, is a
copy of $\R^4\times S^5$. More generally, when the $i$th set of
$N_i$ D3-branes are placed at a conical singularity of $X$ with
Sasaki-Einstein link $(Y_i,g_i)$, this boundary is replaced by
$\R^4\times Y_i$. The restriction of form fields on $\mathcal{M}$ to this ``internal'' boundary
thus naturally determines the IR superconformal field theory.
We should therefore
regard the spacetime $\mathcal{M}$ as having $m+1$ boundary components:
the UV boundary $\R^4\times Y$, and the $m$ components of the IR boundary,
which if $X$ is smooth are all diffeomorphic to $\R^4\times S^5$.

The dynamical fields of interest in this section are the RR fields and the NS $B$ field.
Consider a generic $p$-form
field strength $G$ with $(p-1)$-form potential $C$ on a
spacetime $\mathcal{M}$. This means that  locally
 $G=\diff C$. We assume that $G$ is a fixed
field strength on $(\mathcal{M},g_{10})$, satisfying the relevant
equation of motion. We may then pick a particular potential
$C^{\circ}$, defined locally in coordinate patches, such that $G=\diff C^{\circ}$.
Any other $C$ field giving rise to the same field strength $G$ is then given by
\bea\label{Cfluc}
C = C^{\circ} + C^{\flat}~,
\eea
where $C^{\flat}$ is a \emph{flat} $C$ field, {\it i.e.} it is closed.
The small gauge transformations on $C$ are of the form
\bea\label{Cgauge}
C \rightarrow C +\diff \lambda
\eea
where $\lambda$ is a $(p-2)$-form. This immediately leads to the cohomology
group $H^{p-1}(\mathcal{M};\R)$, classifying the space of $C$ fields
modulo gauge equivalence. Of course, inclusion of large gauge transformations
typically leads to $U(1)$ coefficients instead, and for RR fields
there is a twisting by the $B$ field, as discussed in section \ref{anomaloussection}.

In the present situation $H^{p-1}(\mathcal{M};\R)\cong H^{p-1}(X;\R)$
for the cases $p=3$, $p=5$ of interest. That is, deleting a finite number of
points from a smooth manifold $X$ does not affect the cohomology in these degrees, as one easily
proves using a simple Mayer-Vietoris sequence. However, we do not
want to think of $H^2(X;\R)$ as classifying, say, flat $C_2$ field moduli
of the backgrounds.
The reason is that the restriction $H^2(X;\R)\rightarrow H^2(Y;\R)$
gives the marginal couplings of the UV theory, which should be regarded
as fixed boundary data. We would like to instead classify fields on $X$ that induce the same
field at infinity.
Thus  we are interested in the kernel of the map
$H^{p-1}(X;\R)\rightarrow H^{p-1}(Y;\R)$, which is the same as the image of the map
$H^{p-1}(X,Y;\R)\rightarrow H^{p-1}(X;\R)$ by the long exact cohomology sequence for $(X,Y)$.

In fact, as we explain further below, and also in section \ref{condsection}, we would like to interpet the
form field moduli as living in $H^{p-1}(X,Y;\R)$ itself, rather than its
image in $H^{p-1}(X;\R)$. The elements of $H^{p-1}(X,Y;\R)$ that map to zero in $H^{p-1}(X;\R)$ are,
again by the long exact sequence for $(X,Y)$, images of $H^{p-2}(Y;\R)$.
This may be realised concretely as follows. Take an element $\lambda \in H^{p-2}(Y;\R)$.
By the Hodge theorem we may represent $\lambda$ by a harmonic form. Let
$f$ be a smooth function on $X$ that is equal to 1 on $Y$ and is identically
zero outside a tubular neighbouhood of $Y$ in $X$. Then
$\diff (f\lambda)=\diff f\wedge \lambda$ makes sense as a closed compactly supported $(p-1)$-form on $X$.
In fact, such forms precisely represent the image of $H^{p-2}(Y;\R)$ in $H^{p-1}(X,Y;\R)$.
Although such an expression is exact, and thus a pure gauge mode, the gauge generator
$\lambda$ is \emph{non-zero} on $Y$. Such gauge transformations are always associated
with global symmetries in gauge theory, and indeed later we will  identify these with the
$b_3(Y)$ non-anomalous global $U(1)$ symmetries associated to the RR four-form (so $p=5$ in the
above discussion).

A more refined treatment of the form field moduli thus treats them as compactly
supported cohomology classes. We may describe this explicitly
by requiring $C^{\flat}$ in (\ref{Cfluc}) to be zero
on $Y$, and $\lambda$ in (\ref{Cgauge}) to also be zero on $Y$. The gauge
for $C\mid_{\partial\mathcal{M}}$ is thus held fixed.
This leads to the relative/compactly supported cohomology group
\bea\label{classifyC}
H^{p-1}(\mathcal{M},\partial\mathcal{M};\R)~,\eea
modulo large gauge transformations, which will be twisted by $B$ for RR fields.
Again, in the present situation one may replace $\mathcal{M}$ by $X$,
since deleting a finite number of points from a smooth $X$ will not affect
the cohomology in the degrees of interest.
Note that the result (\ref{classifyC}) is independent of the choice of $C^{\circ}$.

For the $B$ field and $C_2$ field this leads to
the group $H^2(X,Y;\R)/H^2_{\mathrm{free}}(X,Y;\Z)$, classifying flat $C_2$ and $B$ fields
on $X$ with fixed value on the boundary.
The RR four-form $C_4$ is slightly more involved. This has a non-trivial
background flux $G_5$ given by (\ref{Gflux}). This field strength is not exact
since the flux of $G_5$ over $Y$ is equal to $N$, the number of D3-branes.
There is thus no globally defined potential $C^{\circ}_4$ on $\mathcal{M}$
with $\diff C^{\circ}_4=G_5$. However, this doesn't change the discussion much:
we may instead define $C^{\circ}_4$ in an open covering of spacetime by coordinate patches, glued by transition
forms across overlaps.
Such a choice also fixes
a gauge choice $C_4\mid_{\partial \mathcal{M}}$ at infinity. Again, this
cannot be a globally defined four-form, either on the UV boundary
or on any connected component of the IR boundary. More importantly,
the choice of background $C^{\circ}_4$ depends on the positions
$x_1,\ldots,x_m$ of the D3-branes and also on the metric $g_X$ on $X$.
Recall that the latter is conjecturally fixed by a choice of K\"ahler class
$[\omega_X]\in H^2(X;\R)$.
Thus we should more correctly write $C^{\circ}_4(\{x_i\},[\omega_X])$,
and fix a gauge choice $C^{\circ}_4(\{x_i\},[\omega_X])\mid_{\partial\mathcal{M}} =
C_4\mid_{\partial \mathcal{M}}$ at infinity. This shows that the
$C_4$ field is naturally fibred over the mesonic moduli space, whereas
the other supergravity gauge fields are not.

The space of RR field moduli may thus be described as follows.
By Poincar\'e duality we have
\bea
H^2_{\mathrm{free}}(X,Y;\Z)\,\cong \, H_4^{\mathrm{free}}(X;\Z) \ , \qquad \mathrm{and} \qquad H^4_{\mathrm{free}}
(X,Y;\Z)\, \cong \, H_2^{\mathrm{free}}(X;\Z)~.
\eea
The ranks of these groups are thus $b_4(X)$ and $b_2(X)$, respectively.
Thus, before taking into account large gauge transformations,
the different (non-torsion) RR fields may be described by a vector
\bea
([C_2^{\flat}],[C_4^{\flat}])\in \R^{b_4(X)}\oplus \R^{b_2(X)}\cong \R^{\chi-1}
\eea
where $\chi=\chi(X)$ is the Euler number given by (\ref{Euler}).
The lattice of large gauge transformations is
\bea
\Lambda^{X,Y}_B &= &\Bigg\{\left(\frac{2\pi}{\mu_1}\sigma,\frac{2\pi}{\mu_1}\sigma\wedge B +\frac{2\pi}{\mu_3}\kappa\right)\mid
\sigma\in H^2_{\mathrm{free}}(X,Y;\Z), \nn\\
&&
\kappa\in H^4_{\mathrm{free}}(X,Y;\Z)\Bigg\}\subset \R^{\chi-1}~.
\eea
The space of RR field moduli, modulo discrete torsion fields, is then described by the twisted torus
\bea\label{baryonicgroup}
\label{RRXtorus}
([C_2^{\flat}],[C_4^{\flat}]) \in \R^{\chi-1}/\Lambda^{X,Y}_B\, \cong \, U(1)^{\chi-1}~.
\eea
One should compare this to (\ref{RRYtorus}).


\subsection{Comparison: gauge theory and gravity vacua}\label{comparesection}

We conclude this section by comparing the supergravity backgrounds to the classical vacuum moduli space
structure described in section \ref{VMSsection}. In order to construct a gravity background we must first
pick a complex manifold $X$ that resolves $Z=C(Y)$. Since there are $N$ units of $G_5$ flux through
$Y$ at infinity, to preserve Poincar\'e symmetry we must choose where to put $N$ pointlike D3-branes
on $X$. This naturally leads to the symmetric product $\mathrm{Sym}^NX$ as moduli space, precisely
as one expects
for a mesonic moduli space in the gauge theory. Although, as we noted,
once one includes the backreaction of the D3-branes on the geometry, the supergravity
approximation breaks down unless these D3-branes are in large ``clumps''. Thus this
matching is perhaps rather better than one might have expected.

As explained in section \ref{VMSsection}, the gauge theory moduli space $\mathscr{M}$
may be viewed (\ref{pimap}) as a fibration over $\R^{\chi-1}$. The latter is divided
into chambers, and over each chamber $C\subset\R^{\chi-1}$ the fibres are all isomorphic.
In particular, each fibre is a $U(1)^{\chi-1}$ bundle over the mesonic moduli space
$\mathscr{M}_C$. In the case at hand, one expects $\mathscr{M}_C=\mathrm{Sym}^N X$
for some crepant resolution $X$ of $C(Y)$.
A point $\zeta\in C$ in particular determines a classical K\"ahler
class on $\mathscr{M}_C$, with the K\"ahler class varying linearly with $\zeta$.

It should now be clear how one matches this to the parameters of the supergravity backgrounds.
By our conjecture in section \ref{metricsection}, there is a $b_2(X)$-dimensional
space of asymptotically conical Ricci-flat K\"ahler metrics on $X$,
determined by their K\"ahler class in the K\"ahler cone in $H^2(X;\R)$.
These may be identified with $b_2(X)$ of the coordinates of $\zeta \in C$.
We identify the remaining $b_4(X)$ ``FI parameters'' with the $B$ field periods,
which live in $H^2(X,Y;\R)/H^2_{\mathrm{free}}(X,Y;\Z)$.
On the other hand, the periods of $B$ in $H^2(Y;\R)/H^2_{\mathrm{free}}(Y;\Z)$
partly determine the marginal couplings of the UV SCFT.  One puzzle here is that $B$ is periodic
in string theory, whereas in the classical gauge theory the FI parameters and
marginal gauge couplings
are real numbers. However, this is a somewhat standard issue. Indeed, in some cases
the periodicity of $B$ is
known to be related to Seiberg duality -- see, in particular, \cite{carlos} and \cite{strassler}.
Thus one would not expect to see this periodicity in the classical gauge theory,
which in particular involves choosing a fixed Seiberg phase.
The RR field moduli in (\ref{RRXtorus}), which indeed form a torus
$U(1)^{\chi-1}$ due to large gauge transformations, are then identified with the
$U(1)^{\chi-1}$ fibres over $\mathscr{M}_C$.
Supersymmetry pairs the K\"ahler class with
$C_4$, and the $B$ field with $C_2$. In the classical VMS, this is reflected by the
complexification $(\C^*)^{\chi-1}$ of the global baryonic symmetry group. This appears
in the GIT description of obtaining the mesonic moduli spaces $\mathscr{M}_C$ as a quotient of
$\mathscr{M}$. We thus obtain
a surprisingly good matching between the classical gauge theory moduli space
and the space of supergravity backgrounds described in this section.

Notice also that, for fixed choice of smooth Ricci-flat K\"ahler background $(X,g_X)$, positions of the $N$
D3-branes on $X$ and $B$ field modulus,
the space of  RR field moduli form a group under addition, and that this group is
isomorphic to $U(1)^{\chi-1}$. 
In this way we obtain an action of $U(1)^{\chi-1}$
on the moduli space of gravity backgrounds. Given that we are identifying the latter
with the symmetry-breaking vacua in the dual field theory, it is natural to interpet\footnote{One may construct
gravity backgrounds in which only part of
the global symmetry group is spontaneously broken by taking $X$ to be singular. For
simplicity we shall not consider this here.}
this $U(1)^{\chi-1}$ with the group of baryonic symmetries in
the dual field theory, described in section \ref{section2}. In
fact this group has a natural $U(1)^{b_3(Y)}$ subgroup.
Specifically, the $C_4$ moduli in $H^4(X,Y;\R)$
that are images of $H^3(Y;\R)$ are, as explained in the previous
subsection, naturally related to global symmetries on $Y$.
Since these global symmetries come from gauge symmetries of
RR fields, in particular they cannot be anomalous. This identifies the RR gauge symmetries coming from
$H^3(Y;\R)/H^3_{\mathrm{free}}(Y;\Z)\cong U(1)^{b_3(Y)}$ with the non-anomalous baryonic $U(1)$
symmetries in the field theory.
This is a very satisfying check that the picture we have outlined
so far is consistent.

\begin{table}[ht!]
\centerline{
\begin{tabular}{|c|c|c|c|}
\hline
number & gravity  &  gauge theory \\
\hline
\hline
1 & $\tau = C_0+i\e^{-\phi} $   &  marginal coupling \\
\hline
$b_3(Y)$ & $ \tau B- C_2 $ & marginal coupling\\
\hline
\hline
$b_4(X)$ & $ \tau B- C_2 $   & anomalous $U(1)$\\
\hline
$b_4(X)$ & $ \omega_X+iC_4 $    & anomalous $U(1)$ \\
\hline
$b_3(Y)$ & $ \omega_X+iC_4 $ &  non-anomalous $U(1)$\\
\hline
\end{tabular}
}
\caption{Gravity moduli and their interpretation in the dual quiver gauge theory.}
\label{summa}
\end{table}

It would be interesting to study the global structure of these supergravity moduli spaces
in more detail. For example, one could try to relate the Chern classes of the
torus bundle $U(1)^{\chi-1}$ over a mesonic moduli space $\mathscr{M}_C$ in the classical
VMS to the fibration structure of the RR field moduli (\ref{RRXtorus}) over the corresponding supergravity
moduli space of D3-brane positions, which is naturally isomorphic to $\mathscr{M}_C$.
As we have already remarked, the construction of the $C_4$ field certainly depends on
position in this moduli space via (\ref{Gflux}). One approach to this would be to
investigate the induced K\"ahler metric on the supergravity moduli space.
A similar situation was studied in \cite{mal}, where a RR modulus field is indeed fibred
over a mesonic moduli space, with the curvature of the corresponding line bundle being a K\"ahler
form on the mesonic moduli space. For this to make sense globally, the K\"ahler class should be quantised
(although this point was not addressed in \cite{mal}). This is precisely what happens in
the classical GIT description of the mesonic moduli space, where $\zeta=q$ is a lattice point
and is thus ``quantised''. It would also be interesting to investigate how different
resolutions $X_1$ and $X_2$ are glued together across the walls between chambers,
and in particular what happens to the RR fibres in this process.


\section{Linearised fluctuations}
\label{fluctuationsection}

In this section
we consider certain linearised fluctuations of the background fields, by allowing them to depend on position
in $\R^4$.
We shall
argue that the relevant modes require the existence of certain $L^2$ harmonic forms, with respect to appropriate metrics,
where the $L^2$ condition is required in order for the fluctuations to be normalisable (have finite kinetic energy).
We then appeal to
mathematical results on the existence and asymptotic expansions of such forms.
The AdS/CFT interpretation of these modes is postponed
to the next section.

The gauge-invariant form fields of Type IIB supergravity may be obtained by expanding
the RR multi-form (\ref{multiRR}) in forms of definite degree:
\bea
\tilde{G_3} & = & G_3 - H_3\, C_0~,\\
\tilde{G_5} & = & G_5 - H_3\wedge C_2~,\\
H_3 &=&\diff B~,
\eea
where
\bea
G_3&=&\diff C_2~,\\
G_5&=&\diff C_4~.
\eea
These expressions automatically solve the relevant Bianchi identities. The five-form field
strength $\tilde{G_5}$ is required to be self-dual
\bea
\tilde{G_5} & = & * \tilde{G_5}~;\eea
the equation of motion is then implied by the Bianchi identity. The
equations of motion for the remaining fields are
\bea\label{phiEOM}
\nabla^2 \phi & = &\e^{2\phi}|\diff C_0|^2 - \frac{1}{2}\e^{-\phi}|H_3|^2 + \frac{1}{2}\e^{\phi}|\tilde{G}_3|^2\\ \label{CEOM}
\diff^{\dagger}(\e^{2\phi}\diff C_0) &=& \e^{\phi}\langle H,\tilde{G}_3\rangle\\ \label{HEOM}
\diff (\e^{-\phi}* H_3)  & = & - \tilde{G_5} \wedge \tilde{G_3}+ \e^{\phi}\,  \diff C_0 \wedge * \tilde G_3\\ \label{G3EOM}
\diff (\e^{\phi} * \tilde{G_3})  & = &  ~ \tilde{G_5} \wedge H_3\\ \label{metricEOM}
R_{mn}&=&\frac{1}{2}\partial_m\phi\partial_n \phi + \frac{1}{2}\e^{2\phi}\partial_m C_0\partial_n C_0
+\frac{1}{96}\tilde{G}_{5mpqrs}\tilde{G}_{5n}^{\ \ pqrs}\\ \nn
&&+\frac{1}{4}\left(\e^{-\phi}H_{3mpq}H_{3n}^{\ \ pq} + \e^{\phi}
\tilde{G}_{3mpq}\tilde{G}_{3n}^{\ \ pq}\right)-\frac{1}{8}g_{mn}\left(\e^{-\phi}|H_3|^2+\e^{\phi}|\tilde{G}_3|^2\right)~.
\eea
where recall that $\phi$ is the dilaton and $C_0$ is the RR axion. The angle brackets and modulus signs denote
the natural pointwise inner products and norms for $p$-forms, respectively. Thus if $a_{m_1\cdots m_p}$, $b_{m_1\cdots m_p}$
denote the components of two $p$-forms $a$, $b$ then $\langle a,b\rangle = \frac{1}{p!}a_{m_1 \cdots m_p}b^{m_1 \cdots m_p}$,
and $|a|^2=\langle a,a\rangle$. The operator $\diff^{\dagger}=-*\diff *$ denotes the codifferential, the formal adjoint
to the exterior derivative $\diff$.

It will turn out that, for the linearised fluctuations of interest, it is consistent to vary  $C_2$, $C_4$ and $B$,
while keeping the metric, the dilaton and the axion fixed. Noting that our backgrounds have $G_3=H_3=0$, and
constant $\phi$, $C_0$, it is straightforward to  obtain the linearised equations of motion.
The linearisations of the first two equations of motion (\ref{phiEOM}), (\ref{CEOM}) are trivially satisfied
in our backgrounds. The remaining linearised equations of motion give
\bea
\delta  \tilde{G_5} &= & * \delta \tilde{G_5} \label{selfpert}\\
\e^{-\phi}\diff * \delta H_3 & = & - {G_5} \wedge (\delta G_3-C_0\delta H_3)\label{fluc1}\\
\e^{\phi}\diff * (\delta G_3 - C_0\delta H_3) & = & ~ {G_5} \wedge \delta H_3\label{fluc2}\\
0 & =& G_{5mpqrs}\delta \tilde{G}_{5n}^{\ \ pqrs} + \delta\tilde{G}_{5mpqrs}{G}_{5n}^{\ \ pqrs}\label{fluc3}
\eea
where
\bea
\delta G_3 &= &\diff\delta C_2\\
\delta \tilde{G_5} &=& \diff\delta C_4 - C_2\wedge\diff\delta B\\
\delta H_3 & = & \diff \delta B~.
\eea

In section \ref{metricfluc} we  examine the linearised equation of motion for the metric separately. The metric modes in warped non-compact backgrounds
are considerably more complicated than for Calabi-Yau compactifications \cite{GM}, and we shall only give a partial
treatment. In the next three subsections we shall allow $C_2$, $B$ and $C_4$ to fluctuate in turn, imposing a
natural ansatz and then solving the resulting linearised equations of motion. Having done this, it will be immediately clear
that all of these modes may be turned on simultaneously, and that this leads to the same equations. Thus the modes
are completely decoupled from each other.

\subsection{$C_2$ field moduli}\label{C2field}

We begin with the RR two-form $C_2$, since this is technically the simplest.
Let $\psi^A$, $A=1,\ldots,b_4(X)$,
be representatives for a basis of $H^2_{\mathrm{free}}(X,Y;\Z)$.
That is, the  $\psi^A$ are closed two-forms on $X$ which have
integral periods and vanish when restricted to $Y=\partial X$. Then we may write the $C_2$ moduli in (\ref{RRXtorus}) as
\bea
C_2^{\flat} \,=\, \frac{1}{\mu_1} \varphi^A\psi^A~.
\eea
As in section \ref{anomaloussection}, a sum over repeated indices is understood.
The $\varphi^A$, which are periodic constants, determine the $C_2$ moduli.

Consider now a fluctuation of $C_2$, where $\varphi^A$ may depend on the coordinates of $\R^4$.
Thus we write
\bea\label{varyC2}
\delta C_2 \,=\, \frac{1}{\mu_1}\delta\varphi^A\psi^A
\eea
where $\delta\varphi^A$ are functions on $\R^4$. One must check whether such a perturbation
satisfies the linearised Type IIB supergravity equations (\ref{fluc1}), (\ref{fluc2}).
The right hand side of (\ref{fluc1}) is identically zero, as one sees by noting the form of the background
$G_5$-flux given by (\ref{Gflux}). Thus the $B$ field is not sourced by the fluctuation
(\ref{varyC2}). The equation of motion (\ref{fluc2}) for $G_3$, on the other hand, requires
\bea
\diff (*_4\diff\delta\varphi^A\wedge *_6 \psi^A)\, =\, 0
\eea
where $*_6$ denotes the Hodge dual operator on $(X,g_X)$. Notice that the warp factor
$H$ has dropped out of the computation. Assuming the $\delta\varphi^A$ are linearly independent this equation implies that
\bea
\diff *_6\psi^A\,=\, 0
\eea
for all $A$, and the resulting equation for $\delta\varphi^A$ is the equation of motion for a massless
scalar field on $\R^4$.
Since $\psi^A$ is both closed and co-closed on $(X,g_X)$, it is a harmonic two-form
$\psi^A\in\mathcal{H}^2(X,g_X)$.

The variation of the ten-dimensional kinetic term is proportional to
\bea\label{G3kinetic}
\frac{1}{2}\int_{\mathcal{M}} \delta G_3\wedge *_{10} \delta G_3&\propto& e_{AB}\int_{\R^4} \diff\delta\varphi^A\wedge *_4 \diff\delta\varphi^B~,
\eea
where
\bea
{e_{AB}} \,=\,  \int_X \psi^A \wedge *_6 \psi^B~.
\eea
Note that $e_{AB}$ is a symmetric matrix. It may therefore be diagonalised by an orthogonal
change of basis for the $\psi^A$, accompanied by a corresponding change of basis for the fields $\delta\varphi^A$.
In such a basis one obtains canonical kinetic terms on the right hand side of (\ref{G3kinetic}), with
\bea
{e_{AB}} \,=\, \delta_{AB}\int_X \psi^A \wedge *_6 \psi^A\qquad \mathrm{(no}\ \mathrm{sum)}~.
\eea
Notice again that the warp factor $H$ has essentially dropped out of the calculation.
The constants $e_{AB}$ are finite precisely when the $\psi^A$ are
$L^2$ normalisable on $(X,g_X)$. Thus $\psi^A\in\mathcal{H}^2_{L^2}(X,g_X)$.
Using (\ref{tamas}) we see that there are indeed precisely
$b_4(X)$ $L^2$ harmonic two-forms $\psi^A$ on $(X,g_X)$, as required by the analysis above.

Finally, let us consider the asymptotics of the forms $\psi^A$ for large $r$. By construction
we require the $\psi^A$ to be closed and co-closed; this of course implies they are harmonic,
but in general the converse is not true. However, provided $(X,g_X)$ is complete and one
considers  $L^2$ harmonic forms, harmonic is indeed equivalent to closed and co-closed.
On any asymptotically conical manifold, a closed and co-closed form $\psi$
has an asymptotic large $r$ expansion\footnote{We thank T. Pacini for discussions on
the existence of this expansion.} \cite{melrose} of the form
\bea\label{tom}
\psi \,=\, \psi_0 + o(r^{\gamma})~.
\eea
Here $\psi_0$ is closed and co-closed on the \emph{cone} $C(Y)$, and has $L^2$ norm, with respect to the
cone metric,
\bea
\|\psi_0\|\, =\, r^{\gamma}~.
\eea
Even more precisely, $\psi_0$ is one of the homogeneous modes listed in appendix \ref{appendixA},
and thus $\gamma$ takes only a countable set of special values. The notation $o(r^{\gamma})$ in (\ref{tom}) denotes
those forms whose norms are $o(r^{\gamma})$ in the limit $r\rightarrow\infty$.

In the case at hand, we have $p=2$, $n=3$,
in the notation of appendix \ref{appendixA}. Table \ref{modes} implies that the only modes that
are $L^2$ (denoted $L^2_{\infty}$ in the appendix) are of type II and III$^{-}$. Modes of type II
require a harmonic one-form on $(Y,g_Y)$, and since $b_1(Y)=0$ we see that there are no modes of type II.
Thus
\bea
\label{modesofC2}
\psi_0 \,=\,  r^{-1-\sqrt{1+\mu}}\diff\beta_\mu^{(1)}-(1+\sqrt{1+\mu})r^{-2-\sqrt{1+\mu}}\diff r\wedge\beta_{\mu}^{(1)}
\eea
where $\beta_\mu^{(1)}$ is a co-closed  one-form on $(Y,g_Y)$ which is an eigenfunction of the Laplacian
$\Delta_Y$ with eigenvalue $\mu>0$. In particular, this gives
\bea
\gamma\, =\, -3-\sqrt{1+\mu}~.
\eea
Note also that
\bea
\psi\mid_Y \,=\, \lim_{r\rightarrow\infty} \psi\mid_{Y_r} \,=\,  \lim_{r\rightarrow\infty} \left(r^{-1-\sqrt{1+\mu}}\diff\beta_\mu^{(1)} + o(r^{\gamma})\right)\,=\, 0~.\eea
This is consistent with the fact that we require the fluctuations to preserve the boundary conditions
at infinity.
Of course, this analysis is not sufficient to determine which particular mode $\beta_\mu^{(1)}$ is associated
to each $\psi^A$.

\subsection{$B$ field moduli}
\label{Bfield}

The fluctuations of the $B$ field are rather similar. One added complication, however, is
that $\tilde{G_5}$ is no longer invariant. If we write
\bea\label{Bfluc}
\delta B \, =\,  \frac{1}{\mu_1}\delta\sigma^A\psi^A
\eea
then we may keep $\tilde{G_5}$ invariant, and thus in particular also self-dual, if we also vary
\bea\label{inducedC4}
\delta C_4 \,=\,  \frac{1}{\mu_1^2} \varphi_A\delta\sigma^A \psi^A\wedge\psi^A~.
\eea
Note that
the form $\psi^A\wedge\psi^A$ represents a class in $H^4(X,Y;\R)$.  Apart from these minor
differences, the analysis is identical to that in the previous subsection, with similar conclusions. In fact, supersymmetry
pairs the $C_2$ field with the $B$ field, and thus this is expected.

\subsection{$C_4$ field moduli}
\label{c4field}

In order to satisfy the self-duality condition (\ref{selfpert}) we
take the following ansatz, essentially as in \cite{Klebanov:2007cx}
\bea
\delta G_5 \,= \, \frac{1}{\mu_3} (1+*_{10}) \left(\diff \delta\vartheta^M\wedge\Psi^M \right) ~.
\label{G5ansatz}
\eea
The fluctuation of $C_4$ that leads to this will be described below. Here the
$\delta\vartheta^M$, $M=1,\ldots,b_2(X)$, are $b_2(X)$ functions on $\R^4$, and the $\Psi^M$
are representatives of $H^4_{\mathrm{free}}(X,Y;\Z)$. Thus the $\Psi^M$
are closed four-forms on $X$ with integral periods that vanish on $Y$.
The linearised Bianchi identity implies that the scalars $\delta\vartheta^M$ satisfy the
equation of motion
\bea\label{nomass}
\diff *_4 \diff  \delta\vartheta^M\, = \, 0~,
\eea
together with the requirement that
\bea
\diff (H^{-1} *_6 \Psi^M ) \, = \, 0~.
\eea
Recall that $H(x_i)^{-1}=0$
at the locations $x_i$, $i=1,\ldots,m$, of the $m$ stacks of D3-branes.
Since the $\Psi^M$ are closed and co-closed on $(X\setminus\{x_1,\ldots,x_m\},H g_X)$  they  define harmonic four-forms
$\Psi^M\in\mathcal{H}^4(X\setminus\{x_1,\ldots,x_m\},H g_X)$. Equivalently, their duals
\bea
\Phi^M\, \equiv \,  H^{-1}*_6\Psi^M
\eea
define harmonic two-forms $\Phi^M\in\mathcal{H}^2(X\setminus\{x_1,\ldots,x_m\},H g_X)$.

In \cite{Klebanov:2007cx} the equations for such a
harmonic form on the warped resolved conifold  were written down, and it was argued
that there exists a unique solution such that the two-form (denoted $W$ in \cite{Klebanov:2007cx}) is $L^2$ normalisable
in the warped
metric $H g_X$. Below we show that the results of \cite{Klebanov:2007cx} may be generalised.
After dualising the scalar fields $\delta\vartheta^M$ to a corresponding set of
two-forms $\delta a^M$ on $\R^4$,
\bea
\diff \delta a^M \,=\, *_4\diff\delta\vartheta^M~,
\eea
the fluctuation of $C_4$ that gives rise to (\ref{G5ansatz}) is
\bea\label{C4fluc}
\delta C_4\, =\, \frac{1}{\mu_3}\left(\delta\vartheta^M \Psi^M+\delta a^M\wedge\Phi^M\right)~.
\eea
Since $\tilde{G_5}$ is a self-dual five-form its kinetic term vanishes identically.
Moreover, since the fluctuation (\ref{G5ansatz}) is self-dual it automatically solves
(\ref{selfpert}) and (\ref{fluc3}). Following \cite{Klebanov:2007cx} we impose
a normalisablity condition obtained by inserting
the variation in $C_4$ due to $\delta \vartheta^M$ into the ten-dimensional action. This gives
the four-dimensional kinetic term
\bea
f_{MN} \int_{\R^4} \diff \delta\vartheta^M \wedge *_4 \diff \delta\vartheta^N
\eea
where the constants $f_{MN}$ are defined by
\bea
f_{MN} \, =\,  \int_X  \, H^{-1}  \Psi^M \wedge *_6 \Psi^N~.
\eea
As before, an orthogonal change of basis leads to
\bea
f_{MN}\, = \, \delta_{MN} \, \int_X  \, H^{-1} \Psi^M \wedge *_6  \Psi^M\qquad \mathrm{(no}\ \mathrm{sum)}~.
\eea
The constants $f_{MN}$ are therefore finite when the $\Psi^M$ are
$L^2$ normalisable
on $(X\setminus\{x_1,\ldots,x_m\},H g_X)$, or equivalently  $\Phi^M \in\mathcal{H}^2_{L^2}(X\setminus\{x_1,\ldots,x_m\},H g_X)$.

Remarkably, it turns out that  one may argue that precisely $b_2(X)$ such $L^2$ harmonic forms exist on
 $(X\setminus\{x_1,\ldots,x_m\},H g_X)$. Recall that $(X,g_X)$ is a complete asymptotically conical manifold, asymptotic to a
cone over $(Y,g_Y)$. To construct the function $H$ we pick $m$ points $x_i\in X$, $i=1,\ldots,m$, where near
to each point $H$ behaves as in (\ref{Hpoints}). Thus near to $x_i$ the metric $Hg_X$ looks like
\bea\label{Euc}
\frac{L_i^4}{\rho_i^4}(\diff\rho_i^2+\rho_i^2g_{S^5})~.
\eea
If $x_i$ is a singular point with link $(Y_i,g_{Y_i})$ then obviously one replaces $g_{S^5}$ with $g_{Y_i}$.
Defining $R_i=L_i^2/\rho_i$ one sees that the metric (\ref{Euc}) is flat
\bea\label{Eucregion}
\diff R_i^2+R_i^2g_{S^5}~.
\eea
The point $x_i$ is thus at infinity in $(X\setminus\{x_1,\ldots,x_m\},Hg_X)$. On the other hand, near $r=\infty$ the metric
$Hg_X$ approaches
\bea
\frac{L^4}{r^4}(\diff r^2+r^2g_{Y})~.
\eea
Setting $\rho=L^2/r$ we similarly obtain
\bea\label{cheegercone}
\diff \rho^2+\rho^2 g_{Y}
\eea
where $r=\infty$ is the isolated conical singularity $\rho=0$.

The manifold $(X\setminus\{x_1,\ldots,x_m\},Hg_X)$ thus has an isolated conical singularity, near which the metric
looks like the incomplete cone (\ref{cheegercone}),  and $m$ asymptotically
Euclidean regions of the form (\ref{Eucregion}). In particular, $(X\setminus\{x_1,\ldots,x_m\},Hg_X)$ is
asymptotically conical near to each $x_i$, which is a point at infinity in the metric $Hg_X$. If $(Y,g_Y)$ is the round sphere, $(X\setminus\{x_1,\ldots,x_m\},Hg_X)$ is smooth
 and asymptotically conical and we may apply
the results of \cite{Hausel}, summarised in (\ref{tamas}), to determine
the $L^2$ harmonic forms. The UV conformal field theory is $\mathcal{N}=4$ Yang-Mills, and in this case we find
that there are $b_2(X)=0$
such harmonic forms. More generally, the space of interest
has an isolated conical singularity at $\rho=0$. The $L^2$ harmonic forms
on a \emph{compact} manifold $(\bar{X},g_{\bar{X}})$ with isolated conical singularities were studied
by Cheeger in \cite{cheeger}. If $X$ denotes the smooth part of $\bar{X}$, {\it i.e.} $\bar{X}$ with the point
$\rho=0$ in (\ref{cheegercone}) deleted, then
the result for two-forms in dimension six is \cite{cheeger}
\bea
\mathcal{H}^2_{L^2}(\bar{X},g_{\bar{X}}) \, \cong \, H^2(X;\R)~.
\eea
Of course our manifold is not compact, but instead has $m$ asymptotically Euclidean regions.
However, because both types of behaviour -- asymptotically Euclidean ends and isolated conical singularities --
lead to topological results for the $L^2$ cohomology,
one may put the analytic and topological results of \cite{cheeger} and \cite{Hausel} together
to show\footnote{We are extremely grateful to E. Hunsicker and T. Hausel for discussions on this point.} that the $L^2$ harmonic two-forms on $(X\setminus\{x_1,\ldots,x_m\},Hg_X)$ are given by
\bea\label{hunsicker}
\mathcal{H}^2_{L^2}(X\setminus\{x_1,\ldots,x_m\},Hg_X)\, \cong \, H^2(X\setminus\{x_1,\ldots,x_m\},\cup_{i=1}^m S^5;\R)\, \cong \, H^2(X;\R)~.
\eea
Here the copies of $S^5$ are boundaries around the points $x_i$. Thus there are indeed $b_2(X)$ $L^2$ harmonic two-forms
$\Phi^M$, $M=1,\ldots,b_2(X)$, on $(X\setminus\{x_1,\ldots,x_m\},Hg_X)$.

Finally, we consider the asymptotic behaviour of the forms $\Phi^M$ as $r\rightarrow\infty$.
Replacing $\rho=L^2/r$, this becomes $\rho\rightarrow0$. There is then an asymptotic expansion as $\rho\rightarrow0$
\bea
\Phi \,=\, \Phi_0+o(\rho^{\gamma})
\eea
where $\Phi_0$ is a homogeneous closed and co-closed form on the cone $C(Y)$, $g_{C(Y)}=\diff\rho^2+\rho^2g_Y$, with norm
\bea
\|\Phi_0\|\, =\, \rho^{\gamma}~.
\eea
Since we require the $\Phi^M$ to be $L^2$ with respect to the metric $Hg_X$, we are interested in the case $p=2$, $n=3$
and $L^2_0$ in Table \ref{modes}. The two possible modes are thus of type I and type III$^{+}$.
Thus
\bea\label{naharm}
\mathrm{(I)}&:&\Phi_0=\alpha_0^{(2)}\\
&&\gamma=-2\\
\mathrm{(III)}^{+}&:&
\Phi_0=\rho^{-1+\sqrt{1+\mu}}\diff\beta_{\mu}^{(1)}+(-1+\sqrt{1+\mu})\rho^{-2+\sqrt{1+\mu}}\diff\rho\wedge \beta_\mu^{(1)}\\
&&\gamma = -3+\sqrt{1+\mu}
\eea
where $\alpha_0^{(2)}$ is a harmonic two-form on $(Y,g_Y)$, and $\beta_\mu^{(1)}$ again denotes
a co-closed one-form on $(Y,g_Y)$ which is an eigenfunction
of the Laplacian $\Delta_Y$ with eigenvalue $\mu>0$.

To determine which type of asymptotic behaviour we have we may use a topological argument.
Consider the long exact sequence
\bea\label{exactamundo}
&&0\, \cong \, H^1(Y;\R)\longrightarrow H^2(X,Y;\R)\stackrel{f}{\longrightarrow} H^2(X;\R)\longrightarrow \nn\\
&&\longrightarrow H^2(Y;\R)\longrightarrow H^3(X,Y;\R)\, \cong \, 0~.
\eea
From (\ref{hunsicker}) the $b_2(X)$ $L^2$ harmonic two-forms $\Phi^M$, $M=1,\ldots,b_2(X)$, define a basis for $H^2(X;\R)$. The sequence
(\ref{exactamundo})
implies that we may choose this basis such that $b_3(Y)$ restrict to non-trivial classes in $H^2(Y;\R)$,
while $b_4(X)=b_2(X)-b_3(Y)$ restrict to trivial classes in $H^2(Y;\R)$. Let us denote these by $\Phi^I$, $I=1,\ldots,b_3(Y),$
and $\Phi^{b_3(Y)+A}$, $A=1,\ldots,b_4(X)$, respectively.
We have $Y=\lim_{\rho\rightarrow 0}Y_{\rho}$. Thus
\bea
\mathrm{(I)}&:& \Phi\mid_Y=\lim_{\rho\rightarrow0}\left(\alpha_0^{(2)}+o\left(\rho^{-2}\right)\right)=\alpha_0^{(2)}\\
\mathrm{(III)}^+&:& \Phi\mid_Y=\lim_{\rho\rightarrow0}\left(\rho^{-1+\sqrt{1+\mu}}\diff\beta_{\mu}^{(1)}+o\left(\rho^{-3+\sqrt{1+\mu}}\right)\right)=0~.
\eea
These statements may look slightly odd, given that $\rho^{-2}\rightarrow\infty$ as $\rho\rightarrow0$.
However, recall that the notation $o(\rho^{\gamma})$ refers to forms which have \emph{norms}
of order $o(\rho^{\gamma})$ as $\rho\rightarrow0$. In a neighbourhood
of $\rho=0$ such a form may be written
\bea
\phi(\rho)\,=\, \alpha(\rho)+\diff\rho\wedge\beta(\rho)
\eea
where $\alpha(\rho)$, $\beta(\rho)$ are forms on $Y_{\rho}$. If $\phi(\rho)$ is a $p$-form, its square norm, to leading order as $\rho\rightarrow0$, is
\bea
\|\phi(\rho)\|^2\, =\, \rho^{-2p}\left(\|\alpha(\rho)\|_Y^2+\rho^2\|\beta(\rho)\|_Y^2\right)
\eea
where $\|\cdot\|_Y$ denotes the pointwise norm on $(Y,g_Y)$. For us $p=2$ and thus we see that
$\|\alpha(\rho)\|_Y$ is $o(1)$ for modes of type I and $o\left(\rho^{-1+\sqrt{1+\mu}}\right)$
for modes of type III$^+$. In both cases
\bea
\lim_{\rho\rightarrow0}\|\alpha(\rho)\|_Y\, =\, 0 \, \Longrightarrow \, \lim_{\rho\rightarrow0} \alpha(\rho)=0~.
\eea

In particular note that the harmonic two-forms $\Phi^I$, $I=1,\ldots,b_3(Y)$, are asymptotic to the
$b_3(Y)$ harmonic two-forms on $(Y,g_Y)$. This generalises the warped resolved conifold result of \cite{Klebanov:2007cx}.
Note also that the dual four-forms $\Psi^M$, in either case, satisfy
\bea
\Psi^M\mid_Y=\lim_{\rho\rightarrow 0} \Psi^M\mid_{Y_\rho}\, =\, 0~.
\eea

We summarise the properties of the fluctuations discussed so far in Table \ref{fluctable}.

\begin{table}[ht!]
\centerline{
\begin{tabular}{|c|c|c|c|c|c|}
\hline
number & fluctuations & harmonic & mode & $\mathcal{H}^2_{L^2}(Hg_X)$  & $\mathcal{H}^2_{L^2}(g_X)$  \\
\hline
\hline
$b_3(Y)$ & $\delta C_2, ~\delta B$ & $-$ &  I   & $-$   & no \\
\hline
\hline
$b_4(X)$ & $\delta C_2, ~\delta B$ & $\psi^A$ & III$^-$ &  $-$ & yes  \\
\hline
$b_4(X)$ & $\delta C_4$  &  $\Phi^{b_3(Y)+A}$ & III$^+$ & yes & $-$ \\
\hline
$b_3(Y)$ & $\delta C_4$  &  $\Phi^I$ &  I & yes & $-$  \\
\hline
\end{tabular}
}
\caption{Square-integrability of the moduli fluctuations (\emph{cf}.
Table \ref{modes}). The metric fluctuations, that must pair with $\delta C_4$,
 will be discussed in subsection \ref{metricfluc}.}
\label{fluctable}
\end{table}

\subsection{Metric moduli}
\label{metricfluc}

In this section we consider linearised fluctuations of the metric. In principle one should write the
full set of linearised equations for both metric perturbations and also the form fields
$C_2$, $B$ and $C_4$ discussed thus far. As mentioned earlier, although we have fluctuated
the form fields separately in the previous subsections, it is straightforward to substitute
(\ref{varyC2}), (\ref{Bfluc}), (\ref{inducedC4}), (\ref{C4fluc}) into the linearised
equations of motion and verify that these modes are in fact completely decoupled.
As we discuss in this section, the metric modes are rather more involved.
The first problem is to identify the linearised perturbations of asymptotically conical
Ricci-flat K\"ahler metrics on $X$ {\it i.e.} the tangent space to the latter space. We give a partial treatment of this that will
be sufficient to relate the metric modes to the analysis of form field
modes in the previous subsections. For example, this will allow us to
determine the asymptotic eigenvalues $\mu$ in
(\ref{modesofC2}) for certain examples. The second problem is to understand how to promote
these linearised perturbations of the Ricci-flat K\"ahler metric on $X$ to an ansatz that allows these
modes to depend on position in $\R^4$, as we have done in previous subsections. This is surprisingly
complicated for warped Calabi-Yau geometries -- see, for example, \cite{GM} or the very recent paper
\cite{douglaswarped}. From supersymmetry one naively expects to obtain
$b_2(X)$ functions on $\R^4$ satisfying the equation for massless scalar fields, which pair
with the modes of $C_4$ discussed in the previous subsection. However, to show this
rigorously would require substantially more work, not least since the Calabi-Yau manifolds
here are non-compact. We instead simply summarise some of the issues
involved, and refer to the literature for further details.

Before discussing the metric modes, we note that it is \emph{not} possible
for the massless fields found in the previous subsections to obtain masses by
mixing with additional modes that we may turn on. Indeed, this is rather a general statement.
Suppose one has scalar fields $\varphi_i$, $i=0,\ldots,k$, with equation of motion of the general form
\bea
\nabla^2 \varphi_i \, = \, M_{ij} \, \varphi_j + \, \mathrm{higher} \ \mathrm{order}
\eea
where the form of the higher order terms is irrelevant. The physical masses
are obtained by diagonalising the mass matrix $M_{ij}$.
Indeed, we shall encounter precisely such a phenomenon later in the context
of KK theory on AdS$_5\times Y$, where
the $C_4$ field mixes with metric modes producing a non-trivial $2\times 2$
mass matrix (see equation (\ref{massmatrix})). However, in the case at hand
we have shown that setting $\varphi_i\equiv0$ for all $i=1,\ldots,k$, with
$\varphi_0$ a \emph{massless} scalar in four dimensions, solves the
equations of motion. Here the fields $\varphi_i$, $i=1,\ldots,k$, are any modes that
we have not fluctuated in the previous subsections. This immediately implies that $M_{j0}=0$ for
all $j=0,\ldots,k$. Thus the mass matrix necessarily has a zero eigenvalue, although
note that in the process of diagonalising the mass matrix this massless field
will typically be a mixture of $\varphi_0$ with the other fields $\varphi_i$,
$i=1,\ldots,k$. However, the important point is that there is necessarily a \emph{massless}
combination of the modes.

Our conjecture in section \ref{metricsection} implies that there should be
a $b_2(X)$-dimensional K\"ahler moduli space for asymptotically conical Ricci-flat K\"ahler metrics
on a crepant resolution $X$ of a Calabi-Yau cone singularity $Z=C(Y)$.
We may define the $b_2(X)$ K\"ahler classes as
\bea
\xi_M\, =\, \int_{S_M}\omega_X
\eea
where $S_M$, $M=1, \ldots b_2(X)$, denotes a basis for $H_2^{\mathrm{free}}(X;\Z)$.
Note that the exact sequence
\bea
&&0\, \cong \, H_3(X,Y;\R)\longrightarrow H_2(Y;\R){\longrightarrow} H_2(X;\R)\longrightarrow \nn\\
&&\longrightarrow H_2(X,Y;\R)\longrightarrow H_1(Y;\R)\cong \, 0 \,
\eea
means we may split the $S_M$ into $S_I$, $I=1,\ldots,b_3(Y)$, and $S_{b_3(Y)+A}$,
$A=1,\ldots,b_4(X)$. The former are images of $H_2(Y;\R)$ in $H_2(X;\R)$ {\it i.e.}
two-cycles on $X$ that arise from two-cycles on $Y$.

The tangent space to the space of asymptotically conical Ricci-flat K\"ahler metrics on
$(X,g_X)$ should thus be $b_2(X)$-dimensional. We begin by showing that a $b_4(X)$-dimensional subspace
of these linearised perturbations indeed exist, and may be identified with the $b_4(X)$ $L^2$ harmonic two-forms
$\psi^A$ that enter the $C_2$ field and $B$ field fluctuations of sections \ref{C2field} and \ref{Bfield}, respectively.

We may phrase the equations for a Calabi-Yau metric
in terms of the K\"ahler form $\omega_X$ and the holomorphic volume form $\Omega_X$. These satisfy
\bea\label{volumes}
\frac{1}{3!}\omega_X^3 &=& \frac{i}{8}\Omega_X\wedge\bar{\Omega}_X\\
\omega_X\wedge\Omega_X&=&0\\
\diff\omega_X&=&0\\
\diff\Omega_X&=&0~.
\eea
If we fix $\Omega_X$,
the linearised equations for $\delta\omega_X$ are then
\bea\label{primeqn}
\omega_X^2\wedge\delta\omega_X&=&0\\\label{11eqn}
\delta\omega_X\wedge\Omega_X&=&0\\\label{closedeqn}
\diff\delta\omega_X&=&0~.
\eea
We now show that one may solve these equations using the basis of
two-forms $\psi^A$ for $\mathcal{H}^2_{L^2}(X,g_X)\cong H^2(X,Y;\R)\cong H_4(X;\R)$. That is, we take $\delta\omega_X\in \mathcal{H}^2_{L^2}(X,g_X)$.
Note that the $L^2$ condition on $\delta\omega_X$ is the same as that
for a corresponding change in the metric $\delta g_X$, with the natural norm
\bea\label{metricnorm}
\|\delta g_X\|_{L^2}^2 \, =\,  \int_X \diff^6 y \, \sqrt{\det g_X}\, g^{ii'}_X  g_X ^{jj'} \delta g_{X\, ij} \delta g_{X\, i'j'}~.
\eea
Since $(X,g_X)$ is complete and the harmonic forms are $L^2$, they are also both closed and co-closed,
and thus satisfy (\ref{closedeqn}). Here the co-closed condition is the standard gauge-fixing
condition $\nabla^i\delta g_{Xij}=0$ -- see, for example, \cite{besse}.

Suppose that $\alpha$ is a two-form on $X$, $\alpha\in \Omega^2(X)$. Recall that the complex structure
tensor $J$ acts on a two-form $\alpha$ via
\bea\label{Jaction}
(J\circ \alpha)_{ij}\, = \, J^m_i J^n_j \alpha_{mn}~.
\eea
This action squares to the identity. We may thus introduce the projection maps
\bea
\pi_{\pm}: \Omega^2(X)\rightarrow\Omega^2_{\pm}(X)
\eea
defined by
\bea
\pi_{\pm}\alpha\, =\, \frac{1}{2}\left(\alpha+J\circ\alpha\right)\, =\, \alpha_\pm~.
\eea
The splitting of real two-forms into
the $\pm1$ eigenspaces
corresponds, over $\C$, to the splitting into
forms of type $(1,1)$, and types $(2,0)$, $(0,2)$ respectively.
In particular, a two-form $\alpha_-$ with eigenvalue $-1$ under (\ref{Jaction}) may be written
as the real part of a $(2,0)$-form $\alpha^{2,0}\in\Omega^{2,0}(X)$; so
\bea
\alpha_- = \alpha^{2,0}+\overline{\alpha^{2,0}}\eea
where $\overline{\alpha^{2,0}}\in\Omega^{0,2}(X)$. By a slight abuse
of notation, we will refer to $\alpha_-$ as type $(2,0)$ (or equivalently type $(0,2)$).

On a K\"ahler manifold, if $\alpha$ is harmonic then it is easy to show that $\alpha_{\pm}$ are in fact separately harmonic.
One way to see this is as follows.
We note that for any two-form $\alpha$ we have the Weitzenb\"ock formula
\bea
(\Delta\alpha)_{ij} = -\nabla^2\alpha_{ij}-2R_{ij}^{\ \ mn}\alpha_{mn}-2R^m_{\ [ i}\alpha_{j]m}
\eea
where $\nabla^2=\nabla^m\nabla_m$. It follows that
\bea\label{norbert}
(\Delta ( J\circ\alpha))_{ij}= -\nabla^m\nabla_m\left(J_i^{\ p}J_j^{\ q}\alpha_{pq}\right)-2R_{ij}^{\ \ mn}J_m^{\ p}J_n^{\ q}\alpha_{pq} - 2 R^m_{\ [i} J_{j]}^{\ p}J_m^{\ q}\alpha_{pq}~.
\eea
On a K\"ahler manifold we have
\bea
\nabla J \, =\,  0
\eea
and also the curvature identity
\bea\label{Kahlercurv}
R_{ij}^{\ \ mn}J_m^{\ p}J_n^{\ q} \, =\,  R_{ij}^{\ \ pq}\,  =\,  J_i^{\ m}J_j^{\ n}R_{mn}^{\ \ \ pq}~.
\eea
In fact on any Riemannian manifold $(X,g_X)$ where the holonomy group is $G\subset O(n)$, the Riemann tensor
may be regarded as an element of $\mathrm{Sym}^2\Omega^2_{\mathtt{g}}(X)$, where $\mathtt{g}$ is the Lie
algebra of $G$.  Here the symmetric product in $\mathrm{Sym}^2\Omega^2(X)$ reflects the algebraic identity
$R_{ijpq}=R_{pqij}$. The equation (\ref{Kahlercurv}) is precisely the statement
that the Riemann tensor on a K\"ahler manifold is in $\mathrm{Sym}^2\Omega^2_+(X)$. It follows
from (\ref{norbert}) that when $\alpha$ is harmonic we have $(\Delta ( J\circ\alpha)) = J\circ (\Delta\alpha)=0$,
and thus $\Delta\alpha_{\pm}=0$.

If we take $\alpha\in \mathcal{H}^2_{L^2}(X,g_X)$, then $\alpha_\pm$ are also both $L^2$ since
it is straightforward to show that
\bea
\|\alpha\|^2 \,=\, \|\alpha_+\|^2+\|\alpha_-\|^2~.
\eea
Thus $\alpha_{\pm}\in \mathcal{H}^2_{L^2}(X,g_X)$. As discussed in section \ref{metricsection},
all the cohomology of $X$ in degree two is
of type $(1,1)$. Since $\alpha_-$ is harmonic and of type $(2,0)$,
it represents a cohomology class of type $(2,0)$. But since
any such class is trivial, it follows from (\ref{tamas}) that
$\alpha_-=0$ -- in particular, note that (\ref{tamas}) implies that all non-zero
$L^2$ harmonic forms represent non-trivial cohomology classes. Thus $\alpha$ is necessarily of type $(1,1)$.
Finally, consider $\omega_X^2\wedge\alpha$. This is an $L^2$ harmonic six-form.
However, again by (\ref{tamas}) we see that any such six-form must be zero,
since $\mathcal{H}^6_{L^2}(X,g_X)\cong H^6(X;\R)\cong 0$.

This shows that the $b_4(X)$ $L^2$ harmonic two-forms $\psi^A$
satisfy all of the equations (\ref{primeqn}), (\ref{11eqn}), (\ref{closedeqn}).
To conclude our proof that these are indeed tangent directions to the space of
asymptotically conical Calabi-Yau metrics on $X$,
we must show that taking $\delta\omega_X\in \mathcal{H}^2(X,g_X)$
preserves the asymptotically conical condition -- that is, the $L^2$ forms do not
grow too fast. To do this we may again appeal to the results of appendix \ref{appendixA}.
A closed and co-closed form $\alpha$ has an asymptotic expansion
\bea
\alpha \,=\, \alpha_0 + o(r^{\gamma})
\eea
where $\alpha_0$ is one of the closed and co-closed homogeneous modes listed in
the appendix, and $\|\alpha_0\|=r^{\gamma}$ where the norm is taken with respect
to the cone metric. From Table \ref{modes}, the only possible modes are of type II
(of which there are none since $b_1(Y)=0$) and type III$^-$. We thus have
\bea
\alpha_0 \,=\, r^{-1-\sqrt{1+\mu}}\diff\beta_\mu^{(1)} - (1+\sqrt{1+\mu})r^{-2-\sqrt{1+\mu}}\diff r\wedge\beta_\mu^{(1)}
\label{metricforms}
\eea
with $\mu>0$. It follows that these perturbations are indeed subleading to the cone metric
near infinity. These are of course the same expansions (\ref{modesofC2})
for the harmonic two-forms
entering the fluctuations $\delta B$ and $\delta C_2$.

Thus there is certainly a $b_4(X)$-dimensional space of linearised perturbations of an
asymptotically conical Ricci-flat K\"ahler manifold, where the perturbations are $L^2$
with respect to the natural norm (\ref{metricnorm}). In fact, this may be understood
from supersymmetry in the case without any background D3-branes. Such fluctuations of the metric are then paired by
supersymmetry to fluctuations of the RR $C_4$ potential
on the background $\R^4\times X$, as in section \ref{anomaloussection}. In this case the number
of $L^2$ harmonic four-forms on $(X,g_X)$ is given by (\ref{tamas}), which indeed gives
$\dim H^4(X;\R)=b_4(X)$ $L^2$ modes.

The remaining $b_2(X)-b_4(X)=b_3(Y)$ linearised perturbations are thus
not $L^2$-normalisable. Assuming these exist, their asymptotics may be understood as follows\footnote{This
is essentially taken from \cite{Benvenuti:2005qb}, although here
we make the argument more rigorous by using the asymptotic expansion together with the results of
appendix \ref{appendixA}.}.
Consider integrating the closed two-form $\delta\omega_X$ over a two-cycle $S_I$ that is homologous to
a two-cycle on $Y$. This gives a change in the K\"ahler class
\bea
\delta \xi_I  \, =\, \int_{S_I} \delta \omega_X~.
\label{provolone}
\eea
Near infinity we may write
\bea\label{deltaomega}
\delta\omega_X \,=\, \alpha(r)+\diff r\wedge\beta(r)
\eea
where $\alpha(r)$, $\beta(r)$ are forms on $Y_r$. Since $\delta\omega_X$ is closed, and also co-closed
in the usual gauge $\nabla^i\delta g_{Xij}=0$ which fixes diffeomorphism invariance, the
right hand side of (\ref{deltaomega}) will have an asymptotic expansion with leading
term given by a closed and co-closed mode of appendix \ref{appendixA}.
The cycle $S_I$ in (\ref{provolone}) may be represented by a cycle in
$Y_r$. The only mode for which the integral (\ref{provolone}) is both finite and non-zero
is then mode I {\it i.e.}
\bea\label{harmonicmetric}
\delta\omega_X \,=\, \alpha_0^{(2)}+o(r^{-2})
\eea
where $\alpha_0^{(2)}$ is a harmonic two-form on $Y$. In fact we have
\bea\label{metricmoduli}
\delta \xi_I\, =\, \int_{S_I}\alpha_0^{(2)}~,
\eea
where $S_I$ is regarded as a cycle in $Y$. Note that such fluctuations are indeed not $L^2$ normalisable, as one sees from Table \ref{modes}.
Notice that the K\"ahler perturbation of the resolved conifold, discussed originally in \cite{KW2},
is precisely of this form.

Given the above (partial) understanding of linearised perturbations of asymptotically
conical Calabi-Yau manifolds, one would now like to promote these K\"ahler moduli to
dynamical four-dimensional fields. Recall that the backgrounds of interest are of the form
\bea
g_{10} &=& H^{-1/2}\eta_{\mu\nu} \diff x^\mu \diff x^\nu   +  H^{1/2} g_{Xij} \diff y^i \diff y^j \\
G_5 & = & (1+*_{10}) \diff H^{-1}\wedge\vol_4~.
\label{flucks}
\eea
These are a particular class of the backgrounds considered in \cite{GM} and \cite{douglaswarped}.
The moduli may be parametrised by $b_2(X)$
\emph{constant} parameters $u^M$. Note here that a change in the background
Ricci-flat K\"ahler metric will induce a corresponding change in the warp factor $H$, which satisfies
(\ref{Heqn}).
As emphasised in \cite{GM}, and in sharp contrast to the familiar Calabi-Yau compactifications,
it is not possible to promote straightforwardly the moduli to spacetime-dependent scalar fields $u^M(x)$
in four dimensions. The linearised Type IIB equations of motion cannot be solved by a simple ansatz.
Instead one must introduce certain off-diagonal modes, called compensator fields, which are proportional
to derivatives of the scalar fields $u^M(x)$. The resulting equations, and gauge invariances, are then
rather involved. Note that the non-compactness of our geometries will add to these complications.
Very recently the paper \cite{douglaswarped} has appeared, which claims that
the compensator fields may be effectively removed by choosing a certain ten-dimensional gauge condition.
However, we will postpone a more detailed investigation of these metric modes for future work.
We conclude the section by nevertheless noting that the norm of the metric perturbations induced from the
ten-dimensional
kinetic terms, as studied in
\cite{GM}, \cite{douglaswarped}, is given by the natural \emph{warped} norm
\bea
\|\delta g_X\|^2_{Hg_X} \, = \, \int_X \diff^6 y \, \sqrt{\det g_X}\, H\, g^{ii'}_X  g_X ^{jj'} \delta g_{X\, ij} \delta g_{X\, i'j'}~.
\label{goodmetric}
\eea
Compare this with (\ref{metricnorm}).
In particular, notice that all of the $b_2(X)$ linearised metric perturbations are $L^2$ with respect to this warped norm,
whereas only a $b_4(X)$-dimensional subspace is $L^2$ with respect to the unwarped norm (\ref{metricnorm}).
This implies that all of the metric modes will be normalisable with respect to the physical metric (\ref{goodmetric})
coming from the kinetic terms. Again, this is expected from supersymmetry, since all $b_2(X)$ modes of $C_4$ in
section \ref{c4field} are normalisable.


\section{Spontaneous breaking of baryonic symmetries}
\label{central}

In this section we discuss the dual field theory interpretation of the linearised fluctuations
described in section \ref{fluctuationsection}. For the modes corresponding to non-anomalous
baryonic symmetries our analysis will extend and generalise the resolved conifold result of
\cite{KM}. The holographic interpretation of the
modes corresponding to anomalous baryonic symmetries is less obvious. We will see how
the analysis of these leads us to predict the existence of certain particular
modes in the KK spectrum of AdS$_5\times Y$. We will also speculate on the possibility that
some of these modes correspond to  anomalous baryonic currents.


\subsection{Vacuum fluctuations and Goldstone bosons}

As discussed in section \ref{VMSsection}, the classical gauge theory has
a large VMS $\mathscr{M}$. The potential of the classical theory is identically zero at
any point in this moduli space. One thus expects to find massless scalar fields
associated to these flat directions in field space. In section \ref{section3},
and as summarised in section \ref{comparesection}, we have explained how this
classical vacuum moduli space is realised in the dual gravity description.
Roughly, the mesonic directions correspond to moving the $N$ pointlike D3-braneson $X$.
The ``FI parameters'', which are the image of the map $\Pi$ in (\ref{pimap}),
may be identified with the $b_2(X)$ K\"ahler moduli and the $b_4(X)$ $B$ field
moduli, whereas the $U(1)^{\chi-1}$ fibres over the mesonic spaces may be identified with the RR
torus (\ref{RRXtorus}). In section \ref{fluctuationsection} we have shown
that there do indeed exist massless scalar fields on $\R^4$
associated to linearised deformations of these moduli, at least for the $B$ field
and RR field moduli -- as discussed in section \ref{metricfluc}, the metric moduli
would require more work to make this rigorous. Nonetheless, this is clearly a
very satisfying result. Notice that we have not attempted to describe
massless fields associated with moving the positions of the pointlike D3-branes on
$X$. In principle one could study such deformations, but our main interest in this
paper is with the baryonic symmetries and associated directions in moduli space.
As we have alluded to earlier, the IR theory
is then not simply a SCFT, or even a product of SCFTs, but rather will also include
massless particles corresponding to motion along flat directions in the
field theory. Notice that the description of the fluctuations in terms of
massless fields on $\R^4$ is essentially an application of KK reduction on
the \emph{warped} Calabi-Yau $X$ to $\R^4$.
However,
one may also understand the fluctuations by applying more standard holographic arguments, as
we show later in the section. Thus different aspects of the IR theory may be understood using
both holographic and KK techniques. This is a very interesting aspect of these
gravity backgrounds.

Since the global symmetry group $U(1)^{\chi-1}$ acts on the space of supergravity backgrounds without
fixed points, any choice of vacuum spontaneously breaks this symmetry. This precisely happens
in the classical field theory also, for a generic point in the VMS $\mathscr{M}$.
A spontaneously broken global symmetry generally leads to Goldstone bosons.
These are the same as the massless fields referred to above of course -- they are
flat directions given by acting with a broken symmetry generator. Since the
vacua are supersymmetric, these Goldstone fields will have $\mathcal{N}=1$ superpartners.
The Goldstones are fluctuations of RR fields, and are hence pseudo-scalars. Their
scalar partners come from metric and $B$-field fluctuations
This is precisely the pairing of the RR fields with the K\"ahler moduli and $B$ field moduli.
Thus the linearised fluctuations we have found may be tentatively associated
with these Goldstone bosons and their supersymmetric partners.

However, the above, essentially classical, discussion overlooks an important subtlety:
in the quantum
theory only a $U(1)^{b_3(Y)}$ subgroup of the baryonic symmetry group is non-anomalous,
the remaining symmetries being anomalous and thus broken in the field theory by instantons. Their
corresponding classically conserved currents are thus not conserved in the quantum theory.
By Goldstone's theorem, the massless fields associated to motion in the non-anomalous
directions should be \emph{exactly massless} in the quantum theory.
Thus
the $b_3(Y)$ fluctuations corresponding to modes of $C_4$ of type I in
section \ref{c4field}, and the non-normalisable (with respect to
(\ref{metricnorm})) K\"ahler moduli (\ref{metricmoduli}), should also be exactly massless.
Notice that in both cases these modes are constructed from forms that are asymptotic to the
$b_3(Y)$ harmonic two-forms  on $(Y,g_Y)$ -- see equations (\ref{naharm}) and
(\ref{harmonicmetric}), respectively. That $U(1)^{b_3(Y)}$ is an \emph{exact} symmetry
of the quantum theory is simple to understand in our gravity dual, as we alluded to
earlier: these symmetries
come from gauge transformations of the RR four-form. A gauge symmetry is always non-anomalous, 
otherwise the theory is inconsistent. The relevant gauge transformations are of the form
\bea
C_4 \,  \rightarrow \, C_4+ \diff K
\label{Kgauge}
\eea
where $\diff K \mid_{\partial \mathcal{M}}=0$ but $K\mid_{\partial\mathcal{M}}\neq 0$.
Thus $K\mid_{\partial\mathcal{M}}$ defines a class in $H^3(\partial\mathcal{M};\R)$.
For smooth $X$, this is isomorphic to $H^3(Y;\R)$. The gauge transformation
(\ref{Kgauge}) then changes the compactly supported cohomology class of $C_4$,
which recall we are identifying as part of the background moduli.
In fact the group of global
symmetries generated by such gauge transformations is
$H^3(Y;U(1))$. The group of components $\overline{H}^3(Y;U(1))$ is,
from (\ref{coeff}), isomorphic to $H^4(Y;\Z)\cong H_1(Y;\Z)\cong H_{3,\mathrm{tor}}(Y;\Z)$.
These are discrete non-anomalous baryonic symmetries, and arise only if $Y$
has a non-trivial fundamental group.

Another way to see that the symmetry group $H^3(Y;\R)/H^3_{\mathrm{free}}(Y;\Z)\cong U(1)^{b_3(Y)}$
is completely broken by a background with smooth $X$ is to
note that (\ref{nob3}) implies
$H^3(X;\R)/H^3_{\mathrm{free}}(X;\Z)=0$. Indeed, baryons are interpreted as D3-branes
wrapped on three-submanifolds in $Y$, which are thus charged under the
group $H^3(Y;U(1))$. Since there are no three-cycles on $X$, such
D3-branes may presumably annihilate in the interior of $X$, as discussed in
\cite{KW2}. Thus all non-anomalous baryonic symmetries are broken by a choice of $X$.
So again we expect to find $b_3(Y)$ massless
Goldstone bosons, given by linearised fluctuations of $C_4$, together with their
supersymmetric partners, given by fluctuations of the metric.

The anomalous baryonic symmetries are different, however. The classically conserved
currents are not conserved at quantum level, because of the presence of anomalies, as
we reviewed  in section \ref{anomaloussection}.
Thus Goldstone's theorem
does not apply, and there is {\it a priori} nothing to prevent quantum corrections lifting the
classical massless fields. Correspondingly, in the gravity dual  the anomalous baryonic symmetries are not
associated to gauge transformations. The corresponding massless modes  may in particular be lifted by
corrections to the supergravity backgrounds we have been discussing. For example,
there may well be corrections to the gravity backgrounds of section \ref{section3} coming
from D-brane instantons wrapped on compact even-dimensional cycles in $X$.
These presumably couple to the RR moduli in general, but not to the $b_3(Y)$ modes
associated to gauge transformations of $C_4$. We will not pursue this
line of thought further here, but instead postpone some speculative comments on this topic
to the discussion section \ref{discussionsection}.


\subsection{AdS/CFT interpretation: non-anomalous $U(1)$s}
\label{nasection}

In this subsection and the next we present a holographic analysis of the fluctuations of
section \ref{fluctuationsection}. This requires expanding the fluctuations
at large $r$, which recall involves an asymptotic expansion of closed and co-closed forms on
an asymptotically conical manifold.
We first discuss the interpretation of the $b_3(Y)$ $C_4$ field fluctuations of type I.
The argument generalises the discussion in \cite{Klebanov:2007cx} for the
 non-anomalous\footnote{Recall that the resolved conifold has $b_2(X)=b_3(Y)=1$.}
baryonic $U(1)$ of the conifold model.

Equation (\ref{naharm}) implies that the four-forms $\Psi^I$ Hodge dual under $Hg_X$
 to the $L^2$ harmonic two-forms $\Phi^I\in \mathcal{H}^2_{L^2}(X\setminus\{x_1,\ldots,x_m\},Hg_X)$ satisfy
\bea
\Psi^I\sim r^{-3}\diff r\wedge \alpha_0^{(3)I}
\label{puss}
\eea
to leading order as $r\rightarrow \infty$.  Here $\alpha_0^{(3)I}\equiv*_Y\alpha_0^{(2)I}$  are the $b_3(Y)$ harmonic three-forms
on $(Y,g_Y)$. Making a gauge transformation
\bea
\delta C_4 \, \to \, \delta C_4 +\frac{1}{2\mu_3} \diff K
\label{newgauge}
\eea
where $K$ is a three-form with
\bea
K \, \sim \,  r^{-2}\delta \vartheta^I  \alpha_0^{(3)I}~,
\eea
the first term in the fluctuation (\ref{C4fluc}) may be rewritten
\bea
\delta C_4 \, \sim\,  \frac{1}{\mu_3}r^{-2}\diff\delta\vartheta^I \wedge \alpha^{(3)I}_0~.
\eea
Note that the generator of the gauge transformation in (\ref{newgauge}) vanishes at infinity.

The holographic interpretation of this follows from  comparing to the situation in an
AdS$_5\times Y$ background. Here the harmonic three-forms $\alpha_0^{(3)I}$
lead to  $b_3(Y)$ massless gauge fields $A^I$ in AdS$_5$, via the  ansatz
\bea
\delta C_4 \, = \, \frac{1}{\mu_3}A^I \wedge \alpha_0^{(3)I} \qquad (\mathrm{AdS~background})~.
\label{pureadsb}
\eea
These are dual to $b_3(Y)$ non-anomalous baryonic currents $J^I$. The linearised perturbations of the
non-conformal background we have found therefore
induce, near the UV AdS boundary $r\to\infty$,  a perturbation of  the gauge fields $A^I$ which behaves  as
\bea
\delta A^I \, \sim  \, r^{-2} \diff  \delta\vartheta^I ~.
\label{masslessvec}
\eea
According to the AdS/CFT dictionary, the leading order terms of a perturbation in an (approximately) AdS$_{d+1}$
space  with metric
\bea
\diff s^2_{\mathrm{EAdS}_{d+1}} \, = \, \frac{1}{r^2} \diff r^2+ r^2 g_{\R^d}
\eea
admit  different interpretations in the dual field theory \cite{Balasubramanian:1998de,KW2}.
In general, for a massive $p$-form field ${\cal A}$, obeying
\bea
\diff *_{d+1} \diff {\cal A} - m^2 *_{d+1} {\cal A} \, = \ 0~,
\eea
we have the asymptotic expansion \cite{Mueck:1998iz,Yi}
\bea
{\cal A}_{i_1\dots i_p} \sim B_{i_1\dots i_p} \, r^{p+\Delta-d} + C_{i_1\dots i_p}\,  r^{p-\Delta}
\label{genmap}
\eea
where
\bea\label{conformalformula}
\Delta \, = \, \frac{d}{2} + \sqrt{\left( \frac{d}{2} - p \right)^2 + m^2 }
\eea
is the conformal dimension of the dual operator -- a formula which is perhaps more familiar for scalar fields, $p=0$.
The first term in (\ref{genmap}) is non-normalisable, and therefore corresponds to changing the Lagrangian of
the CFT. If this term vanishes and only the second normalisable term appears, then we are in a vacuum of the theory where
the dual operator has a non-zero expectation value.

Notice that (\ref{masslessvec}) is only computing the leading perturbation of $A^I$
as $r\to\infty$ under the linearised perturbations of the background.
In particular we are not\footnote{Note that computing VEVs in the gravity backgrounds, as opposed to their
linearised variations under a change of vacuum,
would presumably involve finding the general solution to certain (non-linear) ten-dimensional equations
with prescribed behaviour at the UV boundary, substituting this into an appropriately
holographically renormalised action, and then differentiating this once with respect to the
boundary data (sources). Such a computation is clearly well beyond the scope of this paper.
For a discussion in the case of AdS$_5\times S^5$, we refer the reader to \cite{skenderis,skenderis2}.} computing VEVs in the background itself -- this would require  a treatment as in
\cite{skenderis,skenderis2}. However, we may nonetheless naively read off conformal dimensions using these results.
Setting $d=4$, $p=1$ and $m^2=0$ in (\ref{conformalformula}) we see that the currents $J^I$ dual to the vector
fields $A^I$ have conformal dimension $\Delta=3$, which is of course correct for a conserved current.
Equation (\ref{genmap}) then implies that
\bea
\langle \delta J^I \rangle \, =  \, \langle \diff  \, \delta \vartheta^I \rangle~.
\label{dod}
\eea
Standard field theory arguments then allow one to interpret the fields
$\delta \vartheta^I$ as Goldstone bosons of the spontaneously broken $U(1)^{b_3(Y)}$ symmetry,
as in \cite{KM}.  Indeed, the classical Noether current for a complex scalar field
$\phi$ is
\bea
J \,=\, \frac{i}{2}\left(\phi\diff\bar{\phi}-\bar{\phi}\diff\phi\right)~.
\eea
If $\phi$ has a classical VEV $\phi_0$, then we may write the Goldstone fluctuation
as $\phi=\phi_0\e^{i\delta\vartheta}$, and then
\bea
\delta J \,=\, |\phi_0|^2\diff\delta\vartheta~.
\eea
As mentioned in the previous subsection, supersymmetry should pair these Goldstone pseudo-scalars
with scalar superpartners. These are clearly the $b_3(Y)$ non-normalisable K\"ahler deformations
(\ref{metricmoduli}), although as stressed in section \ref{metricfluc} we have not
shown rigorously that these lead to modes satisfying massless scalar field equations in $\R^4$.

Note that the currents $J^I$ are necessarily components of conserved
current multiplets.
The lowest components of these superfields may be identified as follows.
If we follow the
arguments in reference \cite{KW2}, a term of order $r^{-\Delta}$ in the metric,
relative to the cone metric, indicates a VEV for a scalar operator
of conformal dimension $\Delta$. Since (\ref{harmonicmetric}) is order
$r^{-2}$ relative to the cone metric, we see that these metric deformations should be
dual to operators of conformal dimension $\Delta=2$ in the dual field theory. This is precisely
as expected for the scalar component of a massless vector multiplet in AdS$_5$. This leads one, as in the conifold model \cite{KW2}, to identify the $\Delta=2$ scalar operators with
the lowest component of the superfield that contains the non-anomalous baryonic currents $J^I$.
Note these are necessarily axial currents, and thus the associated Goldstone bosons should be
pseudo-scalars, precisely as we have found in the supergravity dual.
The expression for the $b_3(Y)$ scalar operators follows from our discussion of quiver gauge theories in section \ref{section2}:
\bea\label{naU}
\mathcal{U}^I\equiv\mathrm{Tr}\left[\sum_{a\in A} Q_a^I\Phi_a^{\dagger}\Phi_a \right]~.
\eea
Here $Q_a^I$ are the baryonic charges (\ref{baryoniccharges}) with $q=q^I\in \Z^{\chi}$
being the $b_3(Y)$ linearly independent solutions to the anomaly cancellation condition (\ref{triangle}).
Recall that $\Phi_a$ are the bifundamental fields. The superfield version of (\ref{naU}) contains
the conserved currents $J^I$ as the $\theta\bar{\theta}$
components.


\subsection{AdS/CFT interpretation: anomalous $U(1)$s}
\label{asection}

Our aim now is to discuss  a possible holographic interpretation of the moduli fields
associated to the remaining $U(1)^{2b_4(X)}$ flat directions, which correspond to anomalous baryonic symmetries.
A key ingredient in the arguments of the previous subsection was the comparison of the asymptotic
expansion of the fluctuating modes to a background AdS$_5\times Y$ analysis. In particular, it is well-known that
the KK spectrum contains massless vector multiplets (so-called ``Betti'' multiplets)
for each three-cycle in $Y$, which are
dual to conserved currents. On the other hand, there is no general understanding of anomalous baryonic currents
in the context of  AdS$_5\times Y$ backgrounds.

In order to  proceed in analogy with the previous subsection, we will first use the aymptotic expansions
of harmonic forms to describe a set of KK modes in AdS$_5$ to compare with. As we discuss below,
these modes must correspond to the set of lowest
dimension operators in the dual SCFT acquiring non-zero VEVs. Based on supersymmetry considerations,
and in analogy with the non-anomalous $U(1)$s, we will make some comments on the specific form of these
operators.

Recall that in section \ref{C2field} we have shown the existence of $b_4(X)$ harmonic two-forms
 $\psi^A \in  \mathcal{H}^2_{L^2}(X,g_X)$
with the following asymptotic expansion (of type III$^-$, in the notation of appendix \ref{appendixA}) near infinity
\bea
\psi^A ~\sim ~\frac{1}{r^{1+\nu^{(1)A}}} \diff \beta^{(1)A} -(1+\nu^{(1)A})
 \frac{1}{r^{2+\nu^{(1)A}}}\diff r \wedge \beta^{(1)A}~, 
\label{massiveBC}
\eea
where  we have defined $\nu^{(1)A}= \sqrt{1+\mu^{(1)A}}$, and
$\beta^{(1)A}$ are co-closed one-forms on  $Y$ obeying
\bea
\Delta_Y \beta^{(1)A} & = & \mu^{(1)A} \,  \beta^{(1)A}\label{dual2} \qquad \mathrm{(no}\ \mathrm{sum)}~.
\eea

The $L^2$ harmonic two-forms $\psi^A$ are invariant under isometries of $(Y,g_Y)$
that extend to isometries of the resolution $(X,g_X)$. One can prove this using Theorem
3 of \cite{hitchin}. The latter states that Killing vector fields of linear growth
(see \cite{hitchin} for the definition)
leave the $L^2$ cohomology classes of $(X,g_X)$ fixed. Killing vector fields
on $(X,g_X)$ that are Killing on $(Y,g_Y)$ indeed have linear growth (their norms
are $O(r)$), so the
theorem applies. In fact the proof of the theorem shows that
the Lie derivative of
an $L^2$ harmonic form $\psi$ is $L^2$ harmonic with $L^2$ cohomology class zero.
However, again using the results of \cite{Hausel} in (\ref{tamas}),
this implies the Lie derivative is zero, and thus $\psi$ is invariant under
such isometries. Since $r$ is also invariant, we see that $\beta^{(1)}$ is invariant under the isometries of $(Y,g_Y)$ that extend to isometries of $(X,g_X)$.

We must be slightly careful when writing expressions such as (\ref{massiveBC}).
 Here we have picked an arbitrary basis
for $\mathcal{H}^2_{L^2}(X,g_X)\cong H^2(X,Y;\R)$. However, we are clearly free
to choose a different basis. The issue is then that the set of asymptotic
modes $\{\beta^{(1)A}\}$ in one basis is clearly not necessarily the same as in another
basis, since the modes correspond only to the \emph{leading order} behaviour of the harmonic
forms at infinity. For example, by adding some multiple of the harmonic form with
smallest $\mu^{(1)}$ to all the other forms, the leading asymptotic behaviour of all
forms in the new basis will be the same.
On the other hand, one can also clearly take linear combinations in such a way that
the leading terms $\beta^{(1)A}$ are all different (although the set of eigenvalues $\{\mu^{(1)A}\}$ may
of course be degenerate). For, if any two harmonic forms have the same
asymptotic $\beta^{(1)}$, one may simply pick a new basis which uses the difference of
these forms as one of the basis elements; the latter will then have subleading behaviour.
In this way, there exists a
set of $b_4(X)$ distinct eigenfunctions $\{\beta^{(1)A}\}$ on $(Y,g_Y)$.

Next, we will show that massive
co-closed one-forms on $(Y,g_Y)$ are dual to massive co-closed three-forms, in the sense that
given any co-closed one-form $\beta^{(1)}$,  one can construct a co-closed three-form $\beta^{(3)}$
with the \emph{same}
eigenvalue. In particular, we will prove the following:
\bea
*_Y \beta^{(3)} & = &  \diff \beta^{(1)}\label{AY1}\\
*_Y \beta^{(1)} & = & \frac{1}{\mu} \diff \beta^{(3)}~.\label{AY2}
\eea
The argument is rather simple.
Consider a co-closed three-form obeying
\bea\label{dualgen1}
\Delta_Y \beta^{(3)} & = & \mu   \, \beta^{(3)}~.
\eea
We have for any two-cycle $S\subset Y$
\bea
\int_S *_Y\beta^{(3)} \, = \, \frac{1}{\mu} \int_S  \diff *_Y \diff \beta^{(3)} \, = \, 0~,
\eea
where in the first equality we used (\ref{dualgen1}) and the second is Stokes' theorem.
Thus the closed form $*_Y\beta^{(3)}$ is exact, and we may write  (\ref{AY1}) for some
$\beta^{(1)}$.  Note that
by the Hodge decomposition $\beta^{(1)} $  may be taken to be co-closed.
Now define the one-form
\bea
\sigma \, = \, \Delta_Y  \beta^{(1)} - \mu \beta^{(1)}~.
\eea
It is straightforward to verify that this is closed and co-closed, and so must be harmonic. However, since
$b_1(Y)=0$, we have that $\sigma=0$. This shows that
\bea
\Delta_Y \beta^{(1)} = \mu \beta^{(1)}
\eea
and also that the relation (\ref{AY2}) holds. This proves that there exists two sets of $b_4(X)$
one-forms and three-forms $\{\beta^{(1)A}, \beta^{(3)A}\}$ on $(Y,g_Y)$, with pairwise equal
eigenvalues  $\{\mu^{A}\}$.

Given these forms, we can perform a KK reduction and obtain a corresponding set of modes in AdS$_5$.
Let us describe  roughly the types of modes that are obtained from these massive forms.
Consider, for instance, Kaluza-Klein reduction via the ans\"atze
\bea
\begin{array}{c}
\delta C_4 \, = \, \frac{1}{\mu_3}{\cal A}^A \wedge \beta^{(3)A}
\label{bigclaim1}\\[3mm]
\delta C_2 \, = \, \frac{1}{\mu_1}{\cal C}^A \wedge \beta^{(1)A}
\label{bigclaim2}\\
\end{array} \quad  \qquad (\mathrm{AdS~background})~.
\eea
Then, for example, if $\beta^{(3)}$ is a co-closed massive eigenfunction
of $\Delta_Y$ with eigenvalue $\mu$, satisfying (\ref{dualgen1}), we
have
\bea
\delta G_5 \, = \,   \diff {\cal A} \wedge \beta^{(3)} -
{\cal A} \wedge  \diff \beta^{(3)}~.
\eea
Then the linearised equation of motion implies
\bea\label{proca}
\diff *_5 \diff {\cal A} - \mu *_5 {\cal A} &= & 0\\
\diff *_5 {\cal A} & = & 0 ~,
\eea
where $*_5$ is the Hodge operator on AdS$_5$. These are precisely the Proca equations for a massive vector
field. A similar consideration applies for reduction of $\diff \delta C_2 $ on a massive one-form;
the $\mathcal{A}^A$ and ${\cal C}^A$ in (\ref{bigclaim1}) thus obey these equations.
Note that the Proca equations are gauge-fixed. More generally we should write
\bea\label{stuckelansatz}
\delta C_4 \, = \, \mathcal{A}\wedge \beta^{(3)}+\varrho\, \diff\beta^{(3)}
\eea
so that a gauge transformation
\bea
\delta C_4 \, \rightarrow \, \delta C_4 + \diff (f\beta^{(3)})
\eea
leads to the transformations
\bea\label{stuckeltran}
\mathcal{A}\,\rightarrow \,\mathcal{A} + \diff f, \qquad \varrho\rightarrow\varrho +f~.
\eea
The ansatz (\ref{stuckelansatz}) then leads to the St\"uckelberg action in AdS$_5$,
where the scalar $\varrho$ is the St\"uckelberg field. Of course, by a gauge transformation
this scalar may be set to zero, and one recovers the Proca equations.

However, the above analysis is certainly too naive. The reason for this is that
massive modes in AdS$_5$ mix, leading to a non-trivial mass matrix. The physical
masses are then the eigenvalues of this mass matrix. This occurs even for scalars fields,
for example as discussed in the appendix of \cite{KW2}.
In the case at hand, the RR four-form mixes with metric modes, which recall we have not
analysed in any detail. The relevant mass matrix has been worked out for the
$S^5$ case in \cite{kim}, although their results are easily generalised to a general
Sasaki-Einstein manifold $(Y,g_Y)$. The mixing of metric and RR $C_4$ modes due to
massive co-closed three-forms $\beta^{(3)}$ on $Y$ of eigenvalue $\mu$
gives rise to a mass matrix (see equation (2.26) of \cite{kim})
\bea\label{massmatrix}
m^2 \,=\, \left(\begin{array}{cc} \mu+8 & 16\mu \\ 1 & \mu\end{array}\right)~.
\eea
Here the bottom right hand corner corresponds to the mass in the Proca
equation (\ref{proca}),
as one sees explicitly from the analysis in \cite{kim}.
The eigenvalues of this matrix are given by
\bea
m^2_{\pm} \,=\, \mu+4\pm4\sqrt{\mu+1}~.
\eea
On the other hand, there is no such mixing for the $C_2$ and $B$ field modes.
In fact these combine straightforwardly into  complex modes.

The AdS/CFT dictionary maps these $2b_4(X)$ massive vector fields to some
vectorial operators of conformal dimension given by (\ref{conformalformula}) in the dual CFT.
For the vector fields  ${\cal A}^A$, picking the positive branch $m^2_+$ gives rise
to conformal dimensions
\bea
\label{currentdimension}
\Delta ({\cal A}^A)\, =\,  4+\sqrt{1+\mu^A}~.
\eea
Notice that, rather remarkably, the square root for the positive branch factorises, giving
the simple surd in (\ref{currentdimension}).

On general grounds,
these vector fields must of course fit into some supermultiplets in AdS$_5$.
However, to our knowledge, there is no general understanding of the structure of
the KK spectrum for a Sasaki-Einstein five-manifold,
other than  $S^5$ and $T^{1,1}$.
We thus proceed slightly indirectly to gain some intuition on the structure of these multiplets.
As we discussed earlier, supersymmetry naturally pairs $C_4$ with the metric, and $C_2$ with the $B$ field.
Using the argument of \cite{KW2}  one can then show that there are
scalar modes $s^A$ associated to  the metric with conformal dimensions
\bea
\Delta ( s^A) \, = \, 3 + \sqrt{1+\mu^A}~.
\eea
These may be read off from the asymptotic expansion (\ref{metricforms}).
It is satisfying to see that these conformal dimensions differ precisely
by 1 from  (\ref{currentdimension}).
We expect these scalar metric modes to arise from symmetric tensor harmonics on $Y$.

Moreover, as we show in appendix \ref{eigensection}, for each of the one-forms $\beta^{1(A)}$
one can also construct a scalar eigenfunction of the Laplacian on $(Y,g_Y)$, defined as
\bea
f^A \, = \, \beta^{1(A)}  \, \lrcorner \, \eta~.
\eea
Here $\eta$ is the canonically defined Killing one-form on a Sasaki-Einstein manifold, 
metric dual to the Reeb vector field  
-- see, for example, \cite{MSY2}. In particular, 
we recall that the cone metric on $C(Y)$ has K\"ahler form
\bea
\omega_{C(Y)} \, = \, \frac{1}{2}\diff (r^2\eta) ~.
\eea
As shown in appendix \ref{eigensection}, the functions $f^A$ obey
\bea
\Delta_Y f^A \, = \, E^A \, f^A \qquad \mathrm{(no ~sum)}
\eea
where the eigenvalues are
\bea
E^A \, =\, \mu^A -2 -2 \sqrt{1+\mu^A}~.
\label{pretty}
\eea
These may be used in the KK reduction on $Y$. In particular, they give rise to
a  set of $b_4(X)$ scalar fields $\pi^A$, by expanding the trace of the metric on $Y$.
These modes mix with scalar modes of $C_4$, and the eigenvalues of the $2\times 2$
mass matrix are \cite{kim}
\bea
m^2_\pm \, =\, E + 16 \pm 8 \sqrt{E+4}~.
\label{mixmix}
\eea
The AdS/CFT mass-dimension formula then gives
\bea
\Delta (\pi^A)\, =\, 5 + \sqrt{1+\mu^A}~,
\eea
where, again, picking the positive branch in (\ref{mixmix}), the surds have simplified rather remarkably.
Notice that this value differs precisely by $1$ and $2$ with respect to  $\Delta({\cal A}^A)$ and $\Delta (s^A)$.

This structure, and the comments we shall make below, are suggestive that the modes $({\cal A}^A,s^A,\pi^A)$
may be part of \emph{massive vector multiplets} in AdS$_5$.  On the other hand, the reduction of $C_2$ and $B$
naturally leads to  \emph{complex} massive vector fields in AdS$_5$, with conformal dimension
\bea
\label{currentdimension2}
\Delta ({\cal C}^A)\, =\,  2+\sqrt{1+\mu^A}~,
\eea
as there is no mass matrix to diagonalise in this case. These should also be part of AdS$_5$ multiplets, but presently
it is not clear to us of which kind.

We do not know if the eigenvalues $\mu^A$ are computable in practice in general.
{\it A priori}, the set of eigenvalues also depends on the resolving Calabi-Yau manifold
$(X,g_X)$, and thus on the K\"ahler class, although we have not shown this dependence
in the notation\footnote{Notice, however, that
these eigenvalues are simply related to eigenvalues of the scalar Laplacian via (\ref{pretty}). 
In particular, $\mu^A = E^A + 4 +2 \sqrt{E^A+4}$.}.
Nevertheless,
there is a class of Sasaki-Einstein manifolds where one can determine the eigenvalues explicitly.
Note that on any Sasaki-Einstein manifold $(Y,g_Y)$ the contact one-form $\eta$ is a 
massive one-form with
eigenvalue\footnote{In fact it is straightforward to show that all Killing one-forms are co-closed massive
one-forms with eigenvalue $\mu=8$, and that this is a strict lower bound on the spectrum
of such massive one-forms. That is, $\mu\geq 8$ with equality if and only if
the eigenvalue is associated to a Killing one-form. This is similar to the Lichnerowicz bound
used in \cite{obstructions}, and is proven in \cite{duff}.
We note in passing that one can also obtain a strict lower bound on the \emph{second} smallest eigenvalue $\mu_*$ by using the 
Lichnerowicz bound on the smallest non-zero eigenvalue of the scalar Laplacian.
In particular, since $E^A\geq 5$, with equality only for $(Y,g_Y)=S^5$, we have that $\mu_* > 15$. This gives corresponding bounds on 
the conformal dimensions of dual operators. For example, $\Delta (s_*)>7$. }
$\mu = 8$, and gives rise to a $L^2$-normalisable harmonic two-form on any Calabi-Yau cone
\bea\label{etamode}
\psi_0 \, = \, \diff (r^{-4} \eta)~.
\eea

There is a certain class of models for which there is a K\"ahler perturbation
of a Ricci-flat K\"ahler metric on $X$ which is exactly the  $\mu=8$ mode (\ref{etamode}).
The corresponding
Calabi-Yau singularities are complex cones over K\"ahler-Einstein
surfaces. For example, the Fano surfaces $\mathbb{F}_0=\mathbb{CP}^1\times\mathbb{CP}^1$, $dP_0=\mathbb{CP}^2$
and del Pezzo surfaces with between 3 and 8 blow-ups are of this form.
The Calabi-Yau cone singularity may be resolved using the Calabi ansatz \cite{Calabi}.
This is an asymptotically conical Ricci-flat K\"ahler metric on the canonical
bundle over the Fano surface, and is
completely explicit, up
to the K\"ahler-Einstein metric on the Fano. Thus $b_4(X)=1$ in these models. The subleading behaviour
to the K\"ahler form on the cone is given by (\ref{etamode}), and
relative to
the cone metric this is precisely order $r^{-6}$, which indicates a dual scalar operator of
conformal dimension $\Delta (s)=6$.

 These backgrounds were discussed in an AdS/CFT context in \cite{Benvenuti:2005qb}.
In fact, in the latter reference it was shown that
in such backgrounds there exist two universal KK scalar modes, coming from reduction of the metric,
which (in our notation) have masses
$m^2(s)=12$, $m^2(\pi)=32$,  thus giving $\Delta(s)=6$, $\Delta(\pi)=8$, respectively.
In this case there is one massive mode $\mathcal{A}$
and one massive mode $\mathcal{C}$.
These are dual to vector operators with
conformal dimensions given by (\ref{currentdimension}) and (\ref{currentdimension2}), which
give $\Delta ({\cal A})=7$ and $\Delta ({\cal C})=5$, respectively.

We may  now proceed, by analogy with the discussion of the non-anomalous $U(1)$ symmetries,
to  give a holographic interpretation of the asymptotic expansions of the
$C_4$ field modes of type III$^+$ and the normalisable K\"ahler perturbations,
and the $C_2$ and $B$ field modes of type III$^-$.
In terms of the coordinate $r$, the asymptotic expansions
of the harmonic forms used to construct the form field modes take
a similar form, namely
\bea
&& \delta C_4 : \qquad \Psi^{b_3(Y)+A} ~\sim \diff (r^{-1-\sqrt{1+\tilde{\mu}^{A}}} \tilde\beta^{(3)A})
\label{massive}\\
&&\delta C_2 : \qquad \qquad \;\,
\psi^A ~\sim \diff (r^{-1-\sqrt{1+\mu^{A}}} \beta^{(1)A})~.
\label{massiveBC2}
\eea
Here the three-forms $\tilde \beta^{(3)A}$ are \emph{a priori} unrelated to the one-forms $\beta^{(1)A}$.
However, it would certainly be rather natural if $\tilde{\beta}^{(3)A}=\beta^{(3)A}$,
where $\beta^{(3)A}$ are the dual set of forms, in the sense of (\ref{AY1}), (\ref{AY2}):
if indeed the metric modes $s^A$ introduced above are in the same multiplets
as the vectors coming from the asymptotic $C_4$ modes, then this is necessary by supersymmetry.
We do not have a direct geometric proof, but will formally drop the tildes in any case.
Again, like many of the issues raised in this section, we will leave further study for
future work.

We may  perform gauge transformations
analogous to (\ref{newgauge}) to
obtain the following\footnote{In the following we ignore the other term
in $C_4$, which involves the two-forms $\delta a^A$. In fact self-duality of the RR fields
requires a similar fluctuation in $C_6$, but this is not important for our analysis.} leading behaviour at large $r$:
\bea\label{C4rfluc}
\delta C_4 &\sim & -\frac{1}{\mu_3}r^{-1-\sqrt{1+\mu^{A}}} \diff \delta \vartheta^A \wedge \beta^{(3)A}
\eea
and a similar expression for $C_2$ and $B$.
Following the logic of the previous subsection, by comparison with the AdS$_5$ background
one concludes that the massive vector modes  have leading behaviour
\bea
\label{Amode}
\delta {\cal A}^A &\sim &-r^{-1-\sqrt{1+\mu^{A}}} \, \diff  \delta \vartheta^A ~,
\eea
thus indicating VEVs for dual vector operators of
conformal dimensions $2 + \sqrt{1+\mu^{A}}$.
However, this is too naive.
One may correctly read off the conformal dimension from the expansion of $C_2$ and $B$,
whereas one obtains the incorrect answer this way from (\ref{Amode}). As we have explained,
this is because the $C_4$ modes mix with metric modes that we have not fluctuated in the
supergravity solution.

Given the discussion of non-anomalous currents in the previous subsection, it seems
rather natural to speculate that the  massive vector fields
$\mathcal{A}^A$ and $\mathcal{C}^A$ should be  dual to the $2b_4(X)$ anomalous baryonic currents.
At least for the $C_4$ modes, this may be further motivated by the fact that these modes appear to
be part of vector multiplets in AdS$_5$. These, as we discuss below,
are natural candidates to be the gravity dual of anomalous currents \cite{yuji}.
However, the eigenvalues $\mu^A$ are just the \emph{leading} terms in the asymptotic expansions. 
Therefore they correspond to the operators with lowest anomalous dimensions, in an infinite tower of operators
getting VEVs (see {\it e.g.} \cite{KW2} and \cite{KM}).
It is then possible that the anomalous baryonic
currents are part of vector multiplets but might have larger anomalous
dimensions, and thus correspond to subleading terms in the expansions  (\emph{cf}. discussion around 
equations  (\ref{expansionalpha}), (\ref{expansionbeta}) in appendix \ref{appendixA}).

In the quiver gauge theory, the anomalous baryonic currents
may be defined in exactly the same was as the non-anomalous currents,
by taking linear combinations
of bifundamental bilinears\footnote{We drop the terms $\e^V$ from the expressions.} $\Phi^\dagger_a \Phi_a$.
Recall that there are $2b_4(X)$ such linear combinations
that are anomalous.  In particular, the lowest component scalar fields are
given by
\bea\label{Ueqn}
\mathcal{U}_q\, \equiv\, \mathrm{Tr}\left[\sum_{a\in A} Q_a\Phi_a^{\dagger}\Phi_a \right]
\eea
where $Q_a=q_{t(a)}-q_{h(a)}$ is the baryonic charge of $\Phi_a$ under the
baryonic $U(1)_q$ given by the charge vector $q\in\Z^{\chi-1}$.
Then, classically, we have the relation
\bea
\mathcal{U}_q \, = \, \sum_{v\in V}q_v\zeta_v~.
\eea
This follows from multiplying (\ref{ted}) by $q_v$, taking the trace, and summing over the nodes in the quiver
$v\in V$. Thus we see that, classically, the VEV of $\mathcal{U}_q$ is simply an ``FI parameter''.
Of course, this is a completely natural extension of the situation for the non-anomalous currents. However, the
operators (\ref{naU}) are
protected, while {\it a priori} (\ref{Ueqn}) are not known to be protected.
Let us denote the  corresponding Noether current for $U(1)_q$ by
$J_q$. Classically $J_q$ is conserved for all $q$, but in the quantum theory
we have
\bea\label{anomalyeqn}
\partial_\mu J_q^{\mu} \, \propto \, \sum_{v\in V} c^v_q \, \chi_v \, \equiv \, \chi_q
\eea
where the anomaly coefficient is (as in (\ref{triangle}))
\bea
c^v_q \, = \, \sum_{a\in A\mid h(a)=v} n_{t(a)}q_{t(a)}-\sum_{a\in A\mid t(a)=v} n_{h(a)} q_{h(a)}~.
\eea
Here
\bea
\chi_v \, = \, *_4 \mathrm{Tr}\,  F_v\wedge F_v
\eea
is an operator constructed from the curvature of the $SU(n_v)$ gauge field $F_v$
corresponding to the node $v\in V$. As we argued in section \ref{anomaloussection},
there is a $b_3(Y)$-dimensional space of charges $q$ for which $c_q^v=0$ for all $v$.

The currents $J_q$ sit in (non-conserved) current superfields $\mathcal{J}_q$ in an $\mathcal{N}=1$
supersymmetric theory, while the operators $\chi_v$ are part of chiral superfields
$\mathcal{O}_v=\mathrm{Tr} \, W_{v \alpha} W^{\alpha}_v$, where $W_{v\alpha}$ is a
gauge superfield for the node $v\in V$. In particular, the lowest component of
$\mathcal{J}_q$ is given by (\ref{Ueqn}). Then the anomaly equation (\ref{anomalyeqn})
becomes the superfield equation
\bea
\bar{D}^2 \mathcal{J}_q \, \propto \, \sum_{v\in V} c^v_q \, \mathcal{O}_v \, \equiv \, {\cal O}_q~.
\label{ano}
\eea

In the supergravity dual, this equation may be related to a Higgs mechanism in the
bulk\footnote{We are grateful to Y. Tachikawa and F. Yagi for clarifying comments.}
\cite{yuji}.
The current $J_q$ is dual to a massive vector field, whose transverse mode precisely eats the scalar
(the St\"uckelberg $\varrho$ in the previous paragraph) dual to the anomaly term $\chi_q$. There are then
four independent bosonic scalar operators \cite{KonishiShiz}, namely the lowest component of $\mathcal{J}_q$
and three independent components in $\mathcal{O}_q$ (the complex gaugino bilinears and the Tr$F^2_v$ terms).
Together with the massive vector,  these are the correct number to fill out a massive vector multiplet in AdS$_5$.
Notice that this discussion clarifies that the
axions for the anomalous $U(1)$s are not physical degrees of freedom,
mirroring the familiar situation reviewed in section \ref{anomaloussection}. In particular,
it also clarifies
that the RR moduli fields have a very different origin in the gravity set-up and
in the large volume worldvolume  setting \cite{Morrison} of section \ref{anomaloussection}.

One can also heuristically understand  
current non-conservation from a holographic point of view.
As discussed in \cite{yuji}, and also in \cite{KOW} in a different context,
 in the dual gravity description one introduces an AdS$_5$
gauge field $\mathcal{A}$ and a scalar $\varrho$
which couple to $J$ and $\chi$ on the holographic boundary, respectively, via a coupling
\bea
\int_{\R^4} \left(\mathcal{A}_\mu J^\mu + \varrho \,\chi\right)\vol_4~.
\eea
The AdS$_5$ gauge transformation
\bea\label{sym}
\mathcal{A}\,\rightarrow\,\mathcal{A} + \diff f, \qquad \varrho\,\rightarrow \,\varrho + f
\eea
then immediately leads to the anomaly equation (\ref{anomalyeqn}) as the associated
anomalous Ward identity for the symmetry (\ref{sym}). Of course, the gauge field
$\mathcal{A}$ here should be identified with a massive gauge field in
(\ref{bigclaim1}), and the gauge transformation (\ref{sym}) is the same as the
St\"uckelberg transformation (\ref{stuckeltran}) which results from RR gauge transformations.

Finally, we return to the interpretation of the
fluctuations $\delta \theta^A$ and $\delta\varphi^A$. Assuming that the asympotic expansions
 contain modes  ${\cal A}^A$ and ${\cal C}^A$ which are dual to the baryonic currents,
 we may tentatively interpret the $2b_4(X)$ massless (in $\R^4$) modes as ``pseudo-Goldstone'' bosons.
Indeed, classically we precisely expect to find these modes in the spectrum.
However, since the corresponding symmetries are anomalous,
Goldstone's theorem does not guarantee that these modes exist and are massless in the quantum theory, which is why
we refer to them as ``pseudo-Goldstone'' bosons.
As we discuss in the concluding section \ref{discussionsection}, there might well
be corrections to the supergravity backgrounds of section \ref{section3},
namely D-brane instanton corrections, which lift these massless fields.


\section{Baryon condensates}
\label{condsection}


In this section we describe a Euclidean D3-brane calculation that
conjecturally determines the holographic condensates of baryon operators in
AdS/CFT. Some basic parts of this calculation were carried out in \cite{baryonic},
extending and generalising\footnote{For related work on the conifold, see 
\cite{Krishnan:2008kv} and \cite{Krishnan:2008gx}.} the warped resolved conifold example in \cite{KM}.
Our aim here is to abstract this to a fairly general prescription for computing
baryon condensates in AdS/CFT, and demonstrate that the result has
the features one expects.

Recall that, in the quiver gauge theories of section \ref{section2}, the gauge-invariant scalar BPS
operators
may be divided into two sets: the meson operators and the baryon operators.
Classically these may be identified with the holomorphic functions on the VMS $\mathscr{M}$. The meson operators are distinguished by being invariant
under the baryonic group $U(1)^{\chi-1}$ which acts on $\mathscr{M}$, whereas by definition a baryon operator
is charged under this group. In the more mathematical language of
section \ref{VMSsection}, the baryon operators are the regular functions
on the space of F-term solutions $\mathcal{Z}$ that are semi-invariants (but not invariants) under
$G_\C$, whereas the meson operators are the invariants under $G_\C$.  In fact
there are very general theorems
that state that the meson operators in quiver gauge theories may be
written in terms of traces of bifundamental fields \cite{LP}, whereas the baryon operators
may be written in terms of generalised determinants \cite{DZ}.
The classical VEV of an operator $\mathcal{O}$ at the point $p\in\mathscr{M}$
is simply $\mathcal{O}(p)\in\C$.

Of course, the main interest is in the quantum theory at strong coupling.
Using AdS/CFT we may identify the space of vacua of the strongly coupled
theory with the gravity backgrounds of section \ref{section3}.
In principle, one should be able to compute the condensate of
any operator $\langle\mathcal{O}\rangle_p$ in a given vacuum $p$.
It is not clear (to the authors, at least) how one would compute such a condensate directly in
quantum field theory. However, the AdS/CFT correspondence implies that such one-point functions may be 
computed in the gravity dual of section \ref{section3}, essentially as a geometric computation in the 
appropriate large $N$ limit.

The method for computing meson condensates in these gravity backgrounds
is more straightforward \cite{KW2}, at least in principle.
This is essentially because meson operators are dual to supergravity modes, for which the AdS/CFT
correspondence is very well developed. See \cite{skenderis,skenderis2} for a discussion of state-of-the-art techniques
for computing holographic  VEVs of meson operators in asymptotically AdS$_5$ ten-dimensional geometries.
In contrast,
baryon operators are dual to D-brane states, and the method for computing correlation functions of such
operators is both conceptually and technically much harder.
In this section we
elaborate on a method for computing baryonic  condensates in the backgrounds described thus far. However, before proceeding to this proposal,
we first recall the AdS/CFT description of baryons as wrapped D3-branes.


\subsection{Baryon operators in AdS/CFT}
\label{baryonsec}

Various properties of SCFTs with Sasaki-Einstein duals may be
studied in terms of the geometry of the dual background.
Particularly
well-understood are the operators dual to supergravity modes, where
the precise map from geometry to field theory was outlined in the
original papers \cite{Gubser:1998bc,Witten:1998qj}.
In the remainder of the paper we are interested in baryon operators, which are dual
to D-brane states. Consider
a compact three-submanifold $\Sigma\subset Y$. By wrapping a D3-brane on this
submanifold we effectively obtain a particle in AdS. This particle will be
BPS  when the wrapped D3-brane is supersymmetric. In particular,
an argument similar to that in section \ref{instanton} implies that a necessary condition is that the cone
$C(\Sigma)\subset C(Y)$ is a complex submanifold, or divisor. The D3-brane also carries a
worldvolume gauge field with two-form field strength $M=2\pi\alpha'F-B$, as described in section
\ref{anomaloussection}. For a D3-brane wrapping $\R_t\times\Sigma$, supersymmetry
requires this gauge field to be \emph{flat}, so $M=0$. Again, this  essentially follows from the
more general discussion in section \ref{instanton}.
Flat $U(1)$ gauge fields on $\Sigma$ are classified, up to gauge equivalence, by the group $H^1(\Sigma;U(1))$ -- see
the discussion in appendix \ref{appendixB}.
Since $b_1(\Sigma)=0$ for the three-submanifolds of interest, the long exact coefficient sequence
implies that $H^1(\Sigma;U(1))\cong H^2_{\mathrm{tor}}(\Sigma;\Z)\cong H_1(\Sigma;\Z)$.
Thus, as originally pointed out in
\cite{Gukov:1998kn}, if $\Sigma$ has non-trivial first homology group,
one can turn on distinct flat connections on the worldvolume of a
wrapped D3-brane. These flat connections are defined on torsion line
bundles over $\Sigma$, which we generically denote by $L$. Thus $c_1(L)\in H^2_{\mathrm{tor}}(\Sigma;\Z)$.

In \cite{Gukov:1998kn, Witten:1998xy, Berenstein:2002ke} such
wrapped D3-branes were interpreted as \emph{baryonic particles}. The dual operator that
creates such a baryonic particle will be referred to as a
\emph{baryon operator}. We then have a correspondence
\bea
(\Sigma, L ) \quad \longleftrightarrow \quad \mathcal{B}(\Sigma, L) ~,
\eea
where ${\cal B}(\Sigma,L)$ denotes the baryon
operator associated to the pair $(\Sigma, L)$.
This also
leads one to identify the non-anomalous $U(1)$ baryonic symmetries
in the field theory as arising from the topology of $Y$. As we recalled in
section \ref{nasection}, massless
fluctuations of the RR four-form
potential $C_4$ in the background AdS$_5\times Y$ may be expanded in a
basis of harmonic three-forms on $(Y,g_Y)$ via the ansatz (\ref{pureadsb}).
This
gives rise to $b_3(Y)$ massless gauge
fields $A^I$ in AdS$_5$ which are dual to the non-anomalous baryonic currents $J^I$.
The baryonic charge of a baryonic particle, arising from a three-submanifold
$\Sigma$, with respect to the $I$th baryonic $U(1)$
is then given by \bea
 Q^I[\mathcal{B}(\Sigma,L)] ~= ~\int_\Sigma \alpha_0^{(3)I}~.
\label{kubi} \eea
For fixed $\Sigma$ the operators $\mathcal{B}(\Sigma,L)$, where $L$ is a torsion line bundle on $\Sigma$,
thus all have equal \emph{non-anomalous} baryonic charge \eqn{kubi}. They also have
 equal R-charge, where the latter is determined by the volume
of $\Sigma$ via \bea R[\mathcal{B}(\Sigma,L)] = \frac{N\pi\vol(\Sigma)}{3\vol(Y)}~.
\label{racharge} \eea

Several comments are now in order. Firstly, note that we have two geometric definitions, or at least identifications,
of baryon operators:  firstly, as holomorphic functions on $\mathscr{M}$; and secondly, as dual
objects to a pair $(\Sigma,L)$. In particular, we have asserted in this subsection that to every $(\Sigma,L)$
there is a baryon operator $\mathcal{B}(\Sigma,L)$ which we may realise
classically as a holomorphic function on the classical gauge theory moduli space $\mathscr{M}$.
Although both identifications are geometric, the general relation between them is completely unobvious.
Having said that, the level zero mesonic moduli space $\mathscr{M}(0)=\mathrm{Sym}^N C(Y)$ is an affine GIT quotient of $\mathscr{M}$
by the complexified baryonic group $(\C^*)^{\chi-1}$. Thus the complex geometry of $Z=C(Y)$ is certainly contained in $\mathscr{M}$.
In fact, for $N=1$, a baryon operator of definite charge $q\in\Z^{\chi-1}$ under $(\C^*)^{\chi-1}$ defines an ample
divisor in the mesonic moduli space $\mathscr{M}(q)$. This follows from the discussion in section \ref{VMSsection}.
We then know from \cite{craw} and \cite{Ueda} that for orbifold gauge theories and
toric quiver gauge theories described by dimers
$\pi:\mathscr{M}(q)\rightarrow Z$ is a crepant resolution of the cone $Z=C(Y)$. Thus
we may take $X=\mathscr{M}(q)$ as an underlying complex manifold for the gravity backgrounds
of section \ref{section3}.
Despite recent papers \cite{zaff, amizaff, counting} that count baryonic
operators in simple examples, this correspondence is still very poorly understood geometrically. The main
motivations for the identification come from general AdS/CFT arguments and the fact that in examples
one sees that it works. For example, the quiver gauge theories in \cite{tilings} were \emph{deduced}
from the above identification of certain set of special baryon operators with pairs $(\Sigma,L)$.

The second comment to make is that the set of baryon operators of the form
$\mathcal{B}(\Sigma,L)$ is very small. Indeed, there are obvious generalisations of the construction
outlined above. For example, rather than wrap a single D3-brane on $\Sigma$, we may wrap
$n$ D3-branes. The worldvolume gauge theory is then a $U(n)$ gauge theory, and presumably
supersymmetry again requires the connection to be flat. A flat $U(n)$ connection is determined by
its holonomies, which define a homomorphism
\bea
\rho:\pi_1(\Sigma)\rightarrow U(n)~.
\eea
Gauge transformations act by conjugation, and thus the flat $U(n)$ connections
are in 1-1 correspondence with
\bea\label{homo}
\mathrm{Hom}(\pi_1(\Sigma)\rightarrow U(n))/\mathrm{conjugation}~.
\eea
For example, if $\pi_1(\Sigma)\cong\Z_r$ then the number of flat $U(n)$ connections on
$\Sigma$ is given by the number of partitions of $n$ into $r$ non-negative integers:
\bea\label{partition}
n = \sum_{i=1}^r k_i,\qquad k_i\in\{0,1,2,\ldots\}~.
\eea
This is easy to see: there are $r$ irreducible representations $\mathcal{R}_i$ of $\Z_r$,
which are all one-dimensional. If we identify $\Z_r$ with the group of $r$th roots of unity
then a root $\zeta\in \Z_r\subset U(1)\subset\C$ is sent to
\bea
\mathcal{R}_i:\zeta\rightarrow\zeta^i~,
\eea
where we may regard $i\in \{1,\ldots,r\}$.
An $n$-dimensional representation of $\Z_r$ may then be constructed
from the $r$-vector  $\mathbf{k}=\{k_i\}_{i=1}^r$. Specifically,
\bea\label{repM}
\mathcal{R}_{\mathbf{k}}:\zeta\rightarrow \mathrm{diag}(\zeta^1,
\ldots,\zeta^1,\zeta^2,\ldots,\zeta^2,\ldots\ldots,\zeta^r,\ldots,\zeta^r)\in U(n)
\eea
where $\zeta^i$ occurs $k_i$ times. Notice that all orderings of the entries in (\ref{repM}) are equivalent
under conjugation. Indeed, a little thought shows that all flat $U(n)$
connections on $\Sigma$, using the identification (\ref{homo}), may be written in the form (\ref{repM}).

These D3-brane states may be interpreted naturally in the field theory as follows.
We have $r$ torsion line bundles $L_i$ and thus $r$ BPS baryon
operators $\mathcal{B}(\Sigma,L_i)$. The BPS baryon operators form a
ring, and thus we may multiply them. $n$ D3-branes wrapped on $\Sigma$
have $n$ times the non-anomalous baryonic charge and R-charge of
a single D3-brane wrapped on $\Sigma$. The candidate dual baryon operators are thus given by
\bea
\mathcal{B}(\Sigma,\mathbf{k})\,\equiv\,\prod_{i=1}^r \mathcal{B}(\Sigma,L_i)^{k_i}~.
\eea
where in order to have the correct non-anomalous charges we precisely require (\ref{partition})
to hold.

We may also consider $\Sigma$ that have more than one connected component,
say $\Sigma=\Sigma_1\cup\Sigma_2\cup\cdots\cup\Sigma_k$, where each
$\Sigma_i$ is connected. If the $\Sigma_i$ are all pairwise disjoint then
presumably these may be treated precisely as above. However,
we may also consider self-intersecting
D3-branes, where $\Sigma_i\cap\Sigma_j\neq\emptyset$ for $i\neq j$. Understanding the
effective theory on such a D3-brane, and thus counting its supersymmetric configuations,
seems quite challenging. For example, there may be massless degrees of freedom,
coming from massless strings between each component, associated to the intersection.

However, even this does not exhaust all baryonic operators that one may construct in the gauge theory. Presumably, the complete spectrum may be obtained by quantising the moduli space of
all BPS D3-branes \cite{Beasley:2002xv}, which include  \emph{time-dependent}, rather than static, wrapped D3-branes.
In this paper for simplicity we restrict attention to static singley-wrapped D3-branes on a compact smooth
connected $\Sigma$. As we shall see, understanding the one-point functions of such operators in our non-conformal 
backgrounds is already quite challenging.


\subsection{Baryon condensates: outline of the prescription}
\label{condsec}

In the remainder of the paper we present a prescription for computing the VEVs of
baryon operators which may be represented by a pair of data consisting of a smooth supersymmetric
three-submanifold $\Sigma$, and a torsion line bundle $L$ on $\Sigma$. The first step in performing any holographic
computation is to extend the data from the
 boundary ($r=\infty$ in AdS$_5\times Y$) to the ``bulk'' (essentially $\R^4\times X$). One  must then identify an object, depending on the
 boundary data, that has appropriate transformation properties under the symmetries of the problem.  Given that a
  baryonic particle is dual to a D3-brane with worldvolume $\R_t \times \Sigma$, a
natural candidate for computing the VEV of the operator that creates such a particle is a \emph{Euclidean D3-brane}
that wraps a divisor $D\subset X$ with boundary $\partial D=\Sigma\subset Y$. More precisely, one should perform
 a path integral for such a Euclidean D3-brane, in a given background geometry, with fixed boundary conditions:
\bea
\langle {\cal B}(\Sigma,L) \rangle \, =\, \int_{\de D =\Sigma} {\cal D} \Psi \exp (-S_{ED3}) \,  \approx\, \sum \exp (-S_{ED3}^\mathrm{on-shell})~.
\label{rough}
\eea
In other words, we compute the partition function for a
non-compact D3-brane in the background supergravity solution, where
the boundary conditions for the D3-brane are held fixed.

An analogous prescription is applied in the case of computing expectation values of Wilson
loop operators \cite{Rey:1998ik,Maldacena:1998im}. Here one is instructed to compute the on-shell action
of a Euclidean string with  worldsheet whose boundary is the loop itself. For baryon operators,
this idea was first proposed\footnote{See also \cite{Benna:2006ib}.} in the context of a warped
resolved conifold model in \cite{KM}, in which case the worldvolume gauge field is zero.
The purpose of the remainder of this paper will be to make the rough formula (\ref{rough}) more precise.
The calculation that we will describe computes the semi-classical approximation to the
partition function of a Euclidean D3-brane. This leads to the saddle-point sum on the
right hand side of (\ref{rough}). In principle one should also
compute the one-loop contributions to this saddle-point approximation. However, our main focus here is on understanding
the worldvolume gauge field instantons, and also the coupling of
RR fields to the D3-brane. In particular the one-loop terms do
not involve the RR fields.

Since the D3-brane worldvolume is \emph{non-compact}, the
action is not invariant under gauge transformations of the RR fields.
More precisely, the phase is not a
gauge-invariant object since it will depend on the choice of reference gauge on the boundary for the background
RR fields.  However, even in the classical gauge theory the overall phase of a particular baryon operator
is not physical. It clearly makes no sense to ask what the phase is of some baryon operator $\mathcal{O}$,
since by acting with a baryonic symmetry this operator is equivalent to $e^{i\alpha}\mathcal{O}$
for any constant phase $\alpha$. Physically there is no way to fix this ambiguity. However, it \emph{does} make sense to ask what the \emph{relative}
phase of the VEV of a baryon operator is at different points in the VMS $\mathscr{M}$, since the above
ambiguity cancels. In the gravity calculation,
 the condensate in (\ref{rough}) of course depends on the particular
gravity background, which is a point in the gravity moduli space $p\in\mathscr{M}^{\mathrm{grav}}$;
specifying $p$ involves specifying a complex manifold $X$, a K\"ahler class for the asymptotically conical Calabi-Yau metric
$g_X$ on $X$, the positions of the stacks of $N$ D3-branes, and the RR fields and $B$ field.
We are then more precisely computing $\langle {\cal B}(\Sigma,L)  \rangle_p$.
However, since the overall phase is not physical, the object of interest is really the
relative phase
\bea\label{argue}
\arg \langle {\cal B}(\Sigma,L)  \rangle_{p,p_0} \, = \, \arg \langle {\cal B}(\Sigma,L)  \rangle_p - \, \arg
 \langle {\cal B}(\Sigma,L)  \rangle_{p_0} ~,
\eea
where $p_0$ is any fixed choice of generic (smooth) point $p_0\in\mathscr{M}^{\mathrm{grav}}$. One of our
main results is that the quantity (\ref{argue})  is in fact
gauge-invariant. The key point will be to show that
under gauge transformations the ``bare'' condensate $\langle {\cal B}(\Sigma,L)  \rangle_p$ transforms via terms which depend
only on the  boundary data, which then cancel in (\ref{argue}).

In \cite{baryonic} we gathered some preliminary evidence for the validity of the general
prescription (\ref{rough}).
In particular, we showed  that the right hand side transforms with the correct phase
under gauge transformations of $C_4$ of the type (\ref{Kgauge}),
which are dual to the non-anomalous $U(1)^{b_3(Y)}$ baryonic symmetries. Specifically,
\bea
\delta\,  \langle {\cal B}(\Sigma,L) \rangle \, =  \, \exp (i\beta_I Q^I [{\cal B}(\Sigma,L)]) \,
\langle {\cal B}(\Sigma,L) \rangle ~.
\eea
We also explained that the logarithmically divergent part of the  Born-Infeld action
\bea
S_{BI} \, = \, T_3\int_D \diff^4\sigma \sqrt{\det g_D}  H~,
\label{moreaction}
\eea
is in general proportional to the R-charge (\ref{racharge}), and hence also
conformal dimension, of the baryon operator ${\cal B}(\Sigma,L)$,
as expected from the AdS/CFT dictionary. This conformal dimension is given by
\bea
\Delta(\Sigma) \,=\, \frac{N\pi\vol(\Sigma)}{2\vol(Y)}~.
\eea
To obtain a finite contribution from (\ref{moreaction}) one can define
the following quantity
\bea
S_{BI}^\mathrm{finite} \, = \, \lim_{r_c \to \infty} \left[T_3\int_{D_{r_c}}
 \diff^4\sigma \sqrt{\det g_D}  H -  T_3 L^4 \int_\Sigma \mathrm{dvol}[\Sigma] ~\log r_c \right]~,
 \label{SH}
\eea
where $D_{r_c}$ is a cut-off compact four-manifold with boundary, such that  $\lim_{r_c \to \infty} D_{r_c} = D$ and
$\de D = \Sigma$. This definition is in the spirit of holographic renormalisation
(see {\it e.g.} \cite{Karch:2005ms}). Indeed, we have
subtracted a ``counterterm'' that depends only on the boundary data, and in particular is
covariant (it is simply the integral of the volume form of $\Sigma$).
 In the  following we will not discuss (\ref{SH}) any further, but instead focus our attention on the
reminding part of the on-shell action.
As we shall explain, this is finite and therefore
contributes multiplicatively to the baryon condensate.


\subsection{Supersymmetric wordvolume instantons}
\label{instanton}

In order to compute the \emph{on-shell} Euclidean action one must solve the equation of motion
for the gauge-invariant two-form $M=2\pi\alpha'F-B $ on the D3-brane worldvolume.
We focus on the contribution of supersymmetric D-branes for which one obtains certain non-linear
instanton equations for $M$.
These were investigated in \cite{marcos} for Euclidean D-branes in a Calabi-Yau manifold, as well as
other special holonomy manifolds.

In the presence of general fluxes and warp factors, the
analysis becomes significantly more complicated.
However, it was shown in \cite{martucci}
that the resulting  equations are a rather natural extension of the
flux-less equations, when expressed in terms of
generalised calibrations. In the present paper, we are interested in
Type IIB backgrounds that are
warped Calabi-Yau geometries (\ref{deformedmetric}). In this case
the $\kappa$-symmetry analysis for Euclidean D-branes essentially
carries over \cite{landscape}
 from the original treatment in \cite{marcos}. The equations
(for a general Euclidean D$(2n-1)$-brane)
 may be written as
 \bea
 \e^{i\omega-M}|_{2n} & = & \e^{i\theta} \frac{\sqrt{\det (h+M)}}{\sqrt{\det h}} \diff \vol_{2n}
 \label{inst1}\\
 i_k \Omega_X\wedge \e^{i\omega-M} & = & 0 \qquad k=1,2,3~.
 \eea
Here
\bea
\omega \, = \, H^{1/2} \omega_X
\label{warpedomega}
\eea
where $H$ is the warp factor in (\ref{deformedmetric}), and $\omega_X$, $\Omega_X$ are the
K\"ahler form and holomorphic three-form of the Calabi-Yau $(X,g_X)$,
respectively. The symbol $i_k$ denotes contraction with a complex
vector field $\de/\de z^k$.

Moreover, it is shown in \cite{landscape} that in the presence of a
non-trivial warp factor $H$ the phase $\theta$ takes the fixed
value $\e^{i\theta}=-1$.
When $n=2$, the case in which we are interested, $D$ must be a \emph{divisor},
holomorphically embedded in $X$, and the equations for $M$ read
 \bea\label{instantons}
 M_- \,= \, 0~, \qquad \qquad \nn
 \omega \wedge M \,= \, 0~,
 \eea
where recall that $M_-$ is the real part of a type $(2,0)$-form. These are in fact the usual instanton equations. $M$ is a primitive $(1,1)$-form, which on a K\"ahler four-manifold
$(D,g_D)$ is
equivalent to being anti-self-dual on $D$
\bea\label{asd}
*_4 M \,=\, - M~.
\eea
The metric $h$ induced on $D$ via its embedding into
the spacetime (\ref{deformedmetric}) is conformal to the K\"ahler metric $g_D$ on
$D$ induced via the embedding of $D$ into $(X,g_X)$. Specifically,
\bea
h \,=\, H^{1/2}g_D~.
\eea
However,  the Hodge star  operator is conformally invariant when acting on middle-dimensional forms. Thus the above
equations may be viewed as saying that $M$ is harmonic anti-self-dual on $(D,g_D)$.


\subsection{The worldvolume gauge field}
\label{moreonm}

Let us now discuss in more detail the worldvolume gauge field $M$. As explained in
section \ref{condsec}, given a three-submanifold $\Sigma$
we first need to pick an asymptotically conical divisor $D$, such that $\de D = \Sigma$.
We will impose the following topological conditions on
$\Sigma$ and $D$:
\bea\label{assumptions}
b_1(\Sigma)=0~, \quad H_1(D;\Z)=0~, \quad H^2(D;\C)\cong H^{1,1}(D)~.
\eea
These assumptions will simplify our computations later. In fact these
conditions are not too restrictive, since they hold for any toric
divisor $D$, with boundary $\Sigma$, in a smooth toric 3-fold variety $X$.
For example, in this case $\Sigma$ is necessarily a Lens space.
We also assume that
\begin{center}
$D$ is a spin manifold~.
\end{center}
This is certainly more restrictive. We impose it simply so that
the worldvolume gauge field is related to a genuine line bundle\footnote{This assumption may presumably be lifted without
altering our overall conclusions.}
$\mathcal{L}$ on $D$, rather than a Spin$^{\mathrm{c}}$ structure.
Having made such a choice of $D$, one needs to extend the
torsion line bundle $L$ on $\Sigma$ to a line bundle $\mathcal{L}$
over $D$, whilst also solving the instanton equations
(\ref{instantons}) described in the previous subsection.
In the remainder of this subsection we explain how to solve this problem.

The supersymmetry conditions imply that $M|_\Sigma =0$, and thus the worldvolume
 line bundle $L$ is indeed a torsion  line bundle on $\Sigma$.  Notice this implies that
\bea
\label{tortoise}
2\pi\alpha' F\mid_{\Sigma}\, =\, B\mid_{\Sigma}~.
\eea
In section \ref{flatsection} we made a fixed choice of background $B$ field on $Y$, and thus
$B\mid_\Sigma$ is also a fixed closed two-form. The curvature two-form $F\mid_\Sigma$ of $L$ is thus not
flat, but is rather related to the $B$ field via (\ref{tortoise}).
In fact, we shall argue momentarily
that $M$ must be a harmonic two-form that is $L^2$-normalisable on $(D,g_D)$.
It then follows from the asymptotic expansion at large radius
(\emph{cf.} appendix \ref{appendixA}) that indeed $M=0$ on $\Sigma$.
More precisely, this may be rephrased as the statement $\lim_{r_c\rightarrow\infty}M\mid_{\partial D_{r_c}}=0$.

The first problem is whether or not we may extend, topologically, the line bundle
$L$ on $\Sigma=\partial D$ over $D$ itself; if it does not,
the instanton does not exist. The extendability of the line bundle
is determined by the long exact cohomology sequence
for $(D,\partial D=\Sigma)$:
\bea\label{longroad}
\cdots &&\longrightarrow H^1(\Sigma;\Z)\longrightarrow
H^2(D,\Sigma;\Z)\stackrel{f}{\longrightarrow}H^2(D;\Z)
\stackrel{i^*}{\longrightarrow} \nn \\
&&\longrightarrow H^2(\Sigma;\Z)\longrightarrow
H^3(D,\Sigma;\Z)\longrightarrow\quad \cdots~.\eea
Here $f$ is the forgetful map that forgets that a class is relative,
and $i:\Sigma\hookrightarrow D$ is the inclusion map.
Since $b_1(\Sigma)=0$ by assumption (\ref{assumptions}), the universal coefficients theorem
implies that $H^1(\Sigma;\Z)=0$. By Poincar\'e duality, $H^3(D,\Sigma;\Z)
\cong H_1(D;\Z)= 0$, where the latter is again by assumption (\ref{assumptions}).
Exactness of the sequence
(\ref{longroad}) then implies that every element of $H^2(\Sigma;\Z)$
lifts to an element of $H^2(D;\Z)$. In fact,
\bea
H^2(\Sigma;\Z) \, \cong\,  H^2(D;\Z)/f(H^2(D,\Sigma;\Z))~.\eea

Concretely, this means that the line bundle $L$ over
$\Sigma$ always extends over $D$ to a line bundle
$\mathcal{L}$ with first Chern class $c_1(\mathcal{L})\in H^2(D;\Z)$.
Moreover, the extension is unique up to adding to $c_1(\mathcal{L})$ an element
$f(c)$, where $c$ is any element in $H^2(D,\Sigma;\Z)$.
In fact, even more is true. Since $H_1(D;\Z)$ is trivial, again the
universal coefficients theorem says that $H^2(D;\Z)$ is torsion-free,
and is thus a lattice. Similarly, $H^2(D,\Sigma,\Z)\cong
H^2_{\mathrm{cpt}}(D;\Z)$ is also a lattice. The pairing
\bea
H^2_{\mathrm{cpt}}(D;\Z)\times H^2(D;\Z)\rightarrow H^4_{\mathrm{cpt}}(D;\Z)\cong\Z
\eea
given by cup product and integral over $D$ says that
\bea
\Lambda \,= \, H^2_{\mathrm{cpt}}(D;\Z)~, \qquad \Lambda^* \, = \, H^2(D;\Z)
\label{lattices}
\eea
are \emph{dual} lattices.

Having chosen an extension of $L$ to a line bundle $\mathcal{L}$ over $D$,
we have now fixed uniquely the cohomology class of $M$, namely
\bea
\label{toad}
 [M]\, =\, \iota^*[B] +(2\pi)^2\alpha' c_1(\mathcal{L}) \,
 \in \, H^2(D;\R)~.
 \eea
Recall that the background $B$ field is flat and that different $B$ field moduli are described by the group
$H^2(X,Y;\R)$. More precisely, we pick any flat extension $B^{\circ}$ of $B\mid_Y$ over $X$,
and then any other flat $B$ field with the same gauge at infinity is
\bea
B\, =\, B^{\circ}+B^{\flat}
\eea
where $B^{\flat}$ represents a class in $H^2(X,Y;\R)$.
In particular, $[B]\in H^2(X;\R)$, and thus also
$\iota^*[B]\in H^2(D;\R)$, are determined by the moduli.

We must now solve the instanton equations (\ref{instantons}) for $M$
in the cohomology class (\ref{toad}).
Furthermore, $M$ must be chosen to be  square-integrable,
$M \in {\cal H}^2_{L^2}(D,g_D)$. To see this, notice that
for $\kappa$-symmetric configurations,
the BI part of the on-shell  Euclidean  D3-brane action
\bea
\label{barry}
S_{BI}\, =\, T_3\int_D \diff^4\sigma \sqrt{\det(h+M)}~,
\eea
may be simplified upon using equation (\ref{inst1}) \cite{marcos}.
In particular, using the relation (\ref{warpedomega}) and specialising to linear instantons
(\ref{instantons}), the action (\ref{barry}) becomes
\bea
 S_{BI}  \, = \, T_3\int_D
 \diff^4\sigma \sqrt{\det g_D} \left( H + \frac{1}{4}\mathrm{Tr}_{g_D} \, M^2\right)~,
 \label{cuteaction}
 \eea
where we used anti-self-duality of $M$ to rewrite the $M\wedge M$ term as the pointwise
square norm on $(D,g_D)$
\bea
\| M\|^2_{g_D} \, = \, \frac{1}{2}\mathrm{Tr}_{g_D} \, M^2~.
\eea
As we recalled earlier, the first term in (\ref{cuteaction})
is logarithmically divergent at infinity.
This divergence is physical, as it gives information on the conformal
dimension of a baryon operator in the dual CFT \cite{baryonic}.
Therefore, it is natural to require that the integral of Tr$_{g_D}\,M^2$ does not affect this conformal
dimension, justifying the requirement that  $M$ is $L^2$-normalisable.

We may now easily argue that $[M]$ may indeed be represented by an $L^2$ harmonic two-form $M \in
\mathcal{H}^2_{L^2}(D,g_D)$. We apply once again the results (\ref{tamas}) of \cite{Hausel}
 to an asymptotically conical divisor $(D,g_D)$, of real dimension four.
In particular, in the case at hand the long exact sequence (\ref{longroad}), when tensored with the
reals $\R$, implies that
\bea\label{isoboys}
H^2(D;\R)\stackrel{f^{-1}}{\cong} H^2(D,\Sigma;\R)\cong H^2_{\mathrm{cpt}}(D;\R)\eea
is an isomorphism of vector spaces. Concretely, this means that
 every de Rham cohomology class on $D$
is represented by a compactly supported cohomology class. That is,
given $[M]\in H^2(D;\R)$ there is a compactly supported class
$[M]_{\mathrm{cpt}}\in H^2_{\mathrm{cpt}}(D;\R)$ such that
$f([M]_{\mathrm{cpt}}) = [M]$. The middle isomorphism in (\ref{tamas})
then shows that every element of $H^2(D;\R)$ is represented by a \emph{unique}
$L^2$ harmonic two-form.

To conclude, we need to show that
$M \in \mathcal{H}^2_{L^2}(D,g_D)$ is type $(1,1)$ and primitive in order to satisfy
the instanton equations (\ref{instantons}), and that moreover $M|_\Sigma=0$ (notice that
$L^2$-normalisability does not {\it a priori} imply this).
The arguments are analogous to those presented in subsection  \ref{metricfluc}.
Firstly, note that if
$M$ is harmonic and $L^2$-normalisable, then $M\wedge \omega_D$ is an $L^2$
harmonic four-form on $D$. However, from (\ref{tamas})
we see that $\mathcal{H}^4_{L^2}(D,g_D)\cong H^4(D;\R)=0$, which
implies that any such four-form must be zero. This proves that
$M$ must be primitive. We then also require that $M$ be of Hodge type $(1,1)$. This follows from
a similar argument to that presented in subsection \ref{metricfluc}: on a K\"ahler manifold
$M_{\pm}$ are separately harmonic if $M$ is.
 Since all the $H^2$
cohomology of $D$ is of type $(1,1)$ by assumption in (\ref{assumptions}) {\it i.e.}
$H^2(D;\C)\cong H^{1,1}(D)$, it follows from (\ref{tamas}) that $M_-=0$ and thus
$M$ is of type $(1,1)$.
Thus we have proven that there always exists a unique solution to the instanton equations.

Finally, looking at Table \ref{modes} in appendix \ref{appendixA} in the case that $p=n=2$, one learns that
the leading term in the large $r$ expansion of $M$ is a closed and co-closed mode of type III$^-$, namely
\bea
M_0 \,  = \, r^{-\sqrt{\mu}} \diff \beta_\mu - \sqrt{\mu} r^{-1-\sqrt{\mu}} \diff r \wedge \beta_\mu
\label{muon}
\eea
where  $\beta_\mu$ is a one-form on $\Sigma$ obeying
\bea
\Delta_\Sigma \, \beta_\mu \; = \; \mu \, \beta_\mu
\label{kaon}
\eea
with $\mu>0$. This shows that $M=0$ on $\Sigma$.


\subsection{A topological action for $M$}
\label{toponeaction}

Having explained how to solve for a supersymmetric gauge field $M$ in a given cohomology class
$[M]\in H^2(D;\R)$, we now begin our discussion of the on-shell D3-brane action, evaluated on such solutions.
Let us consider  the combined Born-Infeld  and Chern-Simons parts of the action that depend on $M$.
These pair naturally to construct the complex action
\bea
S[M] \, = \, i\mu_3\left[
\frac{\tau }{2} \int_D M \wedge M  +  \int_D M \wedge C_2 \right]~,
\label{topaction}
\eea
where $\tau=C_0+i\exp(-\phi)$ is the axion-dilaton. Recall also from section \ref{flatsection}
that the background $C_2$ field is flat, of the form
\bea
C_2\, = \, C_2^{\circ}+C_2^{\flat}
\eea
where $C_2^{\circ}$ is a fixed flat $C_2$ field on $X$ inducing a fixed gauge choice $C_2^{\circ}\mid_Y=
C_2^Y$ on $Y$, and $C_2^{\flat}$ represents a class in $H^2(X,Y;\R)$.
In particular, $C_2^Y$ determines a choice of marginal coupling in $H^2(Y;\R)/
H^2_{\mathrm{free}}(Y;\Z)$, and a background $C_2$ determines a cohomology  class
$[C_2]\in H^2(X;\R)$.

In this section we will  show that the action
(\ref{topaction}) is  a \emph{topological} invariant:
that is, it depends only on the topological classes $[M], \iota^*[C_2] \in H^2(D;\R)$.
In the following subsection we will investigate more fully the dependence
of the on-shell action on the various background fields.

More precisely, let $[M]_{\mathrm{cpt}}=f^{-1}[M]\in
H^2_{\mathrm{cpt}}(D;\R)$ denote the compactly supported version of
$[M]$, and similarly $[C_2]_{\mathrm{cpt}}=f^{-1}[C_2]\in
H^2_{\mathrm{cpt}}(D;\R)$. Then we will show that for $M$ the $L^2$ harmonic form
constructed above we have
\bea \int_D M\wedge M &= &
[M]_{\mathrm{cpt}} \cup [M]\label{cuppa}\\[3mm]
\int_D C_2 \wedge M  &=& \frac{1}{2}[M]_\mathrm{cpt} \cup [C_2] + \frac{1}{2}[C_2]_\mathrm{cpt} \cup [M]~,
\label{symcup}
\eea
where the right hand side of these formulas denote the cup product
\bea H^2_{\mathrm{cpt}}(D;\R)\times
H^2(D;\R) \, \rightarrow \, H^4_{\mathrm{cpt}}(D;\R)\cong\R~.
\eea

Consider  (\ref{cuppa}) first. Let $\alpha$ denote any closed two-form representing $[M]$,
and let $\alpha_{\mathrm{cpt}}$ denote any closed compactly
supported two-form representing $[M]_{\mathrm{cpt}}$. Consider
the integral
\bea
\int_D M\wedge M - \int_D \alpha_{\mathrm{cpt}}\wedge \alpha \, =\,
\int_D (M+\alpha_{\mathrm{cpt}})\wedge (M-\alpha) + \int_D M\wedge
(\alpha-\alpha_{\mathrm{cpt}})~.
\eea
Now
\bea \alpha-M \,=\,
\diff\lambda~, \qquad \alpha-\alpha_{\mathrm{cpt}}\, =\, \diff\sigma
\eea
since $[M]=[\alpha]$ and $[\alpha]=f([\alpha]_{\mathrm{cpt}})$ by
assumption. Since $\alpha_{\mathrm{cpt}}$ is zero in a neighbourhood
of infinity we have $\alpha = \diff\sigma$ in this neighbourhood. More
precisely, we may define $U=(r_0,\infty)\times \Sigma$; then for
large enough $r_0$ we have $(\diff\sigma-\alpha)\mid_U=0$. We also have
\bea \diff(\sigma-\lambda)\mid_U\, =\, M\mid_U~.
\eea
Recalling the asymptotic expansion (\ref{muon}), (\ref{kaon}), we may thus take
\bea
\label{sonofnorbert}
(\sigma -\lambda)|_{\Sigma_{r_c}} \, =\, r_c^{-\sqrt{\mu}} \beta_\mu~,
\eea
to leading order in $r_c$ as $r_c\rightarrow\infty$.
Note that we may also add $\diff f$ to (\ref{sonofnorbert}), where $f$ is any
 function on $U$ (not
necessarily bounded as $r_c\rightarrow \infty$) -- however, this
drops out of the integral below since $M$ is closed. Indeed,
 we then have
\bea
\int_D M\wedge M - \int_D
\alpha_{\mathrm{cpt}}\wedge \alpha \, = \, \lim_{r_c\rightarrow\infty}
\int_{\Sigma_{r_c}} (\sigma-\lambda)\wedge M\, = \, 0~,
\eea
where the last
equality follows since both $M\mid_{\Sigma_{r_c}}\rightarrow 0$
and $(\sigma-\lambda)|_{\Sigma_{r_c}} \to 0$, as
$r_c\rightarrow\infty$.

Now conisder (\ref{symcup}). The discussion is analogous to that above.
Given any $[C_2]\in H^2(D;\R)$, there is a unique compactly supported class
$[C_2]_\mathrm{cpt}\in H^2_\mathrm{cpt}(D;\R)$
such that $f([C_2]_\mathrm{cpt})= [C_2]$.
Let $\gamma$ and $\gamma_\mathrm{cpt}$ be two-forms representing
$[C_2]$ and  $[C_2]_\mathrm{cpt}$, respectively, and consider the integrals
\bea
&&\int_D M \wedge C_2  - \int_D \alpha_\mathrm{cpt} \wedge \gamma  =   \nn\\[2mm]
&& \qquad \qquad \quad = \int_D (M + \alpha_\mathrm{cpt}) \wedge (C_2 - \gamma) +\int_D  M \wedge \gamma
  - \int_D C_2\wedge \alpha_\mathrm{cpt}  \label{summa1}\\[2mm]
  &&\int_D C_2 \wedge M  - \int_D \gamma_\mathrm{cpt} \wedge \alpha  =   \nn\\[2mm]
&& \qquad \qquad \quad = \int_D (C_2 + \gamma_\mathrm{cpt}) \wedge (M - \alpha) +\int_D  C_2 \wedge \alpha
  - \int_D \gamma_\mathrm{cpt}\wedge M \label{summa2} ~.
\eea
Now we have
\bea
\gamma - C_2   \, = \,  \diff \nu~,
\eea
thus the first terms on the right hand side may be evaluated by parts, giving
\bea
\int_D (M + \alpha_\mathrm{cpt}) \wedge (C_2 - \gamma)
& = & - \int_\Sigma M \wedge \nu\nn\\[2.5mm]
\int_D (C_2 + \gamma_\mathrm{cpt}) \wedge (M - \alpha)
& = & - \int_\Sigma C_2  \wedge \lambda ~.
\eea
As usual, we should understand the integrals on the right hand side of these expressions
as a limit of integrals over $\Sigma_{r_c}$. Summing  \eqn{summa1}  and \eqn{summa2} we obtain
\bea
&& \!\!\!\!\!\!\!\!\!\!\!\!\!\!\!\!\! 2 \int_D C_2 \wedge M  - \int_D \gamma_\mathrm{cpt} \wedge \alpha  -
\int_D \alpha_\mathrm{cpt} \wedge \gamma  = \nn\\[2.5mm]
 && = \int_D M \wedge (\gamma - \gamma_\mathrm{cpt}) - \int_\Sigma M \wedge \nu
 + \int_D C_2 \wedge (\alpha - \alpha_\mathrm{cpt})
 -\int_\Sigma C_2  \wedge \lambda ~.
\label{tapiro}
\eea
Now we have
\bea
 \gamma - \gamma_\mathrm{cpt} \, =\,  \diff \zeta~,
 \eea
thus integrating again by parts, the second line in \eqn{tapiro}  reduces to
\bea
\int_\Sigma M \wedge (\zeta-\nu) + \int_\Sigma C_2 \wedge (\sigma -\lambda) ~.
\label{hastovanish}
\eea
We then use the fact that
\bea
M- \alpha_\mathrm{cpt} ~= ~\diff (\sigma - \lambda)~, \qquad\quad C_2- \gamma_\mathrm{cpt}~=~\diff
(\zeta - \nu)~.
\eea
The argument for each term in (\ref{hastovanish}) being zero is slightly different. Firstly, $\diff
(\zeta- \nu)$ is a well-defined smooth two-form on $\Sigma$, and thus $\zeta - \nu$ may be taken to be
a smooth one-form;
any exact part, divergent or otherwise, drops out of the integral.
Since $M=0$ on $\Sigma$, then the first
integral in (\ref{hastovanish}) is zero.
Secondly, $(\sigma -\lambda)$ vanishes on $\Sigma$, proving that also
the second integral in \eqn{hastovanish} is zero.
In conclusion, we have shown that
\bea
\int_D C_2 \wedge M & =& \frac{1}{2} \int_D \gamma_\mathrm{cpt} \wedge \alpha  +
\frac{1}{2} \int_D \alpha_\mathrm{cpt} \wedge \gamma ~.
\eea
which is \eqn{symcup}.


\subsection{Gauge transformations of the action}
\label{gaugetransf}

We will now discuss the effect of various gauge transformations on the D3-brane action, extending
the exposition in \cite{baryonic}. Because the worldvolume
$D$ is non-compact the discussion of gauge invariance is slightly subtle. Since the Born-Infeld part of the action is manifestly gauge-invariant,
in the following we will focus on
the Chern-Simons action:
\bea
\label{genCSaction}
S_{CS} \,  = \,  i\mu_3\int_D \left[C_4 + M \wedge C_2 + C_0 \frac{1}{2} M \wedge M \right]+ \frac{2\pi i}{48}\int_D
C_0 \left[p_1(R_{TD}) - p_1(R_{ND})\right]
\eea
where $C_{2p}$ are the RR potentials. Recall that $\mu_3$ is given by (\ref{Dcharge}).
The second term in (\ref{genCSaction}) contains the curvature
couplings in (\ref{CScoupling}), where $p_1(R_{TD})$ and $p_1(R_{ND})$ denote
Pontryagin curvature forms for the tangent bundle $TD$ of $D$ and its normal bundle
$ND$ in $\mathcal{M}$. We will postpone a discussion of this term until
section \ref{curvaturesection}.

Recall the discussion of background RR fields from section \ref{flatsection}. We fix a
gauge choice for the RR potentials on  $Y$, which we may pull back to the UV boundary
$\R^4\times Y$. In particular this determines certain marginal couplings of the UV theory.
These RR potentials  are then extended over $X$, or more precisely over spacetime
$\mathcal{M}=\R^4\times (X\setminus\{x_1,\ldots,x_m\})$, to potentials $C_*^{\circ}$ satisfying the relevant equations of motion. Here the subscript $*$ may take any of the values $0$, $2$ or $4$, so $C_*$ denotes
any of $C_0$, $C_2$ or $C_4$.
One may then add to these background fields any compactly supported flat RR field.
 These determine the flat form-field moduli
discussed in section \ref{flatsection}. We thus generally write
\bea\label{splitmyshorts}
C_*\, =\, C_*^{\circ}+C_*^{\flat}~.
\eea
We first show that, for \emph{fixed }
gauge at infinity, the on-shell D3-brane action is a well-defined function of the flat RR field moduli in (\ref{RRXtorus}).
That is, the action is invariant under compactly supported small and large gauge transformations.
It nevertheless certainly depends on $C_*^{\circ}$, and in particular on
the gauge choice this induces at infinity. However,
we will then show that under \emph{any} gauge transformation
\bea\label{changemymind}
C_*^{\circ}\, \rightarrow \, C_*^{\circ}+\diff\lambda~,
\eea
where $\diff\lambda$ is unrestricted at infinity,
the on-shell action changes by terms that depend \emph{only} on the boundary data.
The prescription for computing the relative phase of the condensate in (\ref{argue}) is then that the two terms on the right hand side should be
computed with the same fixed background $C_*^{\circ}$, inducing a fixed gauge choice
at infinity. The two terms then certainly depend on this choice, as well as
on the compactly supported cohomology classes of the flat fields $C_*^{\flat}$ in (\ref{splitmyshorts}). However, if we change
the choice of $C_*^{\circ}$ via a general gauge transformation (\ref{changemymind}),
or similarly by large gauge transformations, the two terms will transform in the same way,
since the change in the action depends only on the boundary data. This way
the relative phase computed in (\ref{argue}) is independent of the background gauge
choice of $C_*^{\circ}$, and is also gauge invariant under compactly supported gauge transformations.
Thus the relative phases, computed in this manner, depend only on the moduli that we
described in section \ref{section3}.
We discuss small and large gauge transformations in turn.

\subsubsection{Moduli: small gauge transformations}

Consider the small gauge transformation
\bea
C_2 \to C_2 + \frac{2\pi}{\mu_1}\diff\lambda
\eea
where $\lambda$ is any one-form on $\mathcal{M}$ that vanishes on the UV boundary
$\R^4\times Y$. We will refer to such gauge transformations throughout as \emph{compactly supported}.
As explained in section \ref{flatsection},
this transformation must be accompanied by a shift of the four-form potential
\bea
C_4 \to C_4 + \frac{2\pi}{\mu_1}B \wedge \diff\lambda
\eea
leading to the change in the action
\bea\label{deltaCS2}
\delta S_{CS} \,=\, i\mu_3\int_D \frac{2\pi}{\mu_1} \diff\lambda\wedge\e^{2\pi \alpha' F} \,= \, i \int_{\Sigma}\lambda_{\Sigma}\wedge F \, =\, 0~.
\eea
Here $\lambda_\Sigma\equiv\lambda\mid_\Sigma=0$ follows since $\Sigma\subset \R^4\times Y$ and $\lambda$ vanishes on the latter.

Now consider compactly supported small gauge transformations of $C_4$ {\it i.e.} such that
the gauge generators vanish at infinity.  A shift
\bea
C_4 \to C_4 + \frac{2\pi}{\mu_3}\diff K ~.
\eea
leads to a change in the action
\bea
\delta S_{CS} \, =\, 2\pi i \int_D \diff K \, = \,   2\pi i \int_\Sigma K_\Sigma ~.
\eea
But this integral vanishes since $K_{ \R^4\times Y}=0$ and so in particular $K_\Sigma=0$.

\subsubsection{Moduli: large gauge transformations}

Now consider the large gauge transformation
\bea
C_2 \to C_2 + \frac{2\pi}{\mu_1}\sigma
\eea
where $\sigma$ represents a class in $H^2_{\mathrm{free}}(\mathcal{M},\partial\mathcal{M};\Z)\cong H^2_{\mathrm{free}}(X,Y;\Z)$. The net effect is the shift in the action
\bea
\delta S_{CS} \, =  \,i\mu_3\int_D  \frac{2\pi}{\mu_1}\sigma\wedge \e^{2\pi \alpha' F} = 2\pi i \int_D
\sigma \wedge \frac{F}{2\pi}~.
\eea
However, since $[F]/2\pi\in H^2(D;\Z)$ is quantised, the last expression may be understood
as the cup product
\bea
H^2(D,\Sigma;\Z)\times H^2(D;\Z)&\rightarrow & \Z\nn\\
\left(\, \, \, \iota^*\sigma\, \, \, , \, \, \, \frac{[F]}{2\pi} \, \, \, \right) \quad \, \, & \mapsto & \int_D \sigma \wedge \frac{F}{2\pi}~.
\eea
Here $\iota^*:H^2(X,Y;\Z)\rightarrow H^2(D,\Sigma;\Z)$. Hence the action is invariant modulo $2\pi i\Z$.

Finally,  consider large gauge transformations of $C_4$ that are compactly supported:
\bea
C_4 \to C_4 + \frac{2\pi}{\mu_3} \kappa ~.
\eea
Here ${\kappa}$ is a closed compactly supported four-form with integral periods; that is, it represents a
class in $H^4_{\mathrm{free}}(X,Y;\Z)$. Thus the exponentiated action is manifestly invariant since
\bea
\delta S_{CS} \, =\, 2\pi i \int_D  {\kappa} \, = \, 2\pi i n \, \cong \, 0 \quad \, \, \mathrm{mod} \, \, 2\pi i\Z~.
\eea

We have thus shown that the exponentiated on-shell D3-brane action
is invariant under compactly supported gauge transformations of the RR fields.

\subsubsection{Background choice: small gauge transformations}

We now analyse the transformation properties of the D3-brane action under general small gauge
transformations. This is completely straightforward.
Consider the small gauge transformation
\bea
C_2 \to C_2 + \frac{2\pi}{\mu_1}\diff\lambda~.
\eea
Taking into account the corresponding transformation of $C_4$, the D3-brane action changes by
\bea
\delta S_{CS} \,=\, i\mu_3\int_D \frac{2\pi}{\mu_1} \diff\lambda\wedge\e^{2\pi \alpha' F} \,= \, i \int_{\Sigma}\lambda_{\Sigma}\wedge F_\Sigma~.\eea
This of course depends only on the boundary data on $\Sigma\subset Y$. Note that
$2\pi\alpha' F_\Sigma = B_\Sigma$.

Similarly,
\bea
C_4 \to C_4 + \frac{2\pi}{\mu_3}\diff K ~.
\eea
leads to a change in the action
\bea
\delta S_{CS} \, =\, 2\pi i \int_D \diff K \, = \,   2\pi i \int_\Sigma K_\Sigma~,
\eea
which again trivially depends only on data at the boundary.

\subsubsection{Background choice: large gauge transformations}

We conclude by analysing the transformation properties of the D3-brane action under general large gauge
transformations. This is less straightforward. Only the \emph{exponentiated} action changes by terms
depending only on the boundary data.

We begin with large gauge transformations of the axion. These may also be thought
of as $SL(2;\Z)$ transformations. Under the shift
\bea
C_0\to C_0+1
\eea
the action changes by
\bea
S_{CS} \,\to \, S_{CS}+i\mu_3\int_D  \e^{2\pi \alpha' F}~,
\eea
so that
\bea\label{deltaCS}
\delta S_{CS} = \frac{i}{4\pi}\int_D F\wedge F~.\eea
As we explain below, the change $\delta S_{CS}$ in $S_{CS}$ is thus given by the level
$k=1/2$ \emph{Chern-Simons action} of the
abelian connection $A_\Sigma$ on the three-manifold
$\Sigma$. This makes sense as an element of $i\R/2\pi\Z$
only when $D$ is a spin manifold\footnote{Recall that
when $D$ is not spin, the ``gauge field'' $A$ is more precisely a Spin$^\mathrm{c}$
connection \cite{FW}. In this case the discussion is slightly modified, and the curvature couplings
that we have ignored would be important in the analysis.}.

Let us briefly recall how the Chern-Simons action of $(\Sigma,A_\Sigma)$ is defined.
Suppose first that $L$ is a topologically
trivial line bundle over $\Sigma$ on which $A_\Sigma$ is a connection one-form. Thus
$A_\Sigma$ may be regarded as a globally-defined
one-form on $\Sigma$, and there is no subtlety in defining the Chern-Simons action at level $k$:
\bea\label{naiveCS}
S_{CS}(\Sigma,A_\Sigma)=\frac{ik}{2\pi} \int_{\Sigma} A_\Sigma\wedge\diff A_\Sigma~.
\eea
When the line bundle $L$ is non-trivial, as it generally is in this paper, the definition of the Chern-Simons action
for a connection $A_\Sigma$ on $L$ is more subtle. Let us begin by rewriting
(\ref{naiveCS}) in the case that $L$ is trivial.
If $D$ is a four-manifold with boundary $\Sigma$, we may always extend
$A_\Sigma$ as a one-form $A$ over $D$, and by Stokes' theorem we may write
\bea\label{stokesCS}
S_{CS}(\Sigma,A_\Sigma) = \frac{ik}{2\pi} \int_D F\wedge F
\eea
where $F=\diff A$ is the curvature of $A$. Of course, the result is independent of
the choice of extension of $A_\Sigma$ over $D$.
This formula, together with a non-trivial result in
cobordism theory, is the key to defining $S_{CS}(\Sigma,A_\Sigma)$ in general.
If $\Sigma$ is an oriented three-manifold with a line bundle $L\rightarrow\Sigma$,
then it is a non-trivial fact that there exists an oriented
four-manifold $D$, with boundary $\Sigma=\partial D$,
together with an extension $\mathcal{L}$ of $L$ over $D$. Thus we may simply
define the Chern-Simons action by the formula (\ref{stokesCS}),
where $F$ is the curvature of any connection on $\mathcal{L}$ that
restricts to the connection $A_\Sigma$ on $\Sigma$. Of course, {\it a priori}
this definition then depends on the choice of $(D,\mathcal{L})$.
However, suppose that $(D^{\prime},\mathcal{L}^{\prime})$ is another
such extension. Then the difference in Chern-Simons actions is given by
\bea
2\pi i k\int_{W} \frac{F}{2\pi}\wedge \frac{F}{2\pi}~.
\eea
where $W$ is the \emph{compact} four-manifold $W=D\cup_{\Sigma}-D^{\prime}$. 
Since $F/2\pi$ is integral, the difference in Chern-Simons actions
is therefore an integer multiple of $2\pi i$, provided that $k\in\Z$.
Thus (\ref{stokesCS}) may be used to \emph{define} the
Chern-Simons action, regarded as an element of $i\R/2\pi\Z$.

When $\Sigma$ is a spin three-manifold, which is always true
when $\Sigma$ is oriented, we may also define
Chern-Simons theory at half-integer levels, $k\in\tfrac{1}{2}\Z$.
Again, a key fact is that $(\Sigma,L)$ always bounds a spin four-manifold
with line bundle $(D,\mathcal{L})$. In this case the integral
\bea
\int_W \frac{F}{2\pi}\wedge \frac{F}{2\pi}\in 2\Z\eea
is always even for $W$
a compact spin four-manifold. Thus $k$ may take half-integer
values.

To summarise, a gauge transformation $C_0\to C_0+1$
results in a change in the Chern-Simons term in the D3-brane
action by the level 1/2 Chern-Simons action of $(\Sigma,A_{\Sigma})$.
This is well-defined as an element of $i\R/2\pi\Z$, and thus
the change in the exponentiated action $\exp(-\delta S_{CS})$
depends only on the boundary data. This analysis
is particularly important when we come to consider summing
over worldvolume instantons in the next subsection. In this case
$D$ is held fixed, but we precisely sum over different
line bundles $\mathcal{L}$ on $D$ extending $L$. The change in
the phases of each term in the sum under $C_0\to C_0+1$
are then all equal, modulo $2\pi i$.

Now consider the large gauge transformation
\bea
C_2 \to C_2 + \frac{2\pi}{\mu_1}\sigma
\eea
where $\sigma$ is a closed two-form on $\mathcal{M}$ with integer periods. Thus
$\sigma$ represents a class $[\sigma]\in H^2_{\mathrm{free}}(\mathcal{M};\Z)$.
In particular, $\sigma$ defines a class $[\sigma]_X\in H^2_{\mathrm{free}}(X;\Z)$.
The net effect is the shift
\bea\label{C2big}
\delta S_{CS} \, =  \,i\mu_3\int_D  \frac{2\pi}{\mu_1}\sigma\wedge \e^{2\pi \alpha' F} = i \int_D
\sigma \wedge F~.
\eea
The embedding $\Sigma\hookrightarrow \mathcal{M}$ gives a two-form $\sigma_\Sigma$
with $[\sigma_{\Sigma}]\in H^2_{\mathrm{free}}(\Sigma;\Z)\cong 0$.
The integral in
(\ref{C2big}) may then be understood as a definition of the boundary quantity
\bea\label{linear}
i\int_\Sigma \sigma_\Sigma\wedge A_\Sigma~.
\eea
The argument is similar to that for the Chern-Simons action above.
For $A_\Sigma$ a globally-defined connection one-form on a trivial line bundle $L$,
the integral (\ref{linear}) is well-defined.
We may then rewrite (\ref{linear}) by choosing any four-manifold $D$ that bounds $\Sigma$,
any extension $\sigma_D$ of $\sigma_\Sigma$ that is closed and has integer periods, and any extension
$A$ of $A_\Sigma$. Notice that $\sigma_D$ exists by the same reasoning that
$\mathcal{L}$ and $F$ exist. Then Stokes' theorem implies that
\bea\label{defnlinear}
i\int_\Sigma \sigma_\Sigma\wedge A_\Sigma \, = \, i\int_D \sigma_D \wedge F~.
\eea
A non-trivial line bundle $L$ may be extended to a line bundle $\mathcal{L}$ over $D$,
with $A$ a connection form on $\mathcal{L}$ extending $A_\Sigma$. Then (\ref{defnlinear})
may be used as a definition of the left hand side. Any other $D^{\prime}$,
$\sigma_{D^{\prime}}$ may of course be used, and the difference
between the two definitions is
\bea
i\int_W \sigma_W\wedge F
\eea
where $W=D\cup_{\Sigma}-D^{\prime}$ and $\sigma_W$ is obtained by gluing together
$\sigma_D$ and $\sigma_{D^{\prime}}$, which recall agree on the gluing locus $\Sigma$.
Since $[F]/2\pi$ and $[\sigma]_W$ are integral classes, this last integral is an integer multiple of $2\pi i$, and thus
(\ref{defnlinear}) is a well-defined definition of the left and side, modulo $2\pi i$.

Finally, consider large gauge transformations
\bea
C_4 \to C_4 + \frac{2\pi}{\mu_3}\kappa
\eea
where $\kappa$ is a closed four-form on $\mathcal{M}$ with integer periods. Of course
\bea\label{shouldgorunning}
\delta S_{CS} \, = \, 2\pi i\int_D \kappa_D~.
\eea
If $D^{\prime}$ is any other extension of $\Sigma$ then the difference
\bea
2\pi i\int_D \kappa_D - 2\pi i\int_{D^{\prime}} \kappa_{D^{\prime}} = 2\pi i\int_W \kappa_W \in 2\pi i
\eea
where as usual $D\cup_{\Sigma}-D^{\prime}$ and $\kappa_W$ is constructed by
gluing $\kappa_D$ and $\kappa_{D^{\prime}}$ along $\Sigma$. This shows that
(\ref{shouldgorunning}) depends only on boundary data, modulo $2\pi i$.

\vskip 2mm

\begin{table}[ht!]
\centerline{
\begin{tabular}{|c|c|c|c|c|}
\hline
 &\multicolumn{2}{|c|}{cpt supported} &\multicolumn{2}{|c|}{non-cpt
supported} \\
\hline
 & small & large & small & large \\
\hline
$C_0$ & -- & -- & --  &  $\tfrac{i}{4\pi}\int_D F\wedge F $ \\
\hline
$C_2$ & 0  & $2\pi i \Z$ & $i \int_\Sigma \lambda_\Sigma \wedge F$ &
$i\int_D \sigma_D \wedge F$ \\
\hline
$C_4$ & 0  & $2\pi i \Z$ & $2\pi i \int_\Sigma K_\Sigma$ & $2\pi i \int_D
\kappa_D $ \\
\hline
\end{tabular}
}
\caption{Variation of the on-shell D3-brane action under gauge
transformations of the RR fields. The integrals in the last column are
invariants of the boundary data modulo $2\pi i \Z$. In particular, the top right hand
entry is the level $1/2$ Chern-Simons action for
$(\Sigma,A_\Sigma)$.}
\label{changes}
\end{table}

To summarise, the last two subsections have shown that $\exp(-S_{CS})$ changes by
a quantity that depends only on boundary data, for \emph{any} gauge transformation of the
RR fields in the bulk. In contrast, the previous two subsections have shown that
$\exp(-S_{CS})$ is \emph{invariant} under any \emph{compactly supported} gauge transformation
of the RR fields in the bulk. This is summarised in Table \ref{changes}.

\subsubsection{Curvature terms}
\label{curvaturesection}

Finally, we turn to the curvature terms in (\ref{genCSaction}). Recall that the
first Pontryagin form of a real vector bundle $E$ with curvature form $R_E$ is given by
\bea\label{pon}
p_1(R_E) = -\frac{1}{8\pi^2}\mathrm{Tr}\, {R}_E\wedge {R}_E~.
\eea
In the case at hand, $E$ is either the
tangent bundle of $D$ or its normal bundle in the spacetime $\mathcal{M}$. The relevant connection
in (\ref{CScoupling}) is then the Levi-Civita connection of the induced metric $h$ on $D$, or the induced connection on the normal bundle, respectively.

In this subsection we note that the curvature couplings evaluated at any two points
$p$ and $p^{\prime}$ in the same component of the supergravity moduli space
({\it i.e.} where the topology of $X$ at these two points is the same) are in fact equal.
Thus when we compute the relative phase of the condensate in (\ref{argue}), the curvature
terms simply drop out.

To compute the on-shell action we have fixed a gauge for $C_0$, which means that
$C_0\in\R$ is a fixed real number. Choose points $p$ and $p^{\prime}$ in the supergravity moduli
space which have the same topology $X$ for the Ricci-flat K\"ahler background -- of course,
the K\"ahler class, positions of the $N$ D3-branes, and $B$ field and RR field moduli may be
different. However, in both cases $\Sigma$ is extended to the same divisor $D\subset X$,
and the difference in curvature couplings is thus
\bea\label{difference}
\frac{2\pi i C_0}{48}\left[\int_D \left[p_1(R_{ND})-p_1(R_{TD})\right]-\int_D  \left[p_1(R^{\prime}_{ND})-p_1(R^{\prime}_{TD})\right]\right]~.
\eea
Here $R$ and $R^{\prime}$ denote the curvature forms in the two corresponding spacetimes
$\mathcal{M}$, $\mathcal{M}^{\prime}$. These depend on
the metric $g_X$ on $X$ and also on the positions of the D3-branes. However,
we may now define the double
\bea\label{doubleD}
\bar{D}\equiv D\cup_{\Sigma} -D~.\eea
The difference (\ref{difference}) is then
\bea
\frac{2\pi i C_0}{48} \int_{\bar{D}} \left[p_1(R_{N\bar{D}})-p_1(R_{T\bar{D}})\right]
\eea
which is manifestly a topological invariant, since $\bar{D}$ is closed without boundary.
One must be slightly careful in this argument, since the boundary $\Sigma$ along which we
glue is at infinite distance. However, one can simply cut off the integral at some
large $r_c$, and glue the metrics and connections (smoothing appropriately) along $\Sigma_{r_c}=
\partial D_{r_c}$. We may conveniently view $N\bar{D}$ as the normal bundle of $\bar{D}$ in the
spacetime double
\bea
\bar{\mathcal{M}} = \mathcal{M}\cup_{\R^4\times Y} -\mathcal{M}^{\prime}~.
\eea

The key observation is that, due to its construction (\ref{doubleD}), $\bar{D}$ has an orientation-reversing
diffeomorphism which sends a point in one copy of $D$ to the corresponding point in the other copy.
The fixed point set of this map is $\Sigma$. However, it is well-known that if $\bar{D}$ admits
an orientation-reversing diffeomorphism, the Pontryagin number
\bea
p_1(\bar{D}) \equiv \int_{\bar{D}} p_1(R_{T\bar{D}})
\eea
is zero. This is easy to see: the definition (\ref{pon}) is independent of orientation,
whereas the fundamental class of $\bar{D}$ (and hence the integral) changes sign under
a change of orientation. But any integral is diffeomorphism-invariant, hence the result.
A similar result is true for the Pontryagin numbers of a vector bundle $E$ over $\bar{D}$, \emph{provided}\footnote{This is certainly not true in general.} the
orientation-reversing diffeomorphism lifts to a bundle isomorphism of $E$. In the case at hand, the first Pontryagin class of $E$,
which lives in $H^{4}(\bar{D};\Z)\cong\Z$, will
then be invariant, and thus the Pontryagin number will change sign. The normal bundles
in the two spacetimes are certainly isomorphic (although they have different
curvature forms). Thus there is a natural bundle isomorphism of the normal bundle of the double
that covers the orientation-reversing diffeomorphism,
and it follows that
\bea
\int_{\bar{D}} p_1(R_{N\bar{D}})\, =\, 0~.
\eea
Note that an alternative proof of the above would have been to use an APS index theorem
argument, as in \cite{GST}. The idea would be to relate the curvature terms to
an appropriate linear combination of indices of operators with APS boundary conditions.
The APS index theorem would then relate the curvature terms to the indices, which would be
topological invariants of $D$ in $X$ and thus fixed integers, and boundary terms.


\subsection{Sum over gauge field instantons: theta functions}
\label{theta}

In section \ref{toponeaction} we showed that $S[M]$ is a topological
invariant, depending only on the cohomology classes $[M], \iota^*[C_2] \in H^2(D;\R)$.
Recall that  we have
\bea
[M] \,=\, \iota^*[B]+(2\pi)^2\alpha' c_1(\mathcal{L})\, \in \,  H^2(D;\R)~.
\eea
where  $\mathcal{L}$ is a line bundle over $D$ that restricts to
$L$ on $\Sigma=\partial D$.

However, for fixed $L$ there are typically countably infinitely many $\mathcal{L}$ that extend $L$ over $D$, and
thus countably infinitely many
instantons $\{M(\mathcal{L})\}$ with different topological classes $[M(\mathcal{L})]\in H^2(D;\R)$.
This infinite set may be characterised as follows.
Recall that $\Lambda = H^2_{\mathrm{cpt}}(D;\Z)$ and $\Lambda^* = H^2(D;\Z)$
are dual lattices under the cup product
\bea
\Lambda\times \Lambda^*\,\rightarrow\, \Z~,
\eea
and  that there is a  natural map
\bea
f:\Lambda\rightarrow\Lambda^*
\label{cantremember}
\eea
that forgets that a class has compact support.
Let $\mathcal{L}_0$ be any fixed extension of $L$ over $D$, with $c_1(\mathcal{L}_0)\in \Lambda^*$.
We then define
\bea
[M]_0 \, = \, \iota^*[B]+(2\pi)^2 \alpha' c_1(\mathcal{L}_0)\, \in\,  H^2(D;\R) ~,
\eea
so that  the set of \emph{all} gauge instantons that are asymptotic to the torsion line
bundle $L$ is given by
\bea
\Big\{[M(n)] = [M]_0 + (2 \pi)^2\alpha' f(n)\mid n\in\Lambda\Big\}~.
\eea
The D3-brane path integral thus naturally produces,
for fixed choice of $L$, an \emph{instanton sum}
\bea\label{somewhere}
\sum_{n\in \Lambda} \exp(-S[M(n)])~.
\eea

In order to obtain a more explicit expression for this sum it is convenient to
introduce bases for the dual lattices.
Let $\{e_i\}$ be a basis for $\Lambda$ and $\{e_i^*\}$ be the dual basis
for $\Lambda^*$, so that
\bea
\int_D e_i \wedge e_j^* & = & \delta_{ij}~.
\label{basin}
\eea
Here $i=1,\ldots,b_2(D)$, where $b_2(D)$ is the second
Betti number\footnote{Notice that our topological assumptions (\ref{assumptions}), together with the
discussion in section \ref{moreonm}, imply that all the degree two cohomology of $D$ is
represented by $L^2$ harmonic anti-self-dual two-forms. The  assumptions (\ref{assumptions}) hold if $D$ is a toric divisor,
for example.} of $D$.
In this basis we may express the map (\ref{cantremember}) in terms of a matrix
\bea
f(e_i) \, =\,  f_{ji} \, e_j^*~,
\eea
where as usual a sum is understood over repeated indices. The matrix $\mathbf{f}=(f_{ij})$ is invertible and has
integer coefficients. We now make some further definitions. Let
\bea
b &=& \frac{1}{2\pi\alpha'}\iota^*[B] \qquad \qquad \qquad \quad \! c ~=~ \frac{1}{2\pi\alpha'} \iota^*[C_2]\label{BCdef}\\[3mm]
a &=&   b + 2\pi c_1({\cal L}_0) \qquad \quad \, a(n) = a+ 2\pi f(n)~.\label{mdef}
\label{defs}
\eea
These are all elements of $H^2(D;\R)$. For fixed $L$ and background fields
$C_2$, $B$ and axion-dilaton $\tau$, the instanton sum (\ref{somewhere}) may be written as
\bea
{\cal I}([C_2],[B],\tau,L) &= & \sum_{n\in \Lambda} \exp(-S[M(n)]) \\
&= &\, \sum_{n\in \Lambda} \exp  \left[-\frac{i}{4\pi}\Big(\tau \, a(n)_{\mathrm{cpt}}\cup a(n) +
 a(n)_{\mathrm{cpt}}\cup c + c_{\mathrm{cpt}}\cup a(n) \Big)\right]\nn
\eea
where
\bea
a(n)_{\mathrm{cpt}} = f^{-1}(a(n)) ~,\qquad \quad c_{\mathrm{cpt}} = f^{-1}(c)~.
\eea
We may now expand the various forms in terms of the basis (\ref{basin}) as
\bea
a \, =\,  a_i \, e_i^*~, \qquad c \,=\,  c_i \, e_i^*~,\qquad n \,=\,  n_i e_i ~,
\eea
where $a_i = b_i + 2\pi c_1(\mathcal{L}_0)_i$ and we may take
\bea
b_i \, = \, \frac{1}{2\pi\alpha'}\int_{S_i} B \quad \qquad c_i \, = \, \frac{1}{2\pi\alpha'}\int_{S_i} C_2~.
\eea
where $\{S_i\}$ are a basis of two-cycles for $H_2(D;\Z)$. A computation then shows that
\bea
a(n)_{\mathrm{cpt}}\cup a(n) \, = \, (f^{-1})_{ij}a_i
a_j + 2\pi (f^{-1})_{ji}f_{jk}n_k a_i +
 2\pi n_i a_i + (2\pi)^2  f_{ij}n_in_j\eea
and
\bea
a_{\mathrm{cpt}} (n)\cup c + c_{\mathrm{cpt}}\cup a(n)    =
(f^{-1})_{ij} (a_i c_j + c_ia_j) +
2\pi  (f^{-1})_{ji} f_{jk} n_kc_i + 2\pi  n_ic_i ~.
\eea
At this point, with a fixed basis and dual basis, we may
view $a$, $c$ and $n$ as vectors in $\R^{b_2(D)}$,
and the cup product as simply a dot product of vectors. In this notation, we may write the
instanton sum as a product of two factors
\bea
{\cal I}([C_2],[B],\tau,L) \, =  \,  {\cal P} (\mbox{\boldmath{$c$}},\mbox{\boldmath{$b$}},\tau,L) \,
{\cal Q} (\mbox{\boldmath{$c$}},\mbox{\boldmath{$b$}},\tau,L)
\label{thetavev}
\eea
defined as
\bea
{\cal P} (\mbox{\boldmath{$c$}},\mbox{\boldmath{$b$}},\tau,L)&\equiv & \exp\left[-\frac{i}{2\pi}\mbox{\boldmath{$a$}}
\cdot \mathbf{f}^{-1}_{\mathrm{sym}}\left(\frac{\tau}{2}\mbox{\boldmath{$a$}}+\mbox{\boldmath{$c$}}\right)\right]\\
{\cal Q}(\mbox{\boldmath{$c$}},\mbox{\boldmath{$b$}},\tau,L) & \equiv & \sum_{\mathbf{n}\in\Z^{b_2(D)}} \exp\left[-i \pi \tau \mathbf{n}\cdot \mathbf{f}\mathbf{n}
- \frac{i}{2} \mathbf{n}  (1+\mathbf{f}^T \mathbf{f}^{-1})   (\tau \mbox{\boldmath{$a$}}+ \mbox{\boldmath{$c$}})\right]~.
\label{fearfactor}
\eea
Here we have defined
\bea
\mathbf{f}^{-1}_{\mathrm{sym}}\equiv \frac{1}{2}\left[\mathbf{f}^{-1}+(\mathbf{f}^{-1})^{T}\right]~.
\eea

The sum in (\ref{fearfactor}) precisely gives rise to a \emph{Riemann theta function}. This is usually defined as
\bea
\label{riemanntheta}
\theta[\mathbf{z},\mathbf{T}] = \sum_{\mathbf{n}\in \Z^r} \exp\left[2\pi i
\left(\frac{1}{2}\mathbf{n}\cdot\mathbf{T}\, \mathbf{n}+\mathbf{n}\cdot\mathbf{z}\right)\right]~.\eea
Here $\mathbf{z}\in\C^r$ is a complex vector and $\mathbf{T}$ is a complex
symmetric $r\times r$ matrix whose imaginary part is positive definite.
The space of such matrices is denoted $\mathbb{H}_r$, and is known as
the Siegel upper half-space. One requires the imaginary part of $\mathbf{T}$ to
be positive definite in order that the sum in (\ref{riemanntheta})
converges. In fact, it then converges absolutely and uniformly on
compact subsets of $\C^r\times\mathbb{H}_r$. Defining
\bea
\mathbf{T} ~= ~ -\tau \, \mathbf{f}_{\mathrm{sym}}\qquad
\quad \mathbf{z} ~= ~ -\frac{1}{4\pi}
(1+\mathbf{f}^{T}\mathbf{f}^{-1})\left({\tau}\mbox{\boldmath{$a$}}+\mbox{\boldmath{$c$}}\right)~.
\label{definitions}
\eea
we have that
\bea
{\cal Q} (\mbox{\boldmath{$c$}},\mbox{\boldmath{$b$}},\tau,L) \, = \, \theta[\mathbf{z},\mathbf{T}]
\eea

At first sight the expression (\ref{thetavev}) seems to depend
on the choice of $\mathcal{L}_0$, which appears in $\mbox{\boldmath{$a$}}$ via equation (\ref{defs}).
Of course, from the original definition of the instanton sum this cannot be true.
Using the transformation properties of the theta function under shifts
$\mathbf{z} \to \mathbf{z} + \mathbf{T}  \mathbf{m} +\mathbf{k}$, with
$\mathbf{m}, \mathbf{k} \in \Z^r$,
one can in fact easily check that the right hand side of \eqn{thetavev} is independent of the choice
of $\mathcal{L}_0$, although each factor is separately not independent.

Note also that $\mathbf{f}_{\mathrm{sym}}$  is indeed negative definite.
The argument for this traces back to the fact that
for any $[M]\in H^2(D;\R)$, which is represented by the vector $\mathbf{M}$ in the above basis, we have
\bea
\mathbf{M}\cdot \mathbf{f}^{-1}\mathbf{M} = [M]_{\mathrm{cpt}}\cup [M] =
\int_D M\wedge M = -\int_D M\wedge *M\leq 0
\eea
where $M\in H^2_{L^2}(D,g_D)$ is the harmonic anti-self dual $L^2$-normalisable two-form that
represents $[M]\in H^2(D;\R)$. The inequality is strict provided
$M\neq0$. This shows that (the symmetric part of) $\mathbf{f}^{-1}$,
and hence also the symmetric part of $\mathbf{f}$, is negative definite.
This is precisely the condition required for the instanton sum to
converge.

Notice that if $\mathbf{f}$ is symmetric
the expressions simplify slightly.
If in addition we formally set $\mbox{\boldmath{$a$}}=0$, one obtains simply
\bea
\label{fracin}
\sum_{\mathbf{n}\in \Z^{b_2(D)}} \exp\Big[-i \pi
\tau\mathbf{n}\cdot\mathbf{f}\, \mathbf{n}
-i\mathbf{n}\cdot{\mbox{\boldmath{$c$}}}\Big]~.
\eea
Interestingly, this sum has appeared recently as the partition function
for fractional instantons \cite{fucito,griguolo}. Indeed, these references obtain
this result by computing a partition function that counts $U(1)$ SYM
instantons with an ``observable insertion''.


\subsection{Coupling to Goldstone and pseudo-Goldstone bosons}
\label{Gcouplingsection}

In this final subsection we collect various pieces together and present an expression for the
gauge-invariant phase of the baryon condensate that we defined in (\ref{argue}).
This will also give us the opportunity
to discuss the dependence of this phase on the RR moduli fields.
The phase of the  ``bare'' condensate, evaluated at a
point\footnote{We have omitted the curvature coupling, which cancels in
(\ref{relphase}) below.}  $p_0 \in \mathscr{M}^\mathrm{grav}$ is
\bea
\arg \langle {\cal B} (\Sigma,L)\rangle_{p_0}  \, =\, - \mu_3\int_D C_4  -\frac{1}{2\pi}\mbox{\boldmath{$a$}}
\cdot \mathbf{f}^{-1}_{\mathrm{sym}}\left(\frac{C_0}{2}\mbox{\boldmath{$a$}}+\mbox{\boldmath{$c$}}\right)
 +\arg \theta[\mathbf{z},\mathbf{T}]~,
 \label{bare}
\eea
where recall that to determine a point $p_0$ in particular means choosing a $B$ field and RR fields. Specifically, these enter into (\ref{bare})
through the definitions (\ref{definitions}), (\ref{BCdef}), (\ref{defs}).
For the relative phase  (\ref{argue}) we then have, with a slight abuse of notation,
 \bea
\arg \langle {\cal B} (\Sigma,L)\rangle_{p_0,p}  \, =\, - \mu_3\int_D [C_4(p)- C_4(p_0) ]
 +\arg \frac{{\cal P}(p)}{{\cal P}(p_0)}
 +\arg \frac{\theta[p]}{\theta [p_0]}~.
 \label{relphase}
\eea
This expression shows that the baryon condensate, as it currently stands,
has a \emph{definite charge} under the
$U(1)^{b_2(X)}$ subgroup of baryonic symmetries associated to $C_4$. On the other hand, the theta function
does not have a definite charge under the remaining $U(1)^{b_4(X)}$ subgroup of baryonic symmetries
associated to $C_2$ (although recall we have shown that
(\ref{relphase}) is \emph{invariant} under small and large gauge transformations of all RR fields,
and in particular is a well-defined function of the $C_2$ moduli).

As we have explained in subsection \ref{VMSsection},
in the classical gauge theory the baryon operators form a ring graded by their charge
under the full baryonic symmetry group $U(1)^{\chi-1}$. One may thus
write a basis of baryon operators which have definite charge (the basis is homogeneous)
under this symmetry group. In all known examples, the classical baryon operators dual to $(\Sigma,L)$
indeed have definite charge under $U(1)^{\chi-1}$. For example, for the
$Y^{p,q}$ theories \cite{quiverpaper} the baryon operators dual to $(\Sigma,L)$ are determinants
of the bifundamental fields, which thus carry charge $\pm 1$ under precisely two $U(1)$ subgroups.
These are simply the $U(1)$s of the head and tail gauge group of the corresponding bifundamental field,
which in general are certainly
anomalous.
However, quantum mechanically, one expects that the vacuum expectation
values of these operators should only have well-defined charges under \emph{exact} global symmetries.
In the gravity dual, the group $U(1)^{\chi-1}$ is identified with the RR field torus (\ref{RRXtorus}),
with a $U(1)^{b_3(Y)}$ subgroup coming from non-compactly supported gauge transformations of the $C_4$ field.
One thus expects the phases of the baryon VEVs to be linear precisely in these $b_3(Y)$ moduli.

However, as our calculation currently stands, the two sets of $b_4(X)$ anomalous symmetries
 enter the condensate calculation rather differently: the $b_4(X)$ moduli
coming from $C_4$ behave in the same way as the $b_3(Y)$ moduli.
We believe this is evidence for also summing over \emph{disconnected}
compact components $D_{\mathrm{cpt}}$ in the full condensate calculation.
{\it A priori}, one should include these as contributions to the Euclidean
path integral with fixed boundary conditions at infinity. The sum over such compact
components would then break the asymmetry we have described above, giving the condensate
a non-linear dependence also on the $b_4(X)$ modes associated to $C_4$.

This discussion may also be phrased in terms of the coupling of the phase of the condensate to the Goldstone
and pseudo-Goldstone bosons. This generalises the discussion in \cite{KM}.
The coupling may be obtained straighforwardly by considering two infinitesimally displaced
points $p_0$ and $p_0+\delta p$ in moduli space. Then the $C_4$ coupling in (\ref{relphase}) gives
\bea
\label{Gcoupling}
\delta\vartheta^M\int_D\Psi^M \qquad M =1,\dots, b_3(Y),b_3(Y)+1,\dots , b_2(X)~,
\eea
while the $C_2$ moduli $\delta \varphi^A$ clearly couple through
 a non-linear ($p_0$-dependent) expression.
Notice that it is straightforward to show that $\Psi^M$ is indeed integrable, using the
boundary behaviour determined in section \ref{fluctuationsection}. Indeed, $\Psi^M$
form a basis for $H^4_{\mathrm{free}}(X,Y;\Z)$, and the coupling $\int_D \Psi^M$
is then topological.

Summing  over \emph{compact} four-cycles,
the $b_4(X)$ pseudo-Goldstone modes of $C_4$, which are associated to classes in
the image $H^4(X,Y;\R)\rightarrow H^4(X;\R)$, would couple differently
to each compact component.
On the other hand, it is simple to see that the $b_3(Y)$ Goldstone modes $\delta\vartheta^I$ do not
couple to the compact components, since
\bea
\delta\vartheta^I\int_{D_{\mathrm{cpt}}} \Psi^I\, =\, 0~.
\eea
This follows since by definition the $\Psi^I$ are exact forms, which thus map to zero in $H^4(X;\R)$.
We thus see that also summing over $D_{\mathrm{cpt}}$ in the condensate
calculation implies that only the $b_3(Y)$ Goldstone bosons couple linearly
to the phase of the condensate, and that the two $b_4(X)$ sets of pseudo-Goldstone
bosons associated to $C_4$  and $C_2$ are then treated more symmetrically.


\section{Summary and discussion}
\label{dissection}

\subsection{Summary}
\label{transection}

In this section we summarise the constructions of the paper. We begin
by recalling how one constructs a symmetry-breaking supergravity background, and
describe the corresponding moduli space. We then summarise the
 prescription for computing baryon condensates in such a background.

\subsubsection{Supergravity backgrounds}

We first summarise how one constructs a supergravity background of section \ref{section3}, and the moduli space of such vacua:
\begin{itemize}
\item The starting point is a Ricci-flat K\"ahler cone $(C(Y),g)$, together with a choice of flat form
fields on the Sasaki-Einstein link $(Y,g_Y)$. The latter means specifying
a flat $B$ field together with a point in the RR torus (\ref{RRYtorus}).
By AdS/CFT, the corresponding AdS$_5$ background determines a dual four-dimensional SCFT, with
the $B$ field and RR fields determining the values of certain marginal couplings.
\item We suppose that the underlying complex variety $Z=C(Y)$ above is such that it admits a crepant resolution
$\pi:X\rightarrow Z$. By the conjecture in section \ref{metricsection},
for each K\"ahler class in the K\"ahler cone of $H^2(X;\R)$ there exists a unique asymptotically conical
Ricci-flat K\"ahler metric $g_X$ on $X$. This is known to be true in some cases, as
discussed in section \ref{metricsection}, and is a
conjectural non-compact version of Yau's theorem.
\item We pick $m$ points $x_1,\ldots,x_m$ on $X$ and place $N_i$ pointlike D3-branes at each point,
such that $\sum_{i=1}^m N_i=N$. Then one can always solve uniquely for the warp factor $H$ in (\ref{Heqn}),
as a sum of Green's functions on $(X,g_X)$. In order that the supergravity approximation
to string theory be valid one requires all $N_i$ to be large.
\item One picks particular differential form representatives of the $B$ field and RR fields on $Y$,
in their appropriate cohomology classes determined by the marginal couplings,
and extends these over $X$ such that they satisfy the supergravity equations of motion.
The only non-flat field is $C_4$, whose field strength is given in terms of the warp
factor $H$ by (\ref{Gflux}). That the flat $B$ field and $C_2$ field on $Y$ may be extended as flat fields over $X$
is a topological fact. The differential forms are denoted $B^{\circ}$, $C^{\circ}_*$.
More precisely, $B^{\circ}$ and $C^{\circ}_2$ may be defined once the resolution $X$ is fixed,
whereas $C^{\circ}_4$ is a function of the D3-brane positions $x_1,\ldots,x_m$ and the metric on $X$.
Thus we should write $C^{\circ}_4(\{x_i\},[\omega_X])$, and choose solutions with fixed gauge
$C^{\circ}_4\mid_{\partial\mathcal{M}}$ at infinity.
\item One may add to these background differential forms any flat field that is compactly
supported, so that the gauge at infinity is fixed. We identify fields iff they differ
by a compactly supported gauge transformation. This leads to the group $H^2(X,Y;\R)/H^2_{\mathrm{free}}(X,Y;\Z)$
classifying the space of such $B$ fields, and the RR torus (\ref{RRXtorus}). These groups are clearly independent
of the choice of fixed background forms $B^{\circ}$, $C^{\circ}_*$. In principle one can also turn on
discrete torsion fields, which should be classified by K-theory.
\item The moduli are then: a choice of crepant resolution $X$, a K\"ahler class in $H^2(X;\R)$,
the choice of where one puts the pointlike D3-branes, and the $B$ field and RR field moduli described
in the last item. The supergravity backgrounds describe\footnote{One might
also consider crepant \emph{partial} resolutions, which if they admit Ricci-flat K\"ahler metrics
with appropriate conical behaviour near the residual singularities would describe
RG flows from the UV SCFT to more interesting (products of) SCFTs in the IR, together with some number of
Goldstone bosons. Such backgrounds, where explicit Ricci-flat K\"ahler metrics are known, were
studied in \cite{baryonic}.} an RG flow from the UV SCFT
dual to $(Y,g_Y)$ to a product of $\mathcal{N}=4$ SYM theories with gauge groups $SU(N_i)$ in the IR,
together with the Goldstone bosons of section \ref{central}.
\end{itemize}

\subsubsection{Baryon condensates}

The computation of baryon condensates in section \ref{condsection} may then be summarised as follows:
\begin{itemize}
\item Our starting point is to pick a smooth\footnote{More generally one can consider multiply-wrapped D3-branes, leading to
flat non-abelian gauge bundles over $\Sigma$, or singular/intersecting $\Sigma$. These form a
larger class of baryon operators, as discussed in section \ref{baryonsec}.} supersymmetric three-submanifold $\Sigma$ together with a
torsion line bundle $L$ over $\Sigma$. A D3-brane wrapped over $(\Sigma,L)$ is dual to a baryon
operator $\mathcal{B}(\Sigma,L)$, whose condensate in one of the above vacua we would like to compute.
\item The conjecture (\ref{rough}) is that the condensate $\langle\mathcal{B}(\Sigma,L)\rangle_p$, in a supergravity vacuum $p$ described above, is given by a path integral over Euclidean D3-branes
in the background with fixed boundary $(\Sigma,L)$, at a (any) point in $\R^4$. In practice we may compute
this semi-classically by evaluating the on-shell worldvolume action of such D3-branes. In this paper we have
focused on the contribution of a particular asymptotically conical divisor $D$ with boundary $\Sigma$.
More generally one should presumably integrate over a moduli space\footnote{For toric geometries note that there is a unique connected toric divisor $D\subset X$ with $\partial D=\Sigma$.} of such
minimal surfaces with boundary $\Sigma$, which also raises the issue of fermion zero modes and
whether one should consider only connected $D$. We shall discuss these matters further in
section \ref{discussionsection}. For now we focus on the contribution to
the semi-classical evaluation of the path integral of a smooth connected divisor $D$.
\item As shown in our previous paper \cite{baryonic}, the part of the Born-Infeld action that
is independent of the D3-brane worldvolume gauge field $M$ has precisely the correct divergence at large
$r$ to interpret $\exp(-S_{D3})$ as the VEV of an operator with conformal dimension equal to that of the
D3-brane wrapped on $(\Sigma,L)$. One may also perform a simple holographic renormalisation
of this part of the action. The condensate is identically zero if $D$ contains any of the
points $x_i$ where the background D3-branes are placed \cite{baryonic}.
\item One must next extend the torsion line bundle $L$ on $\Sigma$ to a line bundle $\mathcal{L}$ on
$D$. Given the topological assumptions (\ref{assumptions}), which for example hold for
toric varieties, this is always possible.
We have then shown that there is always a unique supersymmetric solution for the worldvolume gauge field $M$
which is $L^2$ normalisable, for any extension $\mathcal{L}$. This ensures that the gauge field does not contribute to the conformal
dimension result above (there is no renormalisation required), and that $M$ is flat at infinity. Moreover, the on-shell action is a topological invariant.
\item The imaginary part of the D3-brane action is described by the Chern-Simons terms.
Even classically the overall phase of the VEV of a baryon operator is not physical; but the relative phase
of the VEVs at different points in the moduli space \emph{is} physical, and it is this quantity that we
shall compute. Thus we must pick a base point $p_0$, which is a particular choice of smooth
supergravity vacuum, and compute the phase of the on-shell D3-brane action in a vacuum $p$
relative to the phase evaluated in the background $p_0$. In practice,
we study the case in which both $p$ and $p_0$ both lie in the same
chamber $C$, meaning that $X\cong X_0$ are isomorphic.
\item For fixed $X$ and choice of fixed background fields $B^{\circ}$, $C_0^{\circ}$,
$C_2^{\circ}$ and $C^{\circ}_4(\{x_i\},[\omega_X])$, the invariance of the D3-brane
action under compactly supported gauge transformations in section \ref{gaugetransf}
implies that the Chern-Simons action is a well-defined function of the $B$ field
and RR field moduli. These moduli consist of a point in $H^2(X,Y;\R)/H^2_{\mathrm{free}}(X,Y;\Z)$,
and a point in the RR torus (\ref{RRXtorus}). The value of the Chern-Simons
action certainly depends on the arbitrary choice of fixed background fields above.
However, under \emph{any} gauge transformation of the background fields, the exponentiated Chern-Simons
action changes by terms that depend only on the boundary data. Since the boundary data is fixed
and equal for all points in the moduli space,
if one computes the difference of Chern-Simons
actions, evaluated at any point $p$ in the moduli space and a fixed point $p_0$, respectively,
then this relative value is gauge-invariant modulo $2\pi i \Z$.
\item Finally, the choice of $\mathcal{L}$ is far from unique: for a fixed $L$ on $\Sigma$ there
are infinitely many extensions $\mathcal{L}(n)$ over $D$, labelled by a point in a lattice
$n\in\Lambda\cong H^2(D,\Sigma;\Z)$. Since there is a unique $L^2$ solution to the worldvolume gauge field equations of motion
for each $\mathcal{L}(n)$, in the Euclidean path integral one naturally sums over the lattice
$\Lambda$. This leads to a Riemann theta function, described in section \ref{theta}.
\end{itemize}


\subsection{Discussion}\label{discussionsection}

The results we have described in this paper leave a number of issues open to further study.
In this final subsection we discuss
some of the remaining problems.

Firstly, we encourage geometric analysts to prove the non-compact version of the
Calabi conjecture in section \ref{metricsection}. This 
is vital for the form of the supergravity moduli space we have described. Since submitting the first 
version of this paper to the archive, the conjecture has subsequently been proved 
in \cite{craigvanC} in the case that the K\"ahler class is compactly supported. 
The general case in which $[\omega]_Y$ is non-zero is thus still open, although we believe\footnote{We 
thank C. van Coevering and A. Futaki for discussions on this issue.} solving 
this is now just a technical problem. It would also be
interesting to understand in more detail how the classical VMS of section \ref{VMSsection}
compares to the supergravity moduli space of section \ref{section3}, especially in
its global structure. For example, the $B$ field is 
periodic\footnote{Recall that the conifold
may be realised as the IR fixed point of an RG flow induced via mass perturbation of the $\mathcal{N}=2$
$A_1$ orbifold theory, and the periodicity of the $B$ field
may be understood from a field theory point of view in terms of Seiberg duality of this
theory \cite{strassler}. Whether such an argument can be extended is not clear.}, 
while the FI parameters, over which the classical vacuum moduli space $\mathscr{M}$ fibers, are real numbers.

The metric fluctuations of section \ref{metricfluc} should certainly be studied
properly, giving a more complete fluctuation analysis than we have presented in section
\ref{fluctuationsection}. In particular the recent paper \cite{douglaswarped}, which
appeared whilst this article was being completed, will be very useful. The results
in sections \ref{fluctuationsection} and \ref{central} relate to the
Kaluza-Klein spectrum on general Sasaki-Einstein manifolds. It would
be interesting to undertake a general investigation of these spectra, 
and also to  obtain better control over the eigenvalues $\mu^A$ that arise in 
the asymptotic expansions.
It is also important to study further the identification of
massive vector multiplets in AdS$_5$
with anomalous baryonic currents that we discussed in section \ref{asection}. 
In particular, the key point that needs to be addressed is whether these currents belong to the 
Kaluza-Klein spectrum of AdS$_5\times Y$, as suggested by the results of this paper, or whether 
they correspond to highly massive states, like the Konishi current of ${\cal N}=4$ SYM. 
For instance, it would be interesting to see if it is possible to get a handle on these currents via a
field theory calculation.

An outstanding problem is to understand precisely how baryons in the classical field
theory are related to baryons, realised as wrapped D3-branes, in AdS$_5\times Y$. We have
given some idea of how complicated the latter are in section \ref{baryonsec}, and we refer
the reader back to that section for a reminder of the discussion. Particularly
difficult to understand are D3-branes wrapped on singular (or intersecting) $\Sigma$,
and time-dependent D3-branes. This is essentially a geometric problem. One would also like
to understand how the anomalous part of classical baryonic charge group $U(1)^{\chi-1}$ is
realised in terms of D3-branes wrapping $\Sigma$, with appropriate supersymmetric
gauge bundles, on $Y$: a 1-1 mapping between baryon operators and D3-brane states
implies there is such a realisation. Understanding this problem is probably a necessary prerequisite
to calculating VEVs of more general baryon operators. In section \ref{Gcouplingsection}
we have alluded to the fact that the full condensate probably involves also summing
over compact components. It is also important to address fermion zero modes, which would
give vanishing conditions. Another interesting question is whether there is any
hope that the gravity condensates may be reproduced by a field theory calculation.
In particular, it would be interesting to understand how theta functions may arise.

Finally, perhaps the most interesting remaining issue concerns the
$2b_4(X)$ massless pseudo-Goldstone modes. As we have explained, these massless
modes correspond to flat directions in the classical
moduli space. This is
different from the situation discussed in section \ref{anomaloussection},
where the RR moduli are instead axions which get ``eaten'' by the worldvolume gauge fields, via
a generalised Green-Schwarz mechanism. Notice that the existence of these massless fields
may be also understood from the complementary point of view of Kaluza-Klein reduction on (warped) Calabi-Yau
manifolds.  In particular, $b_2(X)$ of them are  \emph{K\"ahler moduli} of the non-compact Calabi-Yau, complexified by the RR $C_4$ moduli, which are expected to be classically massless.
In general, in Calabi-Yau compactifications a potential for massless modes can be generated by D-brane
instantons wrapping compact cycles in $X$. Thus an instanton-induced superpotential may
lift some of the moduli we have described.
Understanding how such mechanisms may work in the context of AdS/CFT is
clearly very interesting. Since instanton-induced effects are generally
proportional to the on-shell instanton action, the same reasoning as in section
\ref{Gcouplingsection} implies that the $b_3(Y)$ Goldstone bosons, which by
Goldstone's theorem are certainly massless, do not couple to such D-brane instantons.
Thus the $b_3(Y)$ massless fields should be massless after any such instanton effects
are taken into account; the remaining massless modes we have found are not protected, and
it would be interesting to try to understand if and how they may gain a (small) mass
via D-brane instanton effects. Correspondingly, it would be nice to understand the realisation of this mechanism directly in the gauge theory.


\subsection*{Acknowledgments}
\noindent We would like to thank J. Maldacena and E. Witten for several
discussions and insightful comments. We also thank O.~Aharony, C.~van~Coevering, G.~Dall'Agata, 
A.~Futaki, 
M.~Haskins, T.~Hausel, E.~Hunsicker, D.~Joyce, I.~Klebanov,
T.~Pacini, B.~Szendr\"oi, Y.~Tachikawa,  F.~Yagi and S.-T. Yau for useful discussions.
J. F. S. is supported by a Royal Society University Research Fellowship. D. M.
acknowledges support from  NSF grant PHY-0503584.

\appendix

\section{Closed and co-closed forms on cones}
\label{appendixA}

In this appendix we study $L^2$ closed and co-closed forms on cones. If $(W,g_W)$
is a compact Riemannian manifold then its cone $C(W)\cong \R^+\times W$ has metric
\bea
\diff\rho^2 + \rho^2 g_W
\eea
where $\rho>0$. We use the coordinate $\rho$, rather than $r$, since for applications in the main text
we will sometimes have $\rho=r$ but sometimes $\rho=1/r$. We will correspondingly need to study
forms that are $L^2$ on intervals of the form $[\rho_0,\infty)$ and $(0,\rho_0]$ for some (any) $\rho_0$ with
$0<\rho_0<\infty$. We assume that $C(W)$ has
even dimension $2n$, so that $W$ has dimension $2n-1$. The analysis below essentially follows
that in  \cite{cheeger, nagase, cheegerlong}.

Let $\theta$ be a $p$-form on $C(W)$ of the form
\bea
\label{formdecom}
\theta \, = \, g(\rho)\alpha + f(\rho)\diff \rho\wedge \beta~.
\eea
Here $\alpha$ and $\beta$ are pull-backs of forms on $W$, and thus are independent of $\rho$.
One easily computes
\bea
\label{dualtheta}
*\theta \, = \, (-1)^p\rho^{2n-2p-1}g\diff \rho\wedge *_W\alpha + \rho^{2n-2p+1}f*_W\beta~,\eea
and
\bea
\diff\theta &= &g^{\prime}\diff\rho\wedge\alpha+g\diff\alpha-f\diff\rho\wedge\diff\beta\\
\diff^\dagger\theta &=& \frac{g}{\rho^2}\diff_W^\dagger \alpha-\frac{f}{\rho^2}\diff\rho\wedge
\diff_W^\dagger \beta -\left[f^{\prime}+(2n-2p+1)\frac{f}{\rho}\right]\beta~.
\eea
The Laplacian $\Delta=\diff\cod+\cod\diff$ acting on $\theta$ is then
\bea
\label{laplace}
\Delta \theta&=& \left[-g^{\prime\prime}-(2n-2p-1)\frac{g^{\prime}}{\rho}\right]\alpha + \frac{g}{\rho^2}\Delta_W \alpha
-\frac{2g}{\rho^3}\diff\rho\wedge\diff_W^\dagger \alpha \nonumber\\
&+&\left[-f^{\prime\prime}+(2n-2p+1)\left(\frac{f}{\rho^2}-\frac{f^{\prime}}{\rho}\right)\right]\diff\rho\wedge\beta+\frac{f}{\rho^2}\diff\rho\wedge\Delta_W\beta-\frac{2f}{\rho}\diff\beta~.
\eea
Here $*_W$, $\diff_W^\dagger  = (-1)^p *_W \diff \, *_W $ and $\Delta_W$ are the Hodge operator, codifferential and Laplacian on $(W,g_W)$,
respectively.  Note this corrects the formula in \cite{cheeger}.

An arbitrary $p$-form on $C(W)$ may be written
\bea
\theta \, =\, \alpha(\rho)+\diff\rho\wedge\beta(\rho)
\eea
where $\alpha(\rho)$, $\beta(\rho)$ are forms on $W_{\rho}\subset C(W)$. For fixed $\rho$, we may expand
$\alpha(\rho)$ and $\beta(\rho)$ in terms of eigenmodes of
the Laplacian $\Delta_W$
\bea\label{expansionalpha}
\alpha(\rho) &=& \!\!\!\!\sum_{\;\;\mu\in\mathrm{Spec}\Delta_W^{(p)}} \!\!g_{\mu}(\rho) \alpha_{\mu}\\\label{expansionbeta}
\beta(\rho)&=& \!\!\!\!\sum_{\;\;\lambda\in\mathrm{Spec}\Delta_W^{(p-1)}} \!\! f_{\lambda}(\rho) \beta_{\lambda}
\eea
where
\bea
\Delta_W^{(p)} \alpha_{\mu} &=& \mu \, \alpha_{\mu}\\
\Delta_W^{(p-1)}\beta_{\lambda}&=&\lambda\, \beta_{\lambda}~.
\eea
We wish to classify harmonic $p$-forms $\theta$ on the cone that are both closed and co-closed.

Suppose first that $\beta(\rho)=0$. $\diff\theta=0$ immediately gives\footnote{Here and in the rest of this appendix,
a prime denotes derivative with respect to $\rho$.}
$g_{\mu}^{\prime}\alpha_{\mu}=0$ and $g_\mu\diff\alpha_{\mu}=0$ for each mode
$\alpha_\mu$, which implies $g_{\mu}=c_\mu$ is
constant. $\cod\theta=0$ implies that $\diff_W^\dagger \alpha_{\mu}=0$. Thus
$\alpha_{\mu}$ is both closed and co-closed on the link $(W,g_W)$ and thus harmonic, and so $\mu=0$.

Suppose instead that $\alpha(\rho)=0$. $\cod\theta=0$ implies that
$\diff_W^\dagger \beta_{\lambda}=0$, while $\diff\theta=0$ implies that $\diff\beta_{\lambda}=0$.
Thus again $\beta_{\lambda}$ is harmonic on $(W,g_W)$, and so $\lambda=0$. The equation $\cod\theta=0$
also implies
\bea
\rho f_0^{\prime}+(2n-2p+1)f_0 \, =\, 0
\eea
which has general solution
\bea
f_0 \, = \, c\rho^{-2n+2p-1}~.
\eea

More generally, focusing on an eigenmode $\alpha_\mu$ in the equation $\diff\theta=0$  gives $\alpha_{\mu}\propto \diff\beta_{\lambda}$
for some $\lambda$. Without loss of generality we may take $\alpha_{\mu}=\diff\beta_{\lambda}$.
Applying $\Delta_W$ to this relation gives $\lambda=\mu$. We then have the relation
\bea
g_{\mu}^{\prime}=f_{\mu}~.
\eea
The equation $\cod\theta=0$ implies either $f_{\mu}=0$, in which case $g_{\mu}$ is constant
and we reduce to the solution already discussed above, or else $\diff_W^\dagger \beta_{\mu}=0$.
$\cod\theta=0$ then implies
\bea
g_\mu\, \diff_W^\dagger \alpha_{\mu}\, =\, \rho^2\left[f_{\mu}^{\prime}+(2n-2p+1)\frac{f_\mu}{\rho}\right]\beta_{\mu}
\eea
or, equivalently,
\bea
\rho^2g_{\mu}^{\prime\prime}+\rho(2n-2p+1)g_{\mu}^{\prime}-\mu g_{\mu}\, =\, 0~.
\eea
This has general solution
\bea\label{gensoln}
g_\mu \,=\, c_+\rho^{p-n+\nu_p}+c_-\rho^{p-n-\nu_p}
\eea
where $c_{\pm}$ are constants and we have defined
\bea
\nu_p = \sqrt{(p-n)^2+\mu}~.
\eea
One might also worry that there is an additional solution when the two solutions in
(\ref{gensoln}) coincide. This occurs when $\nu_p=0$, which implies (since necessarily
$\mu\geq0$) $\mu=0$, $n=p$, leading
to the equation
\bea
\rho g_0^{\prime\prime}+g^{\prime}_0 \,=\, 0
\eea
which has general solution
\bea
g_0\, =\, c_1+c_2\log\rho~.
\eea
However, note that if $\mu=0$ then $\alpha_0$ is harmonic, and the relation $\alpha_0=\diff\beta_0$
is impossible by the Hodge decomposition on $(W,g_W)$ (alternatively, $\beta_0$ is harmonic
and thus closed, so $\alpha_0=0$).

To summarise,    any closed and co-closed $p$-form on $C(W)$ may be written as a convergent sum
of the following three types of modes
\bea
\mathrm{(I)}&:& \alpha_0\\
\mathrm{(II)}&:& \rho^{-2n+2p-1}\diff\rho\wedge\beta_{0}\\
\mathrm{(III)}^{\pm}&:& \rho^{p-n\pm\nu_p}\diff\beta_{\mu}+(p-n\pm\nu_p)\rho^{p-n-1\pm\nu_p}\diff\rho\wedge\beta_{\mu}~.
\eea
Here $\alpha_0$, $\beta_0$ are harmonic $p$-forms and $(p-1)$-forms, respectively,
while $\beta_\mu$ in mode III is a co-closed $(p-1)$-form which is an eigenfunction
of $\Delta_W$ with eigenvalue $\mu$. Note that $\mu>0$ necessarily for modes of type III.

It is straightforward to compute the pointwise square norms
$\|\theta\|^2=\frac{1}{p!}\theta_{i_1\ldots i_p}\theta^{i_1\ldots i_p}$ of the above  modes. For a general $p$-form $\theta$ as in
(\ref{formdecom}) one obtains
\bea
\|\theta \|^2 \! & = & \!\rho^{-2p} \Big[ g^2 \| \alpha \|^2_W + \rho^2 f^2 \|\beta \|_W^2  \Big]~.
\eea
Here $\| \cdot\|_W$ denotes the pointwise norm on $(W,g_W)$.
The pointwise square norms of the above modes are then given by a non-zero function
on $W$ times the function of $\rho$ given in Table \ref{modes}.
\begin{table}[h!]
\centerline{
\begin{tabular}{|c|c|c|c|}
\hline
mode & $\|\theta\|^2$ & $L^2_0$ & $L^2_{\infty}$\\
\hline
\hline
I & $\rho^{-2p}$ & $p<n$ & $p>n$ \\
\hline
II & $\rho^{-4n+2p}$ & $p>n$ & $p<n$\\
\hline
III$^{+}$ & $\rho^{-2n+2\nu_p}$ & yes & no\\
\hline
III$^{-}$ & $\rho^{-2n-2\nu_p}$ & no & yes\\
\hline
\end{tabular}
}
\caption{Summary of the square-integrability of the various modes.
}
\label{modes}
\end{table}
Using these formulae it is a simple matter to determine which modes are $L^2$ near to $\rho=0$ and $\rho=\infty$.
Fix some $\rho_0$ with $0<\rho_0<\infty$.
If the integral of the pointwise square norm of $\theta$ over $(0,\rho_0]\times W$ is finite
then we shall say that $\theta$ is $L^2_0$. On the other hand, if the integral of the pointwise square norm of $\theta$  over
 $[\rho_0,\infty)\times W$ is finite
then we shall say that $\theta$ is $L^2_{\infty}$. The relevant integrals take the form
\bea
\int F(\rho)\rho^{2n-1}\diff\rho~.
\eea
In particular, if $F(\rho)=\rho^{-2n+\gamma}$ then $F$ is integrable on $(0,\rho_0]$ iff $\gamma>0$ and
is integrable on $[\rho_0,\infty)$ iff $\gamma<0$.


\section{Eigenvalues of Laplacians on $(Y,g_Y)$}
\label{eigensection}

In this section we derive some formulae relating the one-forms $\beta^{(1)A}$ in the main text,
which are eigenforms of the Laplacian $\Delta_Y$, to \emph{scalar} eigenfunctions
on $(Y,g_Y)$. These formulae are used in section \ref{asection}.

Suppose that $\psi$ is an $L^2$ harmonic two-form on $(X,g_X)$. As discussed in the main text, there is
an asymptotic expansion of $\psi$ with leading term
\bea
\psi \, \sim \, \diff \left( r^{-1-\nu}\beta\right)
\eea
where $\beta$ is a co-closed one-form on $(Y,g_Y)$ satisfying
\bea\label{laplaceboy}
\Delta_Y \beta \, = \, \mu \, \beta
\eea
and
\bea
\nu \, = \,\sqrt{1+\mu}~.
\eea
As argued in the main text, $\psi$ is $(1,1)$ and primitive, namely 
\bea\label{contract}
\omega_X\, \lrcorner\, \psi \, = \, 0~,
\eea
where $\omega_X$ is the K\"ahler form on $X$. Since asymptotically
\bea
\omega_X \, \sim \, \omega_{C(Y)}\,=\, \frac{1}{2}\diff(r^2\eta)
\eea
the equation (\ref{contract}) gives, from its leading term,
\bea\label{iden2}
\diff\beta\, \lrcorner \, \diff\eta \, = \, 2(1+\nu)\beta\, \lrcorner\, \eta~.
\eea
Defining the function
\bea
f \, = \, \beta \, \lrcorner\, \eta
\eea
one can also prove the identity
\bea\label{iden1}
\Delta_Y f \, = \, \mu\, \beta\, \lrcorner\, \eta - \diff\beta \, \lrcorner\, \diff\eta~.
\eea
This is proven using (\ref{laplaceboy}), together with the fact that $\eta$ is a Killing one-form, and
the Weitzenb\"ock formula
\bea
\Delta_Y\beta\, = \, -\Delta^i\Delta_i\beta \, + \, \mathrm{Ric}_Y\cdot \beta~.
\eea
On a Sasaki-Einstein five-manifold $\mathrm{Ric}_Y = 4g_Y$. Combining (\ref{iden2}) and
(\ref{iden1}) one obtains
\bea
\Delta_Y f \, = \, E \, f
\eea
where
\bea
E \, = \, \mu - 2 - 2\sqrt{1+\mu}~.
\eea
This last formula is used in section \ref{asection}.


\section{Flat form fields}\label{appendixB}

In this appendix we review the classification of flat form fields, up to gauge equivalence, on
a spacetime $\mathcal{M}$.
Such fields play an important role throughout the paper.

A flat $(p-1)$-form potential $C$ has, by definition, field strength $G=\diff C=0$. Since it
is the field strength $G$ that generally enters the supergravity equations,
 one may typically turn on flat fields
without altering the equations of motion.
The potential $C$ transforms under a form of gauge transformation via
\bea\label{small}
C\rightarrow C+\diff\lambda\eea
where $\lambda$ is any $(p-2)$-form on spacetime $\mathcal{M}$.
In fact, more generally $C$ also transforms under large gauge transformations
\bea\label{large}
C\rightarrow C+ \frac{2\pi}{\mu} a
\eea
where $a$ is any closed $(p-1)$-form with \emph{integral} periods. These reduce
to (\ref{small}) when the cohomology class of $a$ is trivial.
The constant $\mu$
depends on the normalisation of the potential, and may be determined from the Wess-Zumino
couplings of an object coupling electrically to $C$. Specifically, the latter is given by
\bea
S_{WZ}=\mu\int C~.
\eea
The transformations (\ref{large}) then leave the exponentiated  action
$\exp(iS_{WZ})$ invariant.
This leads to the group $H^{p-1}(\mathcal{M};\R)/H^{p-1}_{\mathrm{free}}(\mathcal{M};\Z)$,
classifying the space of closed potentials mod gauge transformations. Here $H^{p-1}_{\mathrm{free}}(\mathcal{M};\Z)$, which is the image of $H^{p-1}(\mathcal{M};\Z)$ in $H^{p-1}(\mathcal{M};\R)$, is the group of large gauge transformations.

However, in general not all flat form fields arise this way. The $p$-form field strength $G$
satisfies a form of Dirac quantisation, and consequently defines an element of
$H^p(\mathcal{M};\Z)$. A flat $p$-form field on a spacetime $\mathcal{M}$ then lies in
the kernel of the map $H^p(\mathcal{M};\Z)\rightarrow H^p(\mathcal{M};\R)$.
But this kernel is by definition the torsion component $H^p_{\mathrm{tor}}(\mathcal{M};\Z)$.
Such a torsion $p$-form field strength is not described \emph{globally} by a closed $(p-1)$-form potential. Indeed,
the short exact coefficient sequence
\bea
0\rightarrow \Z\rightarrow\R\rightarrow U(1)\rightarrow 0
\eea
induces in a standard way the long exact sequence
\bea \label{coeff} \cdots\longrightarrow&&H^{p-1}(\mathcal{M};\Z)\longrightarrow H^{p-1}(\mathcal{M};\R)\longrightarrow
H^{p-1}(\mathcal{M};U(1))\stackrel{\beta}{\longrightarrow} \nonumber \\
&&H^p(\mathcal{M};\Z)\longrightarrow H^p(\mathcal{M};\R)\longrightarrow\cdots\eea
which implies that $H^p_{\mathrm{tor}}(\mathcal{M};\Z)\cong \beta(H^{p-1}(\mathcal{M};U(1)))$.
 Here $\beta$ is the so-called Bockstein map. In fact
it is $H^{p-1}(\mathcal{M};U(1))$ which classifies, up to gauge equivalence, flat form fields with a field strength of
degree $p$. An element of this group may be regarded as
specifying the \emph{holonomy} of the potential over closed $(p-1)$-cycles.
Thus, if $\gamma$ is a chain representing a $(p-1)$-cycle $[\gamma]\in H_{p-1}(\mathcal{M};\Z)$,
we may define the holonomy of the potential $C$ over $\gamma$ to be
\bea
\exp\left(i\mu\int_{\gamma} C\right)~.
\eea
The holonomy of
a flat potential defines a homomorphism $H_{p-1}(\mathcal{M};\Z)\rightarrow U(1)$.
Since $U(1)$ is a divisible group, the group of such homomorphisms is
$H^{p-1}(\mathcal{M};U(1))$. The long exact coefficient sequence (\ref{coeff})
implies that in general the group $H^{p-1}(\mathcal{M};U(1))$
is disconnected, with the number of connected components being the number of
elements in $H^p_{\mathrm{tor}}(\mathcal{M};\Z)$. Thus the discussion in the previous paragraph
misses the flat fields that have torsion fluxes $[G]\in H^p_{\mathrm{tor}}(\mathcal{M};\Z)$, which are not
described globally by a closed $(p-1)$-form potential $C$.

In this paper we shall largely  not include the torsion flat fields in the discussion. An
important exception to this is in section \ref{condsection}.
Another reason  for ignoring the torsion classes is that, although we are treating RR
fields in terms of cohomology in this paper, more precisely they are classified by K-theory \cite{MW}.
These differ in their torsion.


\end{document}